\documentclass[traditabstract]{aa}
\usepackage[varg]{txfonts}
\usepackage{mathrsfs}
\usepackage[latin1]{inputenc}
\usepackage{txfonts}
\usepackage{calc}
\usepackage{graphicx}
\usepackage{natbib}
\bibpunct{(}{)}{;}{a}{}{,}
\usepackage{hyperref}
\usepackage{multirow}
\usepackage{booktabs}
\def\simlt{\lower.5ex\hbox{$\; \buildrel < \over \sim \;$}}
\def\simgt{\lower.5ex\hbox{$\; \buildrel > \over \sim \;$}}
\def\l#1{\left#1}
\def\r#1{\right#1}

\def\bm#1{{\bf #1}}

\def\tell#1{{\tilde{\ell}}^{\,#1}}

\newcommand{\lenstwo}{{\sf lenS$^{2}$HAT}}
\newcommand{\lenspix}{{\sf LensPix}}
\newcommand{\stwo}{S$^{2}$HAT}

\newcommand{\ellmax}{\ell_{max} }
\newcommand{\lensker}[3] {{\cal K}_{\tilde{\ell}^{\,#1}}(\ell^{\,#2}, \ell^{\,#3}) }

\newcommand{\elle}{\ell^{\,E}}
\newcommand{\ellx}{\ell^{\,X}}
\newcommand{\ellphi}{\ell^{\,\Phi}}
\newcommand{\errfield}{\mathscr{E}^{X}_{\ell_{max,1},\ell_{max,2}}}

\titlerunning{High-precision simulations of weak lensing effect on CMB polarization}

\begin{document}
\title{High-precision simulations of the weak lensing effect on cosmic microwave background polarization}
\author{Giulio Fabbian\inst{1}\thanks{\email{gfabbian@apc.univ-paris-diderot.fr}} \and Radek Stompor\inst{1}\thanks{\email{radek@apc.univ-paris-diderot.fr}}}
\institute{AstroParticule et Cosmologie, Univ Paris Diderot, CNRS/IN2P3, CEA/Irfu, Obs de Paris, Sorbonne Paris Cit\'e, France  \label{inst1}}
\date{\today}
\abstract{We studied the accuracy, robustness and self-consistency of  pixel-domain simulations of  the gravitational lensing effect on the primordial cosmic microwave background (CMB) anisotropies
due to the large-scale structure of the Universe. In particular, we investigated the dependence of the precision of the results precision on some crucial parameters of these techniques
and propose a semi-analytic framework  to determine their values so that the required precision is a priori assured and the numerical workload simultaneously optimized.
Our focus was on the $B$-mode signal but we also discuss other CMB observables, such as the total intensity, $T$, and $E$-mode polarization, emphasizing 
differences and similarities between all these cases.
Our semi-analytic considerations are backed up by extensive numerical results. Those are obtained using a code, nicknamed \lenstwo\  -- for lensing using scalable spherical harmonic transforms (\stwo) --
which we have developed in the course of this work. The code implements  a version of the previously described pixel-domain approach and permits performing the simulations at very high resolutions and data
volumes, thanks to its efficient parallelization provided by  the \stwo\ library -- a parallel library for calculating of the spherical harmonic transforms. 
The code is made publicly available.}

\maketitle

\section{Introduction}
The cosmic microwave background (CMB) anisotropies in both temperature and polarization are one of the most studied signals in cosmology and  one of the major available sources of constraints of the early-Universe physics. 
After having decoupled from matter and set free at the time of recombination, CMB photons propagated nearly unperturbed throughout the Universe. The large-scale structures (LSS) emerging in the Universe in the post-recombination period have however left their imprint on them which are referred to as secondary anisotropies. In particular, the gravitational pull of the growing matter inhomogeneities has deviated the paths of primordial CMB photons, modifying somewhat the pattern of the CMB anisotropies observed today. This weak lensing effect on the CMB (see \citet{lewis2006} for an extensive review) therefore offers a unique probe of the matter distribution at intermediate redshift where the forming LSS were still in the nearly-linear regime. Because this depends on the cumulative matter distribution in the Universe, it is expected to be particularly efficient in constraining the properties of all the parameters affecting the growth of LSS, such as neutrino masses and dark energy physics \citep{deputter2009,das2012,hall2012}.\\* 

The first  observational evidence of the CMB lensing signal had been indirect and obtained through cross-correlation  of the CMB maps with high-redshift mass tracers \citep{smith2007,hirata2008}. More recently, more direct measurements have become available, thanks to the latest generation of high-precision-and-resolution ground-based CMB temperature experiments, which have collected high-quality data and made possible a direct reconstruction of the power spectra of this deviation using CMB alone \citep{das2011,vanengelen2012}. Even more recently, this has been further elaborated on by the Planck results  based on the first 15 months of the total intensity data collected by the mission~\citep{PlanckLensing}.\\*
The forthcoming next generation of low-noise CMB polarization experiments such as EBEX \citep{ebex}, POLARBEAR \citep{kermish2012}, SPTpol \citep{sptpol}, and ACTpol \citep{actpol} and their future upgrades \citep[e.g, POLARBEAR-II,][]{tomaru2012} will be able to target a CMB observable most affected by weak lensing -- the B-mode polarization. Indeed, primordial CMB gradient-like polarization ($E$-modes) is  converted into curl-like polarization ($B$-modes) by gravitational lensing
~\citep{ZaldarriagaSeljak1998} and is expected to completely dominate the primordial signal at least at small angular scales. The lensing-generated $B$-modes are interesting because of their sensitivity to the large-scale structure distribution, but also because they are the main contaminant of any primordial $B$-modes signal, which is expected in many models of the very early Universe, and which is one of the major goals of the current and future CMB observations.  Since sensitivities of the CMB polarization arrays are rapidly improving, the experiments aiming at setting constraints on values of the tensor-to-scalar ratio parameter $r \lesssim 10^{-2}$ are expected to be ultimately limited by the lensing signal~\citep[e.g.,][]{Errard2012}. This acts as an extra noise source with a  white spectrum shape on large scales and an amplitude of approximately $5\mu K$-arcmin, which could in principle be separated from the primordial signal with the help of an accurate de-lensing procedure~\citep{kesden2002,seljak2004,smith2012}.

The high quality of forthcoming datasets requires the development, testing and validation through simulations of data-analysis tools capable of fully exploiting the amount of information present there. An important
part of this effort involves simulating very accurate, high-resolution maps of the CMB total intensity and polarization,  covering a large fraction of the sky  and with lensing effects included.
The relevant approaches have been studied in the past~\citep[e.g.][]{lewis2005,basak, wandelt} and  resulted in devising and demonstrating an overall framework for such simulations, as well as in two publicly available numerical codes~\citep{lewis2005, basak}.  Because the 
computations involved in such a procedure are inherently very time-consuming, the  proposed implementations of those ideas unavoidably involve trade-offs between calculation precision and their feasibility, giving rise to a number of problems, practical and more fundamental, which need to be carefully resolved to ensure that these techniques produce high-quality, reliable results. 
The main objective of this paper is  to provide comprehensive answers to some of these problems, with special emphasis on those arising  in the context of  high-precision and-reliability 
simulations  of the B-mode component of the CMB  polarization signal. 

\section{Simulating weak lensing of the CMB}
\subsection{Algebraic background}
The CMB radiation is completely described by its brightness temperature and polarization fields on the sky, $T(\vartheta,\varphi)$ and $P(\vartheta,\varphi)$. Since both fields are (nearly) Gaussian, they are characterized by their power spectra after their harmonic expansion in a proper basis. Temperature is a scalar field and can be conveniently expanded in terms of scalar spherical harmonics,

\begin{equation}
T(\vartheta,\varphi)=\sum_{l=0}^{l_{max}}\sum_{m=-l}^{l}T_{lm}Y_{lm}(\vartheta,\varphi), 
\label{eqn:tempExp}
\end{equation}

\noindent 
while polarization is described by  the Stokes parameters Q and U, which are coordinate-dependent objects, that behave like a spin-2 field on the sphere under rotations~\citep{seljak-zaldarriaga, kks}. The  polarization 
field must therefore be expanded in terms of spin-2 spherical harmonics, $_{\pm2}Y_{lm}(\vartheta,\varphi)$,
\begin{eqnarray}
P(\vartheta,\varphi) &=& (Q+iU)(\vartheta,\varphi)\\ \nonumber 
&=& \sum_{l=0}^{l_{max}}\sum_{m=-l}^{l}-(_{2}E_{lm}+i _{2}B_{lm}) _{2}Y_{lm}(\vartheta,\varphi),
\label{eqn:polExp}
\end{eqnarray}
\noindent
where $_{2}E_{lm}$ and $_{2}B_{lm}$ are the gradient and curl harmonic components of a spin-2 field, whose general definitions for and arbitrary spin-s field are
\begin{eqnarray}
_{|s|}E_{lm} & \equiv& -\frac{1}{2}\left(_{|s|}a_{lm}+(-1)^{s}_{-|s|}a_{lm}\right) \\
i_{|s|}B_{lm} & \equiv& -\frac{1}{2}\left(_{|s|}a_{lm}-(-1)^{s}_{-|s|}a_{lm}\right). \nonumber
\end{eqnarray}

\noindent
Weak gravitational lensing shifts the light rays coming from an original direction $\vec{\hat{n}}$ on the last scattering surface to the observed direction $\vec{\hat{n}^\prime}$, inducing a mapping between the two directions through the so-called displacement field $\vec{d}$ , i.e., for a CMB observable $X\in\{T,Q,U\}$
\begin{equation}
\tilde{X}(\vec{\hat{n}})=X(\vec{\hat{n}}^\prime )=X(\vec{n}+\vec{d}).
\label{eq:lens}
\end{equation}
Hereafter, we use a tilde to denote a lensed quantity, we also use a tilde over a multipole number of a lensed quantity, i.e., $\tell{}$, to distinguish it from a multipole number of its unlensed counterpart.

The displacement field is a vector field on the sphere and can be decomposed into a gradient-free and a curl-free component. In most cases we can neglect the gradient-free component and consider the displacement field $\vec{d}$ as the gradient of the so-called lensing potential $\Phi(\vartheta,\varphi)$, the projection of the 3D gravitational potential $\Psi$ on the 2D unit sphere. This quantity can be computed with Boltzmann codes (e.g. CAMB\footnote{\url{http://camb.info}} or CLASS\footnote{\url{http://lesgourg.web.cern.ch/lesgourg/class.php}}), from galaxy surveys or N-body simulations \citep{carbone2008, das-bode},

\begin{equation}
\Phi(\vec{n})\equiv -2\int_{0}^{\eta_{*}}d_{A}\eta \frac{d_{A}(\eta_{*}-\eta)}{d_{A}(\eta)d_{A}(\eta^{*})}\Psi(\eta,\vec{n}).
\end{equation}

\noindent
Here $\eta_{*}$ is the comoving distance to the last scattering surface, $\eta$ is the co-moving distance, $d_{A}$ is the co-moving angular diameter distance. The lensing potential is expected to be correlated on a large scale with temperature anisotropies and $E$-modes of polarization through the integrated Sachs-Wolfe effect; this correlation mainly affects the large angular scales and is of the order of  $1\%$ at $\ell\approx100$ and will thus be neglected in the following analysis.  \\*
\noindent
Since the lensing potential is a scalar function and can be expanded into canonical spherical harmonics, its gradient (a spin-1 curl-free field) can be easily computed in the harmonic domain with a spin-1 spherical harmonic transform (SHT):

\begin{equation}
_{1}E_{lm}=\sqrt{l(l+1)}\Phi_{lm} \qquad _{1}B_{lm}=0.
\end{equation}

\subsection{Pixel-domain simulations}\label{sect:pixdomainsims}

\subsubsection{Basics}\label{s3ect:basics}

Because typical deviations of CMB photons are on the order of few arcminutes (although coherent over the degree scale), we can work in the Born approximation, i.e., considering this deviation as constant between $\vec{\hat{n}}$ and $\vec{\hat{n}^\prime}$, and evaluate the displaced field along the unperturbed direction.\\*
In practice this means that to compute the lensed CMB at a given point it is sufficient to compute the unlensed CMB at another position on the sky. This observation provides
the basis for the pixel-based approaches to simulating lensing effects of the CMB maps. For every direction on the sky corresponding to a pixel center these methods 
first identify the displaced direction and then compute the corresponding sky signal value, which is used to replace the original value at the pixel center.
The implementations of this approach typically involve the following main steps \citep{lewis2005,basak, wandelt}:
\begin{enumerate}
\item Generating a random realization of the harmonic coefficients of the unlensed CMB map and its synthesis.
\item Generating a random realization of the harmonic coefficients of the lensing potential and then of the spin-1 displacement field in the harmonic domain. Synthesizing the displacement field.
\item Sampling the displacement field at pixel centers and, for each of them, computing the coordinates of a displaced direction on the sky using the spherical triangle identities on the sphere.

Defining $\alpha$ as the angle between the displacement vector and the $\vec{e_{\vartheta}}$ versor, such that $\vec{d}=d\cos\alpha\,\vec{e_{\vartheta}}+d\sin\alpha\,\vec{e_{\varphi}}$, the value of a lensed field, i.e., $T$, $Q$ and $U$, in a direction $(\vartheta,\varphi)$ is given by the unlensed field at $(\vartheta^\prime,\varphi+\Delta\varphi)$ where,
\begin{eqnarray}
\cos\vartheta^\prime &=& \cos d\cos\vartheta-\sin d\sin\vartheta\cos\alpha \\ 
\sin\Delta\varphi &=& \frac{\sin\alpha\sin d}{\sin\vartheta^\prime}.
\end{eqnarray}

\item Computing temperature and polarization fields at displaced positions.\\* 

\item Re-assigning the temperature and polarization from the displaced to new positions to create the simulated lensed map sampled on the original grid.
For the polarization,  we need also to multiply the lensed field by an extra factor taking into account the different orientation of the basis vector at the two points. Calling $\gamma$ the difference between the angles between $\vec{e_{\vartheta}}$ and the geodesic connecting the two points, and defining
\begin{eqnarray}
A&=&\tan\alpha^\prime = \frac{d_{\varphi}}{d\sin d\cot\vartheta + d_{\vartheta}\cos d} \\ 
e^{2i\gamma}&=&\frac{2(d_{\vartheta}+d_{\varphi}A)^{2}}{d^{2}(1+A^{2})} -1 +\frac{2i(d_{\vartheta}+d_{\varphi}A)(d_{\varphi}-d_{\vartheta}A)}{d^{2}(1+A^{2})},
\end{eqnarray}

\noindent
the lensed polarization field becomes 
\begin{equation}
\tilde{P}(\vartheta,\varphi)=e^{2\gamma i}P(\vartheta^\prime,\varphi^\prime).
\end{equation} 
\item Smoothing and, potentially, re-pixelizing the lensed map to match a particular experimental resolution, if needed.
\end{enumerate}

\subsubsection{Challenges and goals}\label{sect:challenges}
There are two main, closely intertwined challenges involved in implementing the approach detailed in the previous section.
The first one is related to the bandwidths of fields used in, or produced as a result of, the calculation, and in particular to the need of
 imposing those on the fields, which are
either naturally not band-limited or are band-limited but have too high bandwidths to make them acceptable from the computational efficiency point of view.
The other challenge arises from step 4 of the algorithm: the displaced directions do not correspond in general to pixel centers of any 
iso-latitudinal  grid on the sphere,  and thus the lensed values of the CMB signal cannot be computed with the aid of a fast SHT algorithm 
and a more elaborated, and computationally costly approach is needed.

We emphasize that both these problems should be looked at from the perspective of the efficiency of the numerical
calculations as well as accuracy of the produced results. We discuss them in some detail below.

\begin{figure*}[!htbp]
\centering
\includegraphics[width=.325\textwidth]{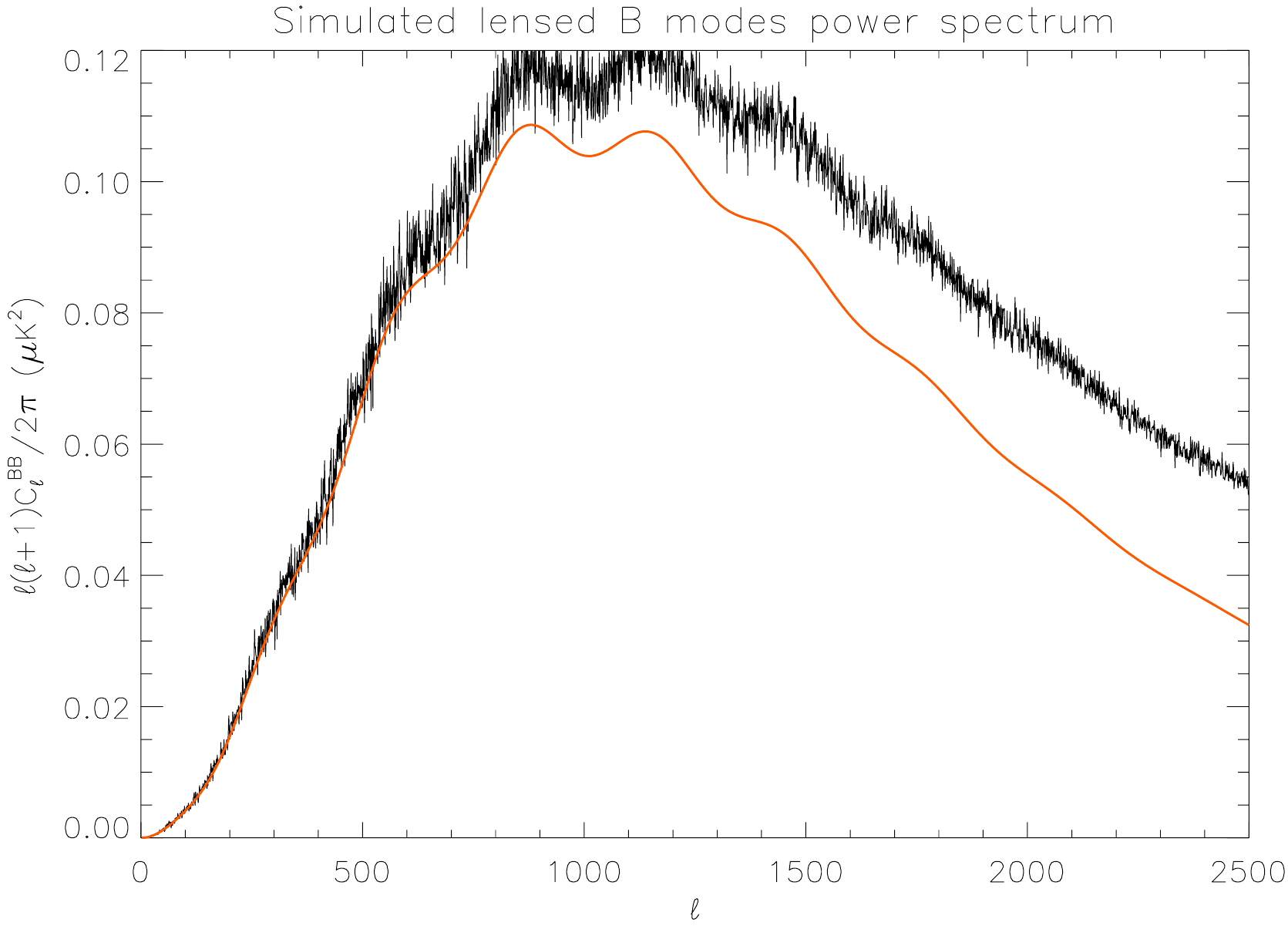}
\includegraphics[width=.325\textwidth]{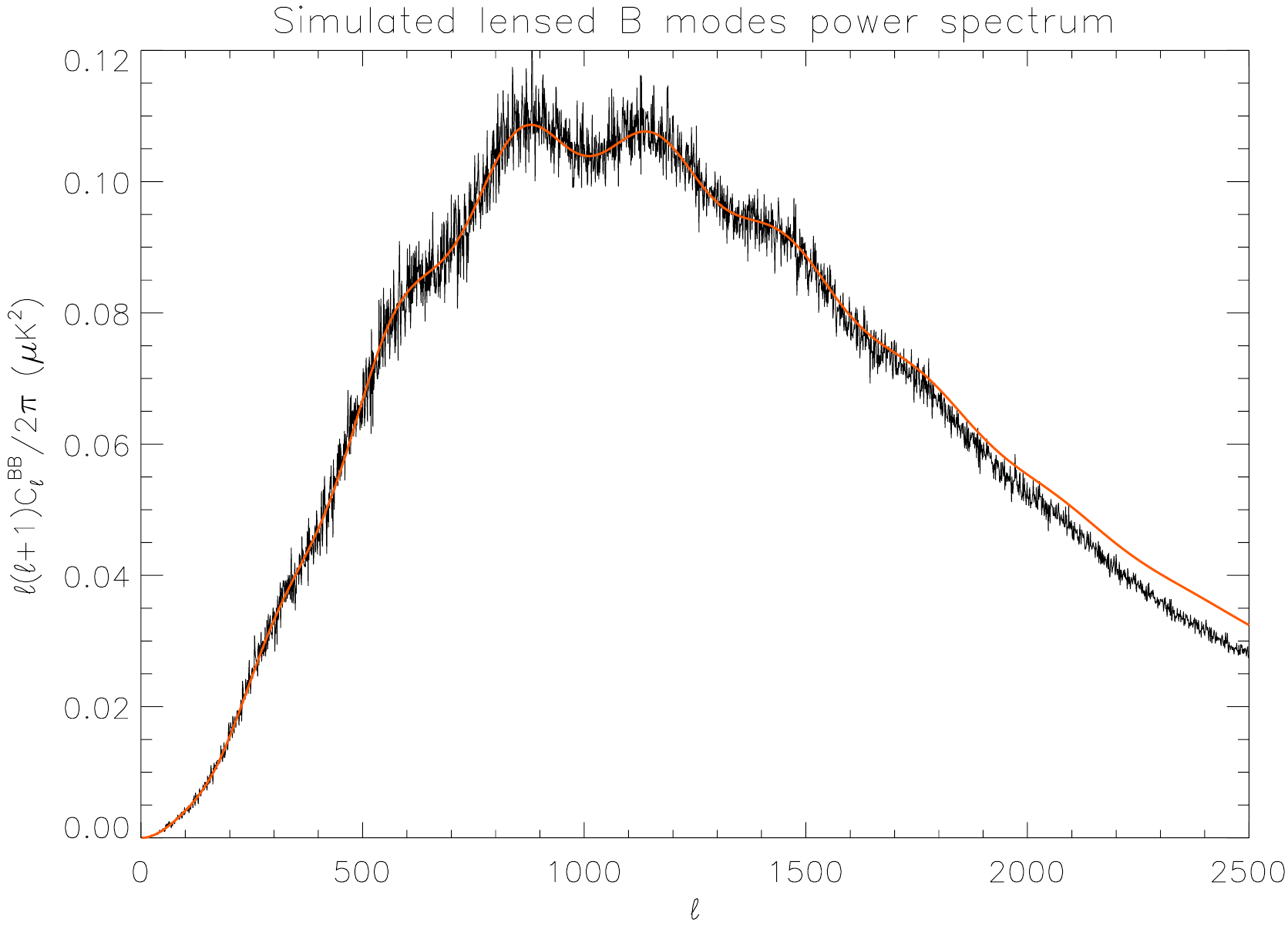}
\includegraphics[width=.325\textwidth]{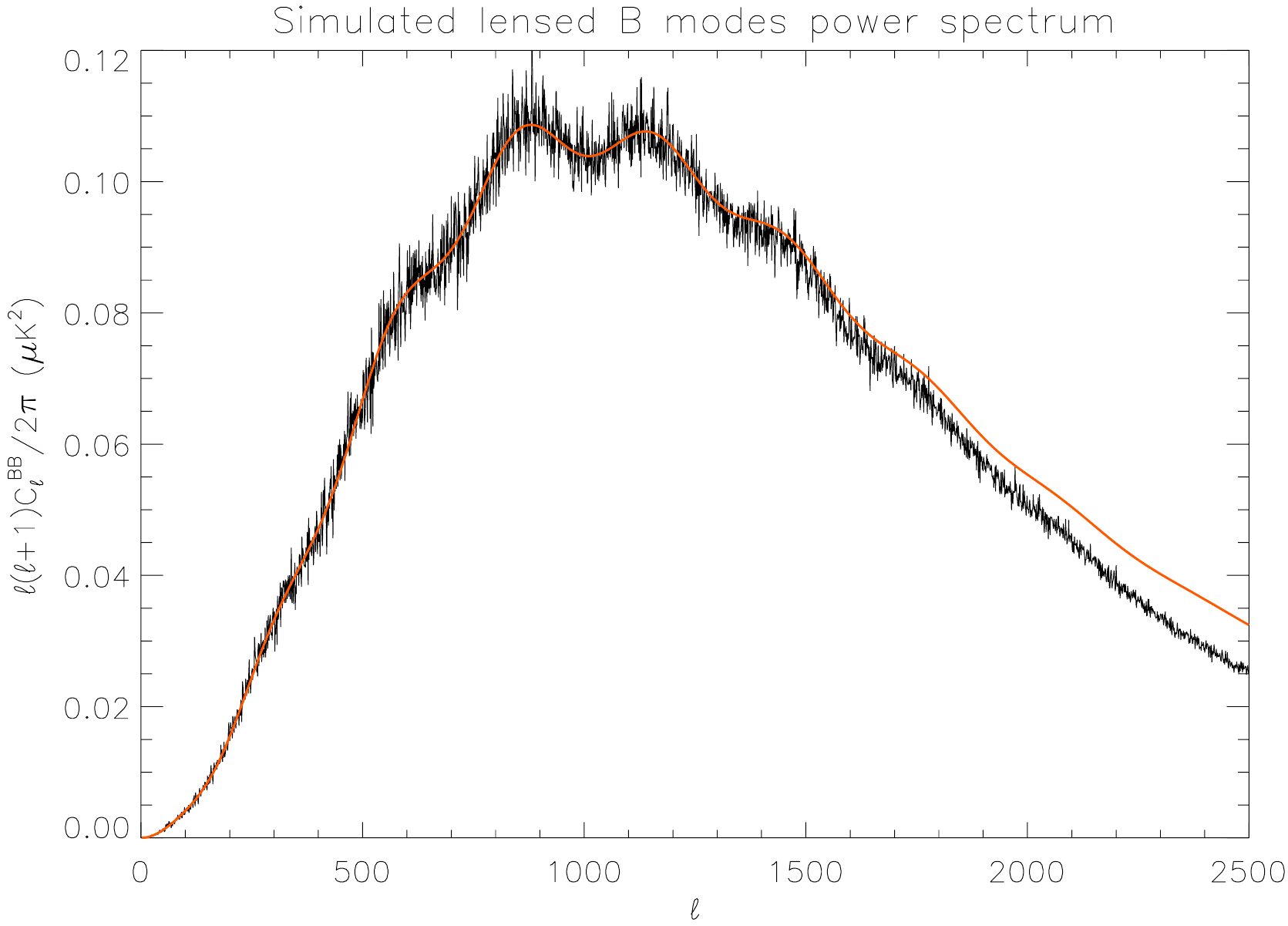}
\caption{Examples of the CMB B-modes lensing calculation and involved numerical effects. All panels show the recovered $B$-modes power spectrum overplotted over the theoretical $B$-mode spectrum computed with CAMB (color line). The bandwidth of the $E$-modes and the potential $\Phi$ is the same in all the panels and set to $2500$, while the resolution of the maps used for simulating the lensed signal  increases progressively from left to right. ECP pixelization has been used in all cases. The recovered $B$-spectrum overestimates the theoretical curve in the left panel due to the power-aliasing effect,
while it underestimates it in the result recovered for much higher resolution as shown on the right. The nearly perfect recovery shown in the middle panel is merely accidental and results from the insufficient signal bandwidth (right panel) that compensates the extra contribution of the aliasing effect (left panel). The spectrum in the right panel is aliasing-free because it does not change anymore with the increasing resolution.
}
\label{bmodesplate}
\end{figure*}

\paragraph{Signal bandwidths.} Because the lensing procedure needs to be applied prior to any instrumental response function convolution, the relevant
sky signals on all but the last steps above require using a resolution sufficient  to support the signal all the way to its 
intrinsic bandwidth, $\ell_{intr}^X,$ where $X$ is either $T$ for the total intensity, or $P$ for the polarization, or $\Phi$ -- for the gravitational potential.  
However, because mathematically the lensing effects can be seen as a convolution in the harmonic domain \citep{hu2000,okamoto2003,hu2002} of the CMB signal --
either the total intensity, $T$, or the polarization, $P$, --  and of the potential, $\Phi$, the bandwidth of the resulting
lensed field will be broader than that of any unlensed fields and is given roughly by $\ell_{intr}^X+\ell_{intr}^\Phi$. Consequently, the lensed map produced in 
step 5 should have its resolution appropriately increased to eliminate potential power aliasing effects. The resolution of the unlensed maps produced in steps 1-5
should then coincide with that of the lensed signal but with the number of harmonic modes set by $\ell_{intr}^X$ and $\ell_{intr}^\Phi$ respectively .

One of the problems arising in this context is related to the fact that the unlensed sky 
signals, $T$, $P$ and $\Phi$, considered here are not truly band-limited even if their power at the small scales decays quite abruptly as a result of Silk damping.
Picking an appropriate value for the bandwidth is therefore a matter of a compromise between the precision of the final products
and the calculation cost, with both these quantities being quite sensitive to the  chosen value, and which will  depend in general on a specific 
application. We emphasize that the presence of the high-$\ell$ power decay plays a dual role in our considerations here. On the one hand,
 it ensures that the lensing
effect at sufficiently large scale can be computed with an arbitrary precision by simply choosing the bandwidth values sufficiently high. On the other hand it does introduce
an extra complexity in defining  a set of sufficient conditions, which ensure required precision, because these will be typically different in the regime of the high signal power and that
of the damping tail. In either case, though, it is clear that whatever the selected bandwidths, the amplitudes of the harmonic modes of the lensed signal close to the highest value of $\tell{X}$
supported by the employed pixelization, i.e., $\tell{X} \sim  \ell_{\, intr}^X+\ell_{\, intr}^\Phi$, will generally be unavoidably misestimated, and satisfactory precision can only be achieved
for harmonic modes lower than some $\tell{X}_{\, ok} < \ell_{intr}^X$.
From the practitioner's perspective the main problem is therefore, given some precision criterion, $\varepsilon$, which we wish to be fulfilled by the harmonic modes of the lensed signal up to some value of $\tell{X}=\tell{X}_{\,ok}$, how to determine the  
required bandwidths of the unlensed signals, $\ell_{\,intr}^X = \ell_{\,intr}^X(\tell{Y}_{\,ok}, \varepsilon)$ where $X$ and $Y$ can be the same, e.g., in the case of the $T$ or $E$ signal lensing, or different, 
e.g., for the potential field or $B$-modes.
  
One effect of these considerations is that if these are maps of the lensed signals, which are of interest as the final product of the calculation, then the biased high-$\ell$ modes should either be filtered out or suppressed before the map
is synthesized from its harmonic coefficients. To ensure that this does not adversely affect the resolution of the final map, the bias should affect only angular scales much smaller than the expected final resolution of the map as produced in step 6 of the algorithm. If the latter is defined 
by the experimental beam resolution, one therefore needs to ensure that no bias is present  for $\tell{X} \simlt \ell_{beam}  \sim \sigma_{beam}^{-1}$, where $\sigma_{beam}$ is an experimental beam width.

\paragraph{Interpolations.}\label{sect:interpolation-discussion}
Interpolation is the most popular workaround of the need to directly calculate values of the unlensed fields for every displaced directions, which typically will not
correspond to grid points of any iso-latitudinal pixelization. 
Three interpolation schemes have been considered to date in the context of the polarized signals. \citet{lewis2005} proposed 
a generic modified bicubic interpolation and demonstrated that it seems to work satisfactorily in a number of cases. This approach together
with the direct summation are both implemented in the publicly available code \lenspix\footnote{\url{http://cosmologist.info/lenspix}}.
Two other methods have been proposed more recently. \citet{basak} implemented the general interpolation scheme, which
recasts a band-limited function on the sphere as a band-limited function on the 2D torus where a non-equispaced fast Fourier (NFFT)  transform algorithm is used to compute the field at the displaced positions. This method would be arbitrarily precise if the sky signals were strictly band-limited. However, the choice of NFFT can become a bottleneck for this algorithm since its numerical workload scales with the number of pixels squared, and its memory requirements are huge. As it is, the NFFT software can be run only on shared-memory architectures, making it more difficult to resolve both these problems. 
Consequently, the issue of the bandwidth values is becoming of crucial importance for the performance and applicability of the method, and its relevance 
 in particular in the context of simulations of upcoming and future high-resolution experiments needs to be investigated in more detail.
\\
\citet{wandelt} proposed a fast pixel-based method using the spectral characteristics of the field to be lensed to compute the weighting coefficient for the interpolation of this field, without using any spherical harmonic algorithm. Its accuracy is set by the number of neighboring pixels used to interpolate the field at a given point.\\*
In addition, \cite{hirata2004} used in their work a polynomial interpolation scheme of arbitrary order and precision, which has been shown to successfully produce temperature maps  \citep{hirata2004, das-bode} 
but has not been tested for the polarized case.\\*
Any interpolation in this context is not without its dangers because interpolations tend to smooth the underlying signals.  For a genuinely band-limited function this
could in principle be avoided as in, e.g., \citet{basak}. However, for the 
actual CMB signals the bandwidth is only approximate and is a function of the required precision and specific application; the sampling density and interpolation scheme
therefore need to be chosen very carefully to render reliable results. Again, the choice of appropriate bandwidth values is therefore central for a successful resolution of this problem.
 
 \paragraph{Numerical workload}
 
 Numerical cost of the direct calculation per direction is given by ${\cal O}(\ell_{max}^2)$ and corresponds
 to the cost of calculating an entire set of all $\ell$ and $m$ modes of associated, scalar, or spin-weighted, Legendre functions. For $N_{pix}$ directions the overall
 cost about ${\cal O}(N_{pix}\,\ell_{max}^2) = {\cal O}(N_{pix})^2$ and is therefore prohibitive for any values of $N_{pix}$ and $\ell_{max}$ of interest. 
 Here, we assumed a relation, $\ell_{max} \propto N_{pix}^{1/2}$, typically fulfilled for the full-sky pixelization with a proportionality coefficient on the order of a few, e.g., for the HEALPix\footnote{\url{http://healpix.sf.net/}} pixelization \citep{gorski2005} we have  $\ell_{max} = 2\sqrt{3\,N_{pix}}$, while  for ECP, $\ell_{max} = 2\,\sqrt{N_{pix}}$.
 The interpolations
 can cut on this load, trimming it to the one needed to compute a representation of the signals on an iso-latitudinal grid, with complexity ${\cal O}(N_{pix}^{1/2}\,\ell_{max}^2) = {\cal O}(N_{pix}^{3/2})$ plus the interpolation with the complexity ${\cal O}(N_{pix})$, or ${\cal O}(N_{pix}\,\ln\,N_{pix})$ in the case of NFFT, in both cases with a potentially large pre-factor. Nevertheless, this is clearly a more favorable scaling than the one of the direct method and, as has been shown in the past, makes such calculations feasible in practice. We note, however, that for the sake of
 the precision of the interpolation one may need to overpixelize the sky, meaning using a higher value of $N_{pix}$ than what would normally be needed to support the harmonic modes all  the way to $\ell_{max}$. Hereafter, we denote the overpixelization factor in each of the two directions, $\theta$ and $\phi$, as $\kappa$. Consequently, the number of pixels used is given by $\kappa^2\,N_{pix}$, where  $N_{pix}$ is the standard full-sky number of pixels as determined by the selected value  of $\ell_{max}$.

\paragraph{Goals and methodology.}

This paper has two main goals. One is to study internal consistency and convergence of the pixel-domain simulations in the context of the currently viable cosmologies. The other is to study the dependence of the precision of these simulations  on some of its most important parameters.

In previous works, analyses of this sort have usually been restricted to comparisons of power spectra of the lensed maps derived by a lensing simulation code and the theoretical predictions  computed via an integration of the Boltzmann equation, as implemented in the publicly available codes, CAMB and CLASS. In these works, the effort has been made to 
find a set of the code parameters for which the resulting spectrum is consistent with the theoretical expectations. 
Such comparisons are without doubt an important part of a code and method validation. However, they are limited to the cases of the gravitational potentials, $\Phi$, derived in a linear theory, and are not applicable in some other cases where the potential is obtained by some other means such as, N-body simulations. In addition, they may on occasion be misleading because the numerical
effects can easily conspire to deliver a spectrum tantalizingly close to the desired one, without any reassurance that the map of the lensed sky characterized by it has correct other statistical properties, such as higher-order statistics. That this is particularly likely and consequential for the $B$-modes spectrum given its low amplitude and the lack of characteristic, fine-scale features.
 An example of such a conspiracy is shown in Fig.~\ref{bmodesplate}, where the power deficit at the high-$\ell$ end caused by the oversmoothing due to the interpolation nearly perfectly compensates  the extra power aliased into the $\ell$-range of interest as a consequence of too crude a resolution of the final map.

We therefore propose to study the robustness of the simulated results by demonstrating their convergence and internal stability with respect to sky sampling and band-limit changes, as expressed by two parameters 
introduced earlier: the upper value of the signal band, $\ell_{max}$, and the overpixelization factor, $\kappa$. Only once the convergence is reached we compare the results to those computed by other means, if any are available.
We note that the convergence tests  do not have to, and should not in general, be restricted to the power spectra comparison only  and could instead  involve other metrics more directly relevant to the simulated maps
themselves.
In all such tests it is typically  required to consider maps with  extreme resolutions, which has been traditionally prohibitive for numerical reasons. We overcome this problem with the help of a high-performance lensing code, \lenstwo, which we have developed for this purpose.  

Our second goal, i.e., to study
the dependence of the calculation precision on the two crucial parameters, $\ell_{max}$ and $\kappa$, is complementary and is aimed at providing meaningful and practically useful guidelines of how to select the
values of these parameters prior to performing any numerical tests given some predefined precision targets. In this context, we present an in-depth semi-analytical analysis of the impact of these parameters on 
the lensed signal recovery.
Though ultimately they may need to be confirmed numerically case-by-case, e.g., using the convergence tests as discussed earlier, they could be of significant help in providing a reasonable starting point for such
tests.

At last we also present a simple, high-performance  parallel implementation of the pixel-domain approach, \lenstwo, which is capable of reaching extremely high sample density on the sphere thanks to its efficient parallelization
and numerical implementation, and which has been instrumental in accomplishing all the other goals of this work. 

\begin{figure*}[!htb] 
\centering
{\includegraphics[width=.5\textwidth]{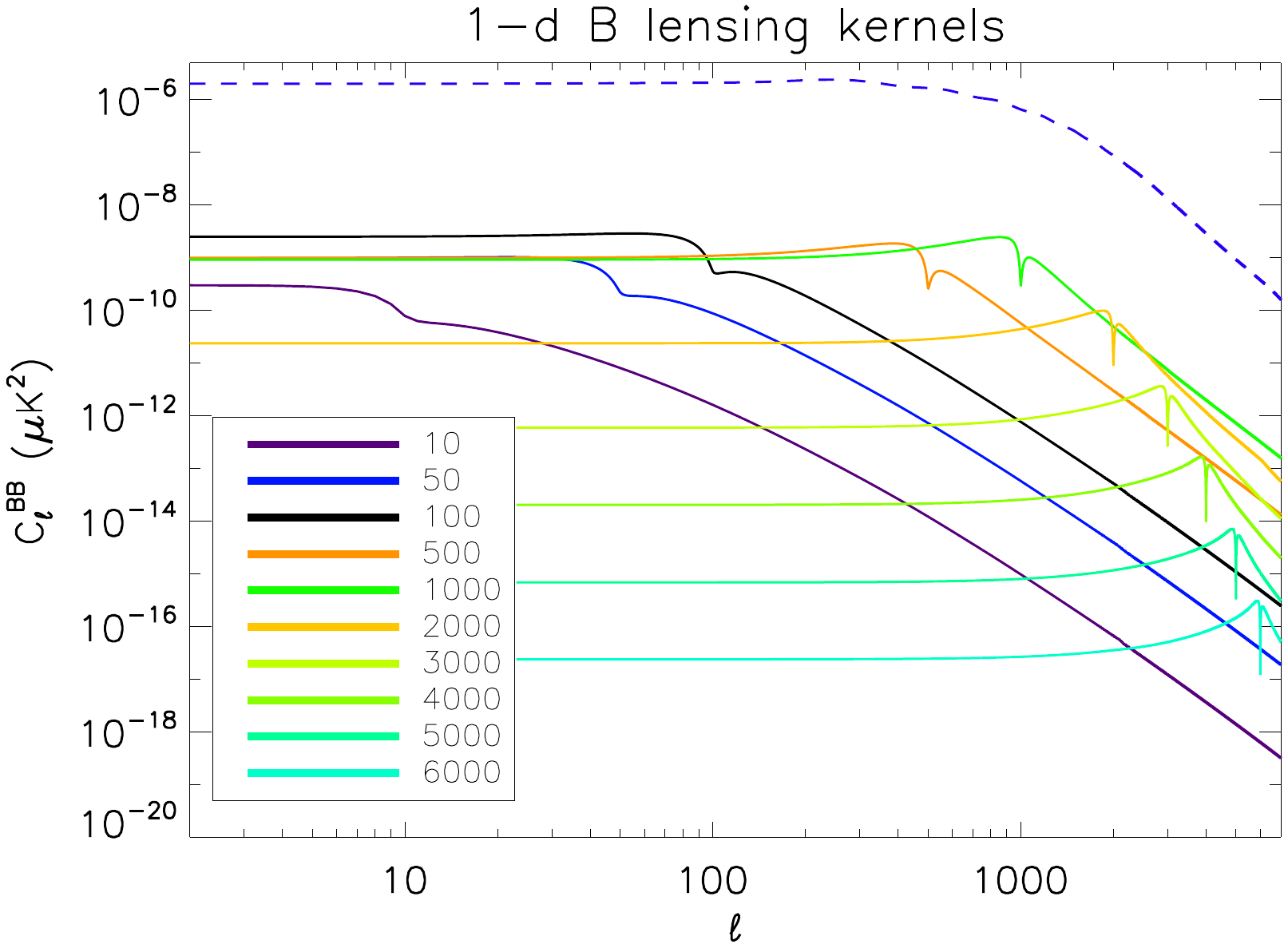}\includegraphics[width=.5\textwidth]{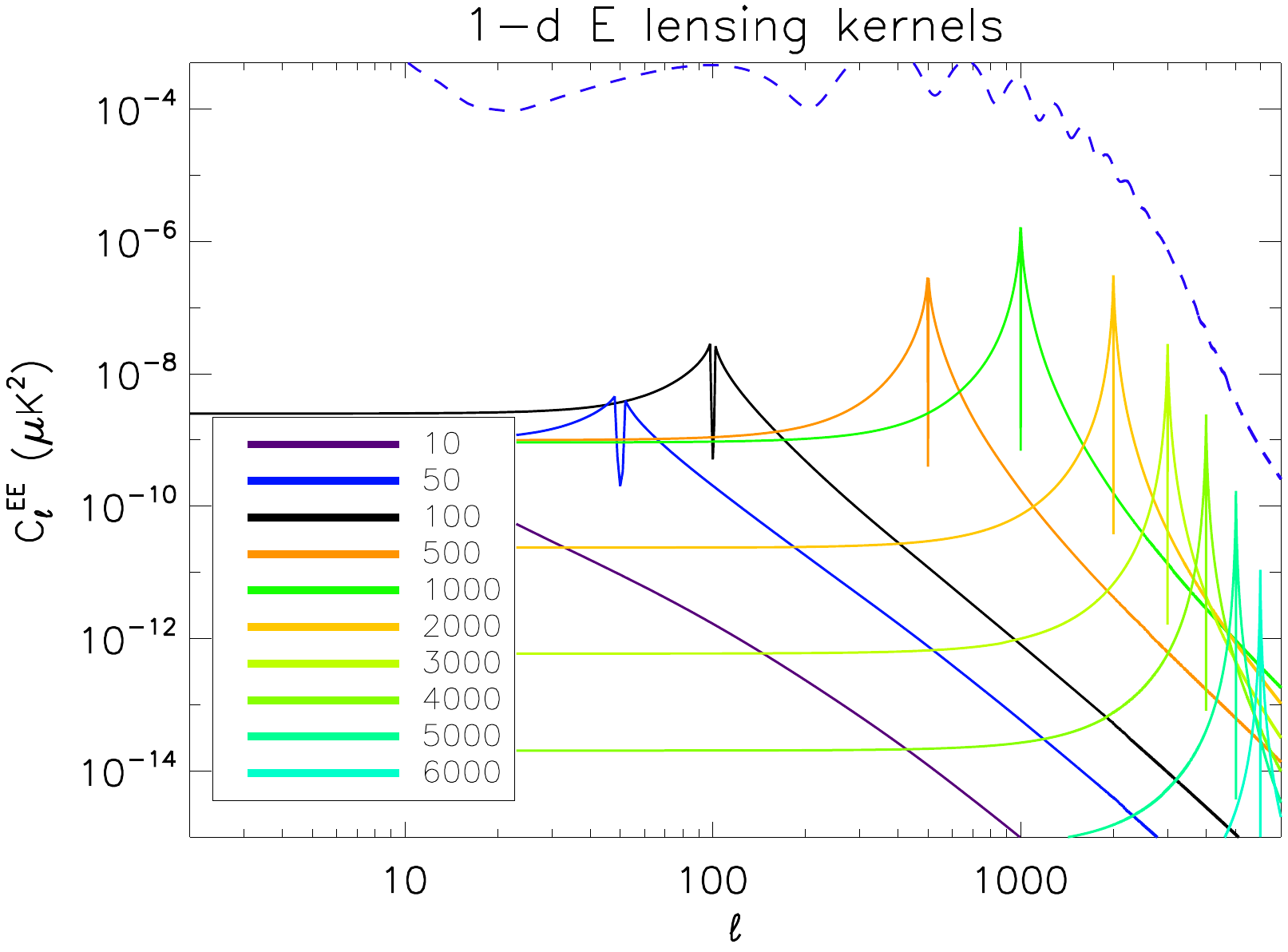}}\\
{\includegraphics[width=.5\textwidth]{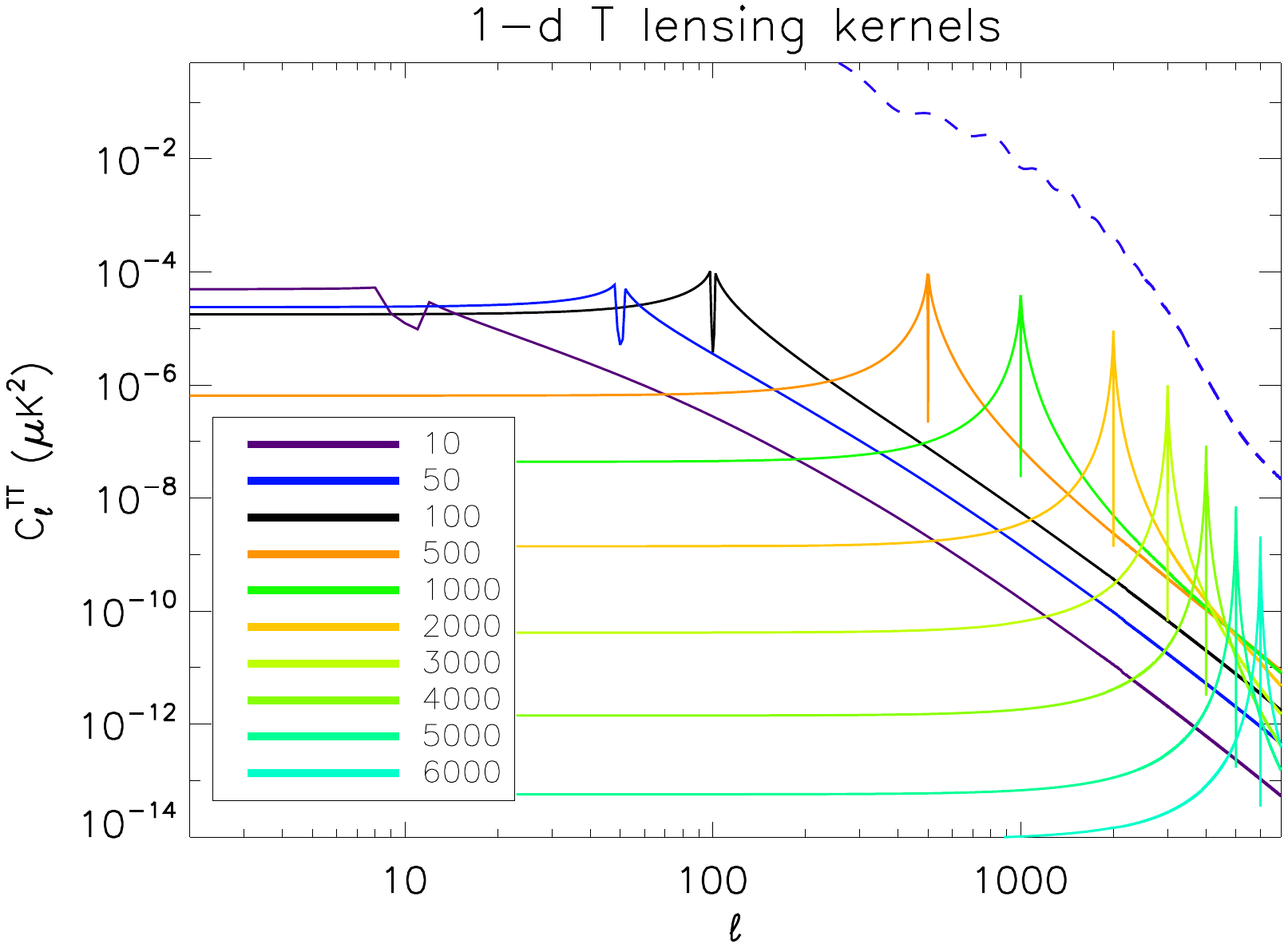}}{\includegraphics[width=.5\textwidth]{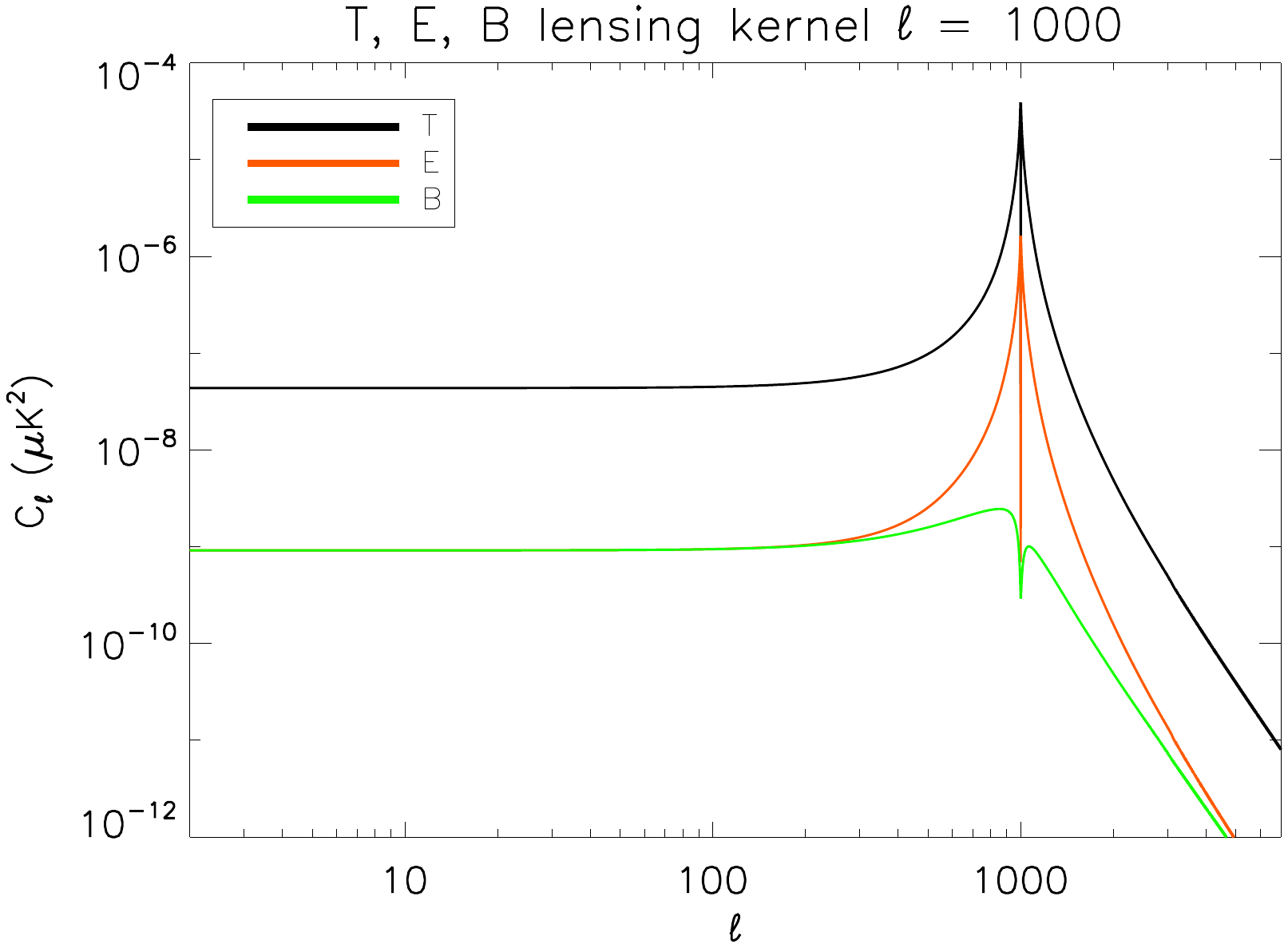}}\\
\caption{1D lensings kernels. The lensed power for $T$, $E$, and $B$ spectra is computed assuming a delta-like spectra with power in a single mode $\ell^{\prime}=$ 10, 50, 100, 500, 1000, 2000,  3000, 4000, 5000 and 6000 in the unlensed CMB spectra. The blue dashed line represents the reference lensed spectra as computed by CAMB. The sum of all single-mode contributions for $\ell^{\prime}\in [0,\infty]$ would reproduce the lensed spectra. For T and E cases, the subdominant contribution of the convolution part only is shown for visualization purposes and offset terms are ignored (see Sect.~\ref{sect:accuracy} and Eq.~\ref{Xaccuracydef}). The comparison of 1D kernel shapes for $T$, $E$, and $B$ for $\ell^{\prime}=1000$ is shown in the bottom-right panel: the peculiar shape of each type of kernel drives the locality and amplitude of the contribution to the lensed spectra.}
\label{fig:1dkers}
\end{figure*}

\section{Exploring the bandlimits} \label{sect:kers}

\subsection{CMB lensing in the harmonic domain}
This section addresses the second of goals, as stated above, and describes a semi-analytic study of the impact of the assumed bandwidth values on the precision of the lensed signal. Our discussion is based on the model of  \citet{hu2000} and focuses on the lensed $B$-mode signal that is obtained obtained as a result of the lensing acting upon the primordial $E$-mode signal, and is the main target of this paper. Similar considerations can be made,  however, for other CMB observable spectra and we present some relevant results calculated for these cases (see Sect.~\ref{sect:accuracy} for some more details).
Using the results of \citet{hu2000}, we represent the lensed $B$-mode signal as
\begin{eqnarray}
\tilde{C}_{\tell{B}}^{BB}&=&\frac{1}{2}\sum_{\ell^{\,\Phi}\ell^{\,E}}\frac{|_{2}F_{\tell{B}\ell^{\,\Phi}\ell^{\,E}}|^{2}}{2\tell{B}+1} C_{\ell^{\,\Phi}}^{\Phi\Phi}C_{\ell^{\,E}}^{EE}(1-(-1)^{L}) 
\label{eq:kerb}
\end{eqnarray}
\noindent
where $_{2}F_{\tell{B}\ell^{\,\Phi}\ell^{\,E}}$ is a spin-$2$ coupling kernel (see \cite{hu2000} for a full expression), $L\equiv\tell{B}+\ell^{\,\Phi}+\ell^{\,E}$ and $C_{\elle}^{EE}$ and $C_{\ellphi}^{\Phi\Phi}$ denote the unlensed power spectra of the $E$ mode polarization and of the gravitational potential, respectively. This formula can be obtained by a second-order series expansion around undisplaced direction, which is expected to be accurate to within 1\%  for multipoles $\tell{B}\simlt 2000$ and 
then for $\tell{B}\gg2000$, where the CMB amplitude is small and can be modeled by its gradient only, while in the intermediate scales its precision degrades to nearly 5\%. The reliability of this analytical model is discussed later in Sec.~\ref{sect:simker}.
We can now introduce 1D kernels, ${\cal H}_{\elle}(\tell{B})$, defined as
\begin{equation}
{\cal H}_{\elle} (\tell{B}) \equiv \frac{1}{2}\,C_{\elle}^{EE}\,\sum_{\ellphi} \, \frac{|_{2}F_{\tell{B}\,\ellphi\,\elle}|^{2}} {2\tell{B}+1} \, C_{\ellphi}^{\Phi\Phi}\,(1-(-1)^{L}).
\label{eq:oneDimKerDef}
\end{equation}
Summed over $\elle$ for a fixed $\tell{B}$, these give the lensed $B$-mode power contained in the mode $\tell{B}$, Eq.~\ref{eq:kerb}, while for a fixed $\elle$ they define the power spectrum of the lensed $B$-modes
signal, generated via lensing from the $E$ polarization signal that contains non-zero power in a single mode $\elle$, and with its amplitude as given by $C_{\elle}^{EE}$. The kernels are displayed in Fig.~\ref{fig:1dkers} together
with their analogs for the total intensity and $E$-polarization signals. We find that the kernels computed for different values of $\tell{B}$ are similar, just shifted with respect to each other accordingly. The change in the
amplitude simply reflects the change in the assumed power of the $E$ signal, which in turn follows that of the actual $E$ power spectrum. The kernels are flat for values $\tell{B}\ll \elle$ and decay
as a power law for $\tell{B} \gg \elle$, displaying a sharp dip at $\tell{B} = \elle$. Similar observations can be made for the $T$ and $E$ kernels, with the exception that
unlike their $E$ and $T$ counterparts, the $B$ kernels are not peaked around the dip. 
This behavior is related to the fact that the lensed $B$-modes signal we discuss here, described by Eq.~\ref{eq:kerb}, is generated by the $E$-polarization, while the main effect of the lensing on  $T$ and $E$  is
imprinted on these signals themselves.
A direct consequence of this is that for any lensed $B$-modes spectrum mode a contribution from local unlensed multipoles will be less dominant, as is the case for the $T$ and $E$ signals, and
nonlocal contributions will be relatively more important and therefore required to be accounted for in high-precision calculations.

Indeed, owing to the flat plateau of the kernels at the low-$\ell$ end, in principle all high-$\ell$ unlensed modes  contribute to the lensed power at the low-$\ell$ end.  The magnitude of their 
contribution is modulated by the shape of the unlensed $E$ spectrum and therefore  eventually becomes negligible only because of the Silk damping, i.e., lack of the power at 
small angular scales in the unlensed fields. Nevertheless, we can expect that nearly all the modes of the unlensed $E$ spectrum up to the damping scale have to be included in the calculation 
of the lensed $B$ spectrum to ensure high-precision recovery of the lensed $B$-modes spectrum with $\tell{B} \simlt 1000$.  
Given some specific target precision, we could and should fine-tune the required $E$-spectrum bandwidth, and whatever is the value selected here, the bandwidth for the potential field will have
to be at least the same.

For high-$\ell$ modes of the lensed $B$-modes spectrum, $\tell{B} \gg 1000$, the non-locality of the power transfer due to lensing is even more striking, as due to the low amplitudes of the $E$ spectrum 
the local contributions are additionally suppressed, and the long power-law tails of the contributions from large and intermediate angular scales, $\elle \simlt 1000$ are evidently dominant. 
Less evident is the fact that also the $E$-power from even smaller angular scales, $\elle \simgt \tell{B}$, may be relevant. The contributions from each of these modes may appear small, Fig.~\ref{fig:1dkers}, but are potentially non-negligible due to the large number of those modes. A high-precision recovery of the high-$\ell$ tail of the lensed $B$-modes spectrum will therefore need a careful assessment of the importance of all these
contributions, nevertheless, a generic expectation would be that the bandwidth of the unlensed $E$-modes spectrum will have to be higher than the highest value of the lensed $B$-modes signal multipole for
which high precision is required, and potentially higher than the scale of Silk damping. Because these very high multipoles of the lensed $B$ spectrum are expected to have a significant contribution from relatively low multipoles of the unlensed $E$ signal, i.e., for which $\elle \ll \tell{B}$
given the triangular relations, Eq.~\ref{eq:trRel}, and the definition of the kernels, Eq.~\ref{eq:oneDimKerDef}, we can conclude that the bandwidth of the potential field used in the simulations will
have to be at least as large as $\tell{B}$.

There are two main conclusions to be drawn here. First,  it is clear that a high-fidelity simulation of the $B$-polarization power spectrum even in a restricted range of angular scales will require broad bandwidths, potentially all the way up to the scale of Silk damping, for both the unlensed $E$-mode polarization signal and the gravitational potential. However, these bandwidth values are not expected to depend very strongly on the maximal 
$B$-mode multipole that we want to recover, at least as long as it is in the range $\tell{B} \simlt 2000$. Second, because the expected bandwidths are broad, it is important to optimize them to ensure efficiency of the
numerical codes without affecting precision of the results.

Thanks to the peaked character of the respective kernels, the lensed modes for the lensed $T$ and $E$ spectra are typically dominated by a local contribution coming from the immediate vicinity of the mode.
This in general permits setting the bandwidth for the potential shorter than the mode of the lensed spectrum to be computed. 
By contrast, the unlensed $T$ and $E$ spectrum have to be known at least up to the 
multiple of  interest of the lensed spectrum, $\tell{X},$ $(X=T$ or $E)$, augmented by the assumed bandwidth of the potential.
These observations reflect the usual
rule of thumb, \citep[e.g.,][]{lewis2005},  indicating that  lower bandwidth values can be used in these two cases for the same required accuracy. 

\begin{figure*}[!htb] 
\centering
\includegraphics[width=\textwidth]{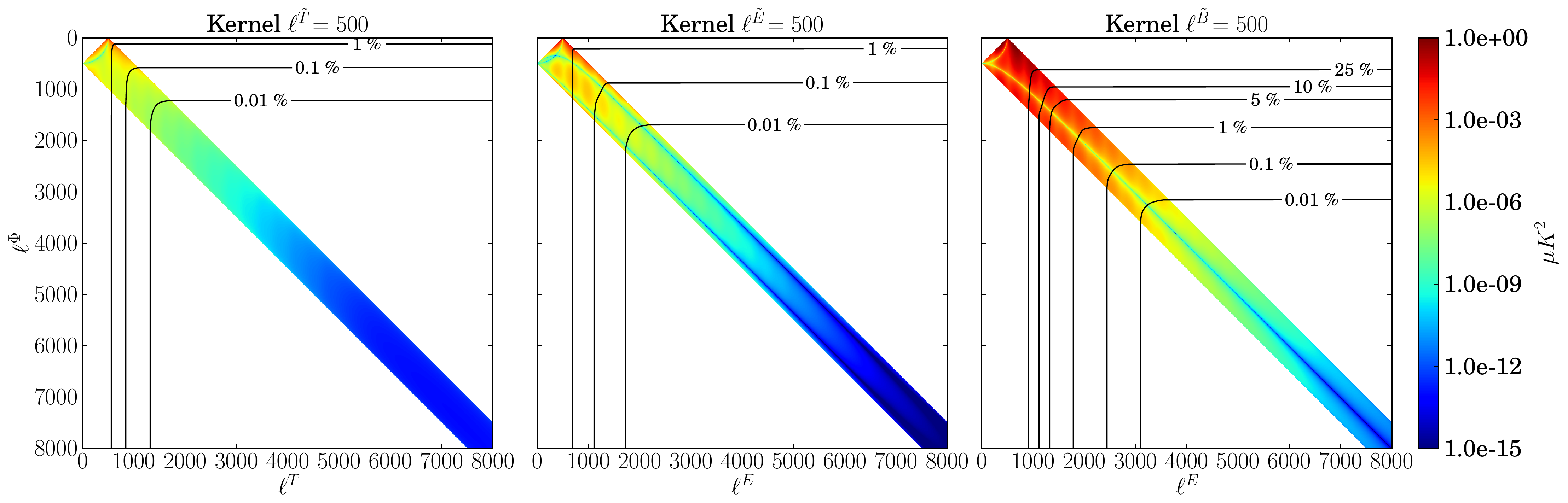}
\includegraphics[width=\textwidth]{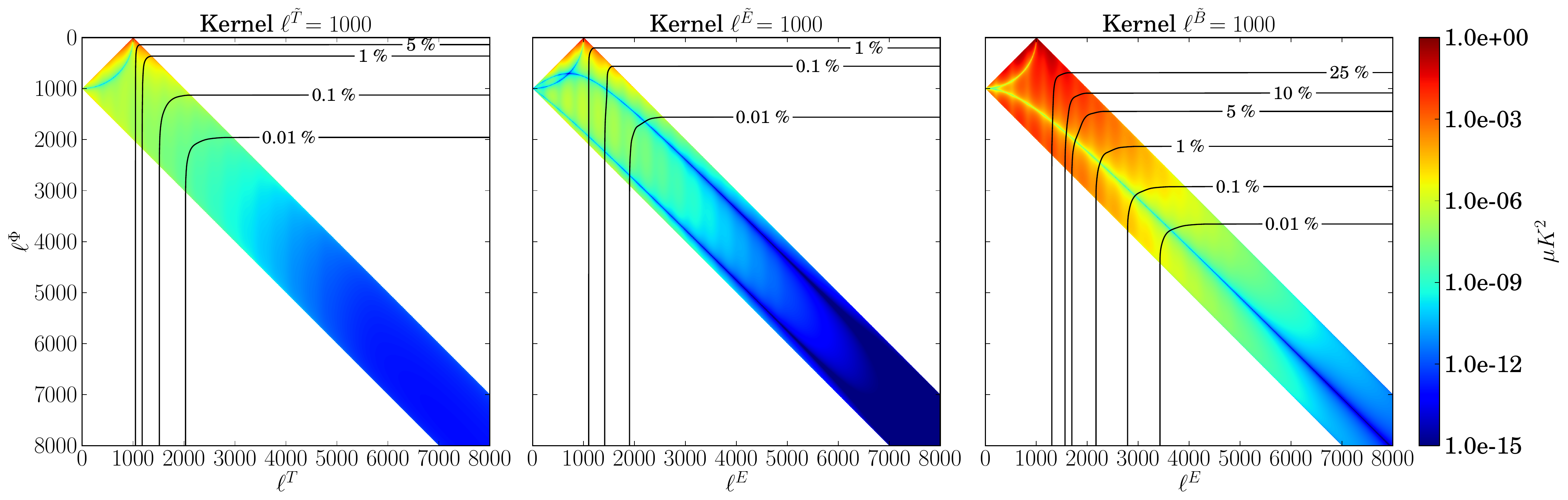}
\caption{Lensing kernels $\lensker{X}{Y}{\Phi}$ for $X=Y=T$, left column, $X=Y=E$, middle, and $X=B, Y=E$, right, for two different values of the multiple number of the lensed signal, $\tell{X} = 500, 1000$, top to bottom. 
The color scale shows the logarithm of the kernel elements and ranges from dark blue $\sim 10^{-15}$ to $\sim 1$, dark red. The solid-line contours show the best achievable precision of the estimated lensed spectrum, that can be obtained if the bandwidths of the $E$ and/or $\Phi$ unlensed spectra are truncated to $\elle$ and $\ellphi$. The contours range from $25$\% to $0.01$\% from left to right. The precision is computed with respect to the lensed multipoles 
calculated with $\elle_{max} = \ellphi_{max} = 8000$.
}
\label{fig:kerlog}
\end{figure*}

\begin{figure*}[!htb] 
\centering
\includegraphics[width=\textwidth]{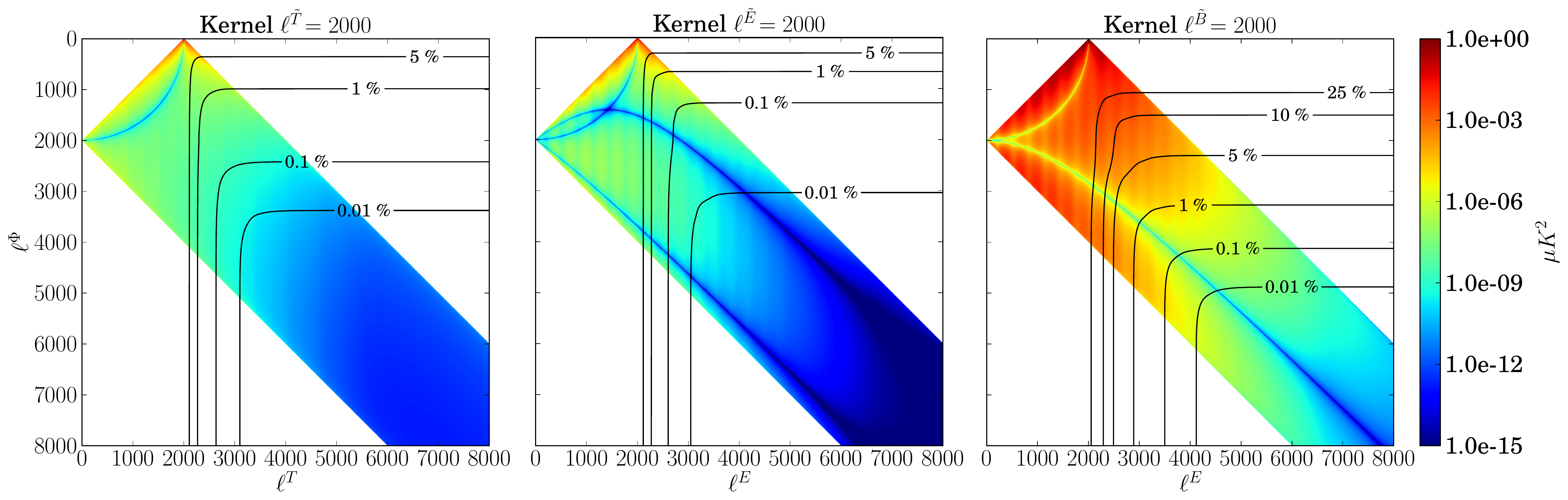}
\includegraphics[width=\textwidth]{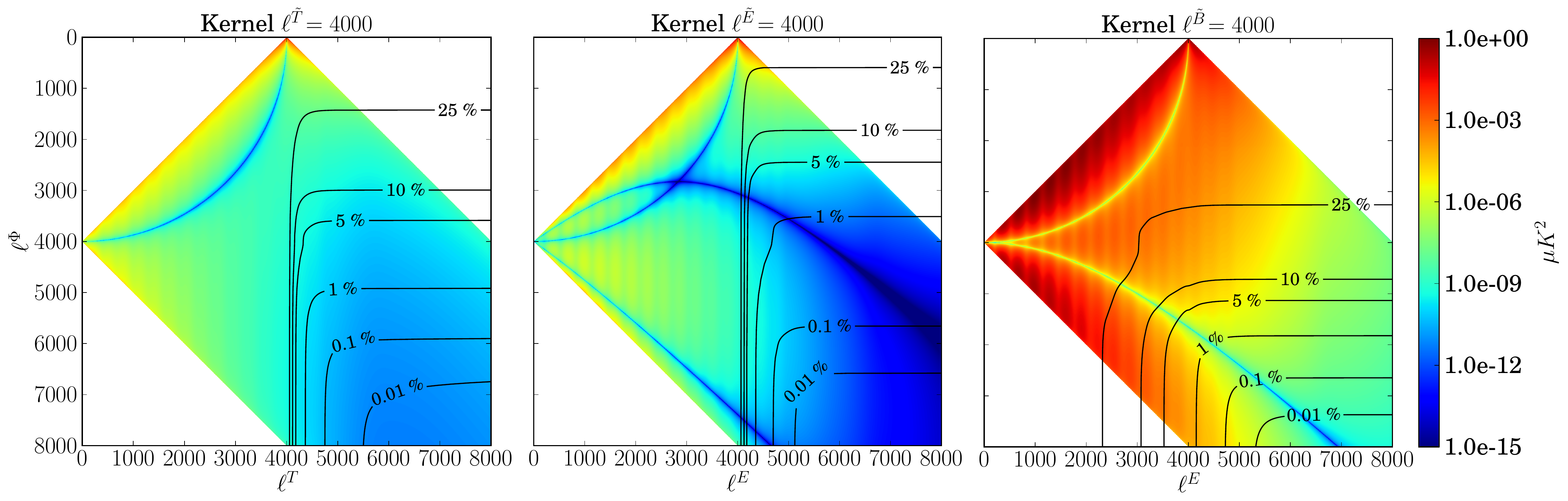}
\caption{Lensing kernels $\lensker{X}{Y}{\Phi}$ for $X=Y=T$, left column, $X=Y=E$, middle, and $X=B, Y=E$, right, for two different values of the multiple number of the lensed signal, $\tell{X} = 2000, 4000$, top to bottom. See Fig.~\ref{fig:kerlog} for additional details.
}
\label{fig:kerlog2}
\end{figure*}

\subsection{Accuracy}\label{sect:accuracy}
In this section we aim at turning the consideration presented above into more quantitative prescriptions concerning the bandwidths of the input fields used in the simulations.
For this reason we introduce 2D kernels, $\lensker{B}{E}{\Phi}$, defined as,
\begin{equation}
\lensker{B}{E}{\Phi}
\equiv \frac{1}{2} \, \frac{|_{2}F_{\tell{B}\,\ellphi\,\elle}|^{2}} {2\tilde{\ell}_B+1} \, C_{\ellphi}^{\Phi\Phi}\,C_{\elle}^{EE}\,(1-(-1)^{L}).
\label{eq:twoDimKerDef}
\end{equation}
These define for a given
 value of  $\ell_B$ a contribution of  the $E$ power at $\ell = \elle$ and the $\Phi$ power at $\ell = \ellphi$  to the amplitude of the lensed $B$-modes spectrum at that $\ell = \tell{B}$, which can then be computed 
 by summing over $\elle$ and $\ellphi$, i.e.,
\begin{equation}
\tilde{C}_{\tell{B}}^{BB}=\sum_{\ellphi,\, \elle} \, \lensker{B}{E}{\Phi}.
\label{eq:ker2dSum}
\end{equation}
 The sum in this equation involves in principle an infinite number of terms and therefore would have to be truncated in any numerical work, either explicitly, e.g., by setting finite limits in the formula above, or
 implicitly, e.g., by selecting the bandwidths, pixel sizes, etc, in the pixel-domain codes. We therefore used these kernels to study the precision problems involved in this type of calculations. As the expressions for
 the kernels are approximate, so will be our conclusions. However, as our goal is to provide guidelines on how to select the correct values for the simulations codes, this should not pose any problems. We will return
 to this point later in this section.
 
We show a sample of the  kernels, $\lensker{B}{E}{\Phi}$ in Fig.~\ref{fig:kerlog}. These are computed for selected values of $\ell_{\tilde B}$ for which the approximations involved in their computation are expected to be
valid. We note that all elements of the kernel, $\lensker{B}{E}{\Phi}$, vanish if the quantity $L$, defined in the previous section, is even, as do those for which the triangular relation
\begin{eqnarray}
{\displaystyle \left |\elle-\ellphi\right |  \leq} &{\displaystyle  \tell{B} \leq \elle+\ellphi}
\label{eq:trRel}
\end{eqnarray}
is not satisfied.
This last fact is a consequence of the Wigner 3-j symbols in the expressions for $_2F_{\tell{B}\, \ellphi\, \elle}$, \citep{hu2000}.
Within these restrictions it is apparent from  Fig.~\ref{fig:kerlog}  that each multipole of the lensed $B$-modes spectra $\tilde{\ell}^{B}$ receives contributions from a wide range of harmonic modes of both E and $\Phi$ spectra, extending to values of $\elle$ and $\ellphi$ significantly higher than $\tell{B}$ and roughly independent of the latter value at least for $\tell{B} \simlt 2000$. For its higher values a non-negligible fraction
of the contribution starts to come from progressively higher multipoles of both $E$ and $\Phi$. Clearly, these trends are consistent with what we have inferred earlier with help of the 1-dim kernels.

As also observed earlier, we find the $B$-modes kernels qualitatively different from those computed for the lensed total intensity and $E$-modes polarization signals, Fig.~\ref{fig:kerlog}, and they are
 more localized in the harmonic space with the bulk of power coming mainly from scales for which both $\ell^{T, E}$  are relatively close to the considered lensed multipole, $\tell{T,\, E}$.
 
We note that all the 2D kernels are positive\footnote{This is not true for the $TE$ kernels, which we comment about later.} and therefore including more terms in the sum, Eq.~\ref{eq:ker2dSum}, will always 
improve the precision of the result. 
From the efficiency point of view one may want  to include in the sum preferably the terms corresponding to the largest 2D kernel amplitudes because they provide the largest contribution to the final lensed result
before adding  those with progressively
smaller kernel amplitudes until the required precision is reached.
This approach would in principle ensure that  the best accuracy is achieved with the smallest number
of included terms. This may therefore look as a  potentially attractive option from the perspective of optimizing the calculations. However, in practice, as the recurrence formulae are usually employed
in the calculations, e.g., either those needed to compute spherical harmonics in the case of the pixel-domain codes or those needed to calculate the 3-j symbols as in a direct application of Eq.~\ref{eq:ker2dSum},
and therefore all the terms up to a given bandwidth are  at our disposal at any time, and it therefore seems efficient and useful to capitalize on those by including all of them in the calculation. 
Consequently, we estimated what degree of precision can be achieved by such calculations by including all the contributions up to some specific bandwidth values for the $E$ and $\Phi$ multipoles. 

For the $B$-modes spectrum we therefore hereafter express the precision of the calculations as
\begin{equation}
A_{\tell{B}}^{B}(\ellphi,\elle) \, =\, 1 \, - \, \frac{ \sum_{\ell_{*}^{\Phi}=0}^{\ellphi}\sum_{\ell_{*}^{E}=0}^{\elle}{\cal K}_{\tell{B}}(\ell_*^{\,E}, \ell_*^{\,\Phi})} 
{\sum_{\ell_*^{\,\Phi}=0}^{\ellmax} \, \sum_{\ell_*^{\,E}=0}^{\ellmax}\, {\cal K}_{\tell{B}}(\ell_*^{\,E}, \ell_*^{\,\Phi})},
\label{Baccuracydef}
\end{equation}
where the sums in the denominator should in principle extend over the infinite range of values of $\ell$, but for practical reasons are truncated to $\ellmax = 8000$, which for the range of lensed multipoles of interest
in this work, $\tell{X} \simlt 5000$, should be sufficient.

\begin{figure}[!ht]
\centering
\includegraphics[width=.5\textwidth]{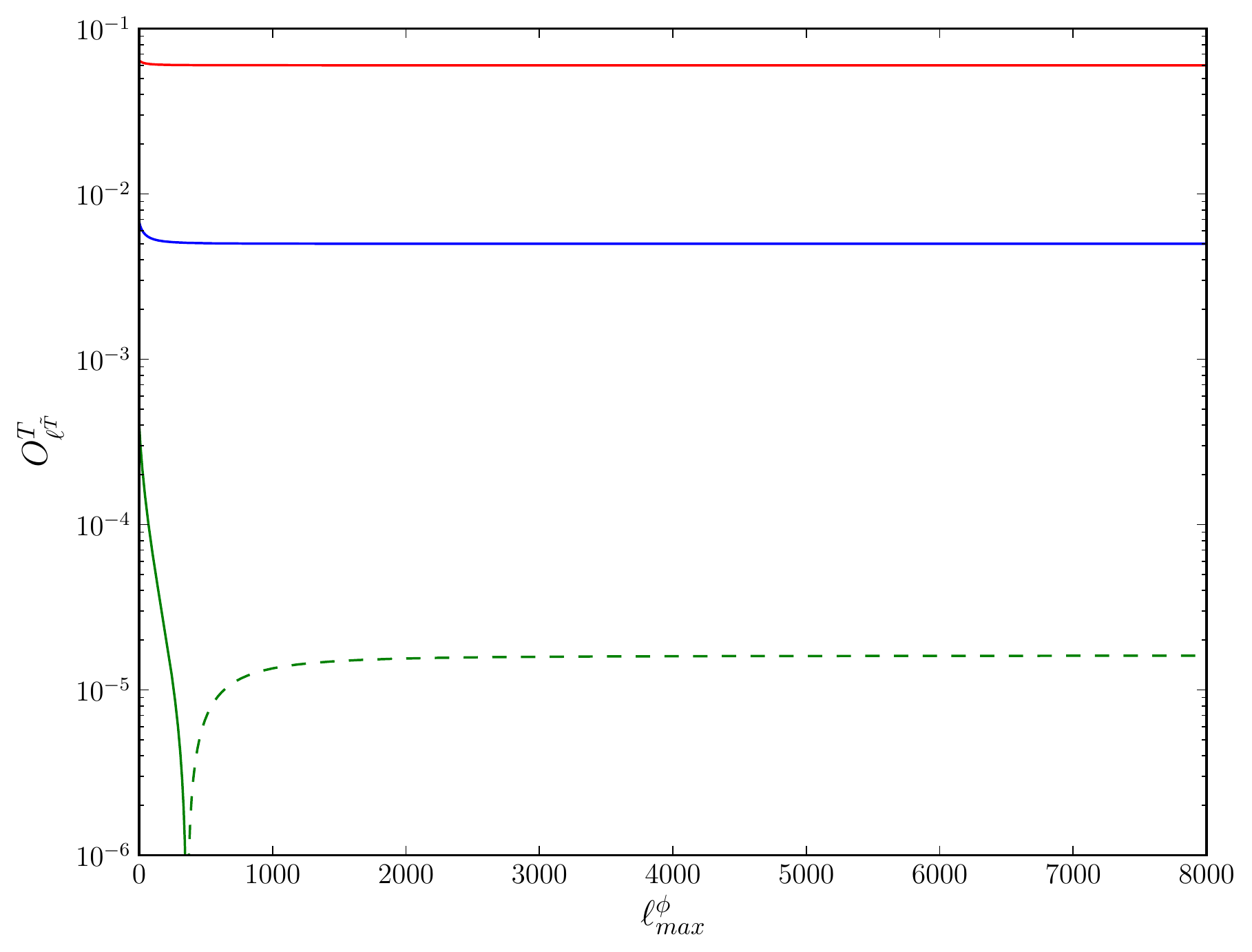}
\caption{Example of the behavior of the offset term, $O_{\tell{T}}^{T}$, as a function of $\ellmax^{\,\Phi}$ for $\tell{T}=500$ (red), $2000$ (blue), $4000$ (green). The dashed part of the green line represents negative values. $O_{\tell{E}}^{E}$ and $O_{\tell{B}}^{B}$ have a similar shape, but a different amplitude.
}
\label{fig:offset}
\end{figure}

This expression can be generalized for all lensed CMB spectra, but in this case our model has to take into account that the main effect due to lensing is to reshuffle the power of the signal and not to convert it into some
other component. Therefore the total variance of the signal has to be conserved \citep[e.g.][]{blanchard1987}.  In this case the lensed power spectra of $X=T$ or $=E$ can be written as

\begin{eqnarray}
&\tilde{C}_{\tell{X}}^{X}&=\left(1-(\tell{X\;2}+\tell{X} - \alpha)\,R\right)C_{\tell{X}}^{X} +\sum_{\ellx, \, \ellphi}\lensker{X}{X}{\Phi}\\ \label{eq:fullmodelcl}
&R&=\sum_{\ell^{\Phi}=0}^{\ellmax^{\Phi}}\frac{\ellphi(\ellphi+1)(2\ellphi+1)}{8\pi}C_{\ellphi}^{\Phi},
\end{eqnarray}
\noindent
where $\alpha$ is an integer that is different for each CMB spectra
\begin{itemize}
\item $\alpha =4$ for X=E
\item $\alpha=0$ for X=T
\item $\alpha=2$ for X=TE.
\end{itemize}
We note that the factor $R$ is a smooth function of the cutoff value of the sum over $\ellphi$, which quickly becomes nearly constant  for $\ellmax^{\,\Phi}\gtrsim 1000$, Fig.~\ref{fig:offset}. Hereafter, we therefore precompute it once
assuming $\ellmax^{\,\Phi}=\ellmax = 8000$ and use it in all subsequent calculations. \noindent
The generalized expression for the accuracy function in Eq.~\ref{Baccuracydef} would then be 
\begin{equation}
A_{\tell{X}}^{X}(\ellphi,\ellx) \, = \, 1 \, - \, \frac{ O_{\tell{X}}^{X}+\sum_{\ell_{*}^{\Phi}=0}^{\ellphi}\sum_{\ell_{*}^{X}=0}^{\ellx}{\cal K}_{\tell{X}}(\ell_*^{\,X}, \ell_*^{\,\Phi})} 
{O_{\tell{X}}^{X}+\sum_{\ell_*^{\,\Phi}=0}^{\ellmax} \, \sum_{\ell_*^{\,X}=0}^{\ellmax}\, {\cal K}_{\tell{x}}(\ell_*^{\,X}, \ell_*^{\,\Phi})},
\label{Xaccuracydef}
\end{equation}
where for shortness we have introduced 
\begin{eqnarray}
O_{\tell{X}}^{X} \equiv \left(1-(\tell{X\;2}+\tell{X} - \alpha)\,R\right)\,C_{\tell{X}}^{X}.
\nonumber
\end{eqnarray}
We note that for cosmological models of the current interest, the factor $R$ is typically found to be on the order of $\mathcal{O}(10^{-6})$ and thus the term $O_{\tell{X}}^{X}$ is expected to be
negative for most of the values of $\tell{X}$ in the range of interest here, see Fig. \ref{fig:offset}.\\*
In Fig.~\ref{fig:kerlog} black solid lines represent the expected error estimates, as expressed by the accuracy function, $A_{\tell{X}}^X(\ellphi,\ell^{Y})$,  for a number of selected values ranging from $25$\% to $0.01$\%. We note that for the shown range of $\tell{}$ only the sub-percent values of the accuracy are likely to be somewhat biased due to the assumed cutoff in the denominator of Eqs.~\ref{Baccuracydef} or~\ref{Xaccuracydef}, 
an effect, which is therefore largely irrelevant for our considerations here.
The fact that our accuracy 
definition is based on an approximate formula is also not a problem because any potential (and small,  \citet{challinor2005}) discrepancy would affect both the numerator and denominator of Eqs.~\ref{Baccuracydef} an~\ref{Xaccuracydef} in 
the same way. It can therefore be shown that to the first order in the discrepancies  amplitude, precision of our accuracy criterion improves progressively when the estimated level of the accuracy, $A_{\tell{X}}(\ellphi,\ellx)$, tends to $0$
 and is degraded to the percent level when $A_{\tell{X}}^{X}(\ellphi,\ellx) \approx 90\%$, i.e., when it is well outside of the region of any interest for the high-precision simulations considered here (see Appendix \ref{appendix:accuracy_precision}).\\*

The differences in the shape of the lensing kernels result in differences in the accuracy contours for different lensed signals and their multipoles as shown in Figs.~\ref{fig:kerlog} and~\ref{fig:kerlog2}. In particular, for lensed $B$-modes,  the contribution of large-scale power of the CMB to the lensed signal is more significant. In spite of these differences, we, however, find that the overall contours seem to share a similar shape made of two lines nearly aligned with the plot axes which meets at a right angle. Consequently, if  one of the two bandwidths is fixed, then the accuracy, which can be reached by  such a computation, will be limited and, moreover, starting from some
value of the other bandwidth, nearly independent on its value. This has two consequences. First, if the attainable precision is not satisfactory given our goals, it can be improved only by increasing the value of the first
bandwidth appropriately. Second, the value of the second bandwidth can be tuned to ensure nearly the best possible accuracy, given the fixed value of the first bandwidth, while keeping it much lower than what the triangular
relation, Eq.~\ref{eq:trRel}, would imply.  This could lead to a tangible gain in terms of the numerical workload needed to reach some specific accuracy.
Turning this reasoning around, we could think of optimizing both bandwidths to minimize the cost of the computation for a desired precision. 
From this perspective, taking the turnaround point of the contour for a given accuracy may look as the optimal choice.
However, this choice
would merely minimize the sum of both bandwidths, (or some monotonic function of each of them) for the given accuracy, which may or may not be relevant for a specific case at hand. 
Instead we may rather select the bandwidths to minimize explicitly actual computational 
cost of whatever code we plan on using. We present specialized considerations of this sort in the next section. 

On a more general level, we find that the standard rule of thumb, interpreting the effects of lensing as a convolution of the unlensed CMB signal with a relatively narrow, $\Delta \ell \sim 500$, convolution kernel due to the
lensing potential, applies only for $T$ and $E$ signals and even in these cases only to low and intermediate values of $\tell{T,\, E} \simlt 2000$ and only as long as a computation precision on the order of $\sim 1$\% is sufficient. 
For higher values of the lensed spectrum multipoles  or higher levels of the desired accuracy in the case of $T$ and $E$ and for all multipoles of the $B$-polarization signal, the required bandwidths of both the respective, 
unlensed CMB signal and the gravitational potential are more similar and indeed the latter bandwidth is often found to be broader.

We note that an analysis of this sort is somewhat more prone to problems  in the case of the $TE$ power spectrum since the lensing kernels $\lensker{TE}{TE}{\Phi}$ are not always positive because they contain the products of two different Wigner 3j coefficients and $TE$ power spectra, which may be non-positive,  rendering the corresponding accuracy function not strictly monotonic. Hereafter, we excluded this spectrum from our
analysis, noting that any band limits prescriptions derived for $T$ and $E$ will also apply directly to $TE$.

\section{Numerical analysis}\label{lens2sec}

In this section, we present results of simulations of lensed polarized maps of the CMB anisotropies and their spectra.
We address two aspects here. First, we numerically study self-consistency of the pixel-domain approach to simulating the lensing effect. Second, we demonstrate how the
consideration from the previous section can be used to optimize numerical calculations involved in these simulations.

We start this section by introducing a new implementation of the pixel-domain algorithm, which we refer to as \lenstwo.

\subsection{\lenstwo}\label{sect:lens2hat}
\noindent
\lenstwo\ is a simple implementation of the pixel-domain algorithm for simulating effects of lensing on the CMB anisotropies. 
The hallmark of the code is its algorithmic simplicity and robustness, with its performance rooted in efficient, memory-distributed  parallelization. 
The code is therefore particularly well-adapted to massively parallel supercomputers. 
Its implementation follows the blueprint described in \citet{lewis2005} that summarized in Sect.~\ref{s3ect:basics}.
The main features of the code are listed below.

\paragraph{Grids.} The code can produce lensed maps in a number of pixelizations used in cosmological applications, but internally it uses grids based on the equidistant 
cylindrical projection (ECP) pixelization where grid points, or pixel centers, are arranged in a number of equidistant iso-latitudinal rings, with points along each ring assumed to be equidistant. This pixelization supports
a perfect quadrature for band-limited functions, which in the context of this work permits minimizing undesirable leakages that typically plague codes of this type.
It can be shown, \citet{driscoll_healy_1994}, that an ECP grid made of $2\,L$ iso-latitudinal rings, each with $2\,L$ points and a weight, as given by
\begin{equation}
w^{j}=\frac{2\pi}{L^{2} }\sin(\theta_{j})\sum_{\ell=0}^{L-1}\frac{\sin\left((2\ell+1)\theta_{j}\right)}{2\ell+1}, \qquad \theta_{j}=\frac{\pi}{2L}j,
\label{weightsdh}
\end{equation}
is required and sufficient to ensure a perfect quadrature for any function with a band not larger than $L$.

\paragraph{Interpolation.}
For the interpolation, the code employs the nearest grid point (NGP) assignment, e.g., we assign to every deflected direction a value of the sky signal computed at the nearest center of a pixel of 
the assumed pixelization scheme, therefore the respective sky signal values are calculable at the fast spherical harmonic speed.
The NGP assignment  is extremely quick and simple, but it requires the computations to be performed at a very high resolution to ensure that the results
are reliable. The sufficient resolution required for this will in general depend on the intrinsic sky signal prior to the lensing procedure, as well as the resolution of the final maps
to be produced, as is discussed in Sect.~\ref{sect:codepars}. 
 As discussed above, in a typical case these are expected to be very high and the computations involved in the problem may quickly 
 become very expensive. Nevertheless, as we show in Sect.~\ref{sec:numPerf}, the overall computational time in this case is only somewhat longer than that involved in some other interpolation 
 schemes, while the memory requirement can be significantly lower. However, the major advantage of this scheme for the purpose of this work is its simplicity and in particular the fact that its
 precision is driven by a single parameter defining the grid resolution.

\begin{figure*}[!htb]
\centering
\includegraphics[width=.333\textwidth]{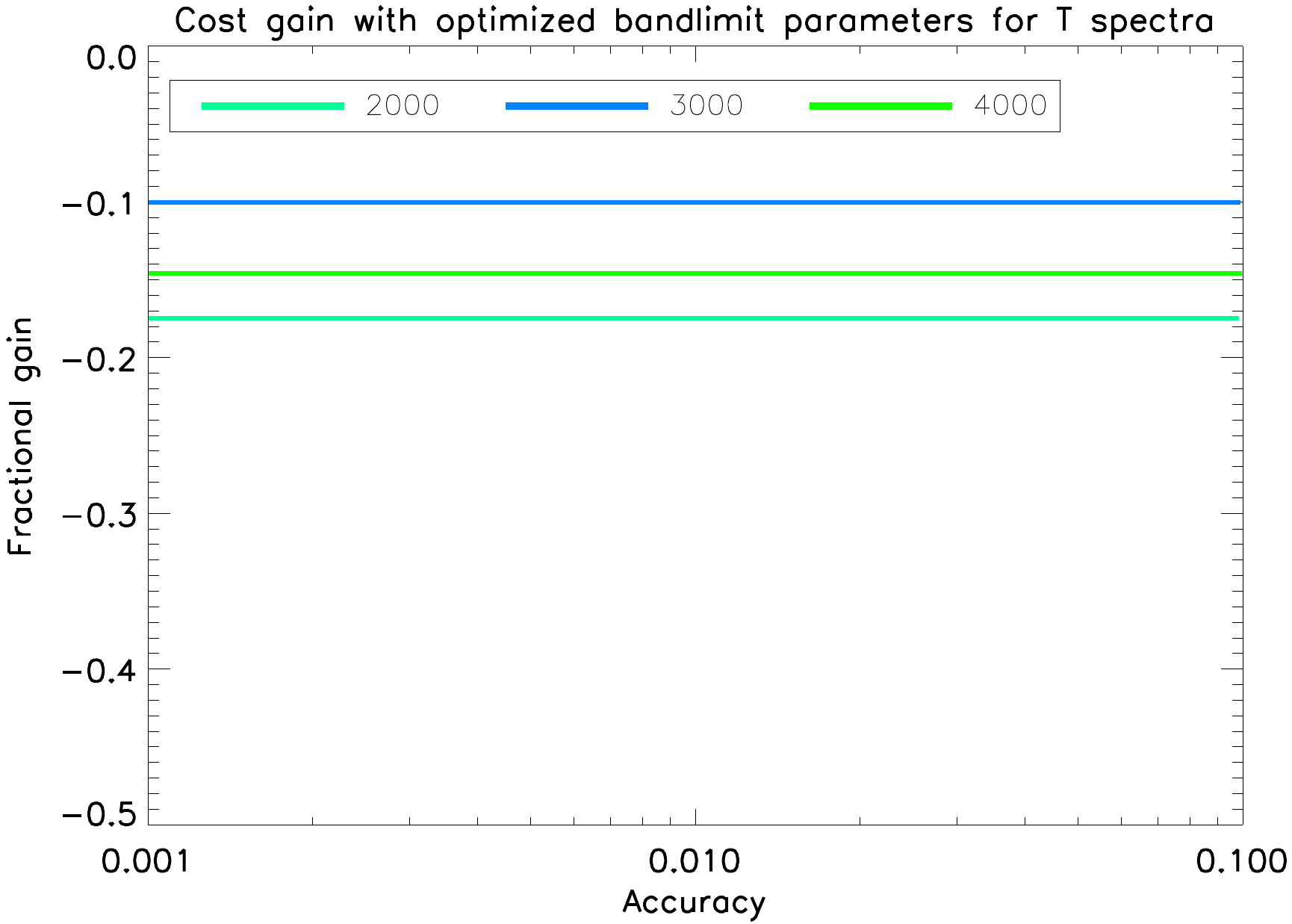}\includegraphics[width=.333\textwidth]{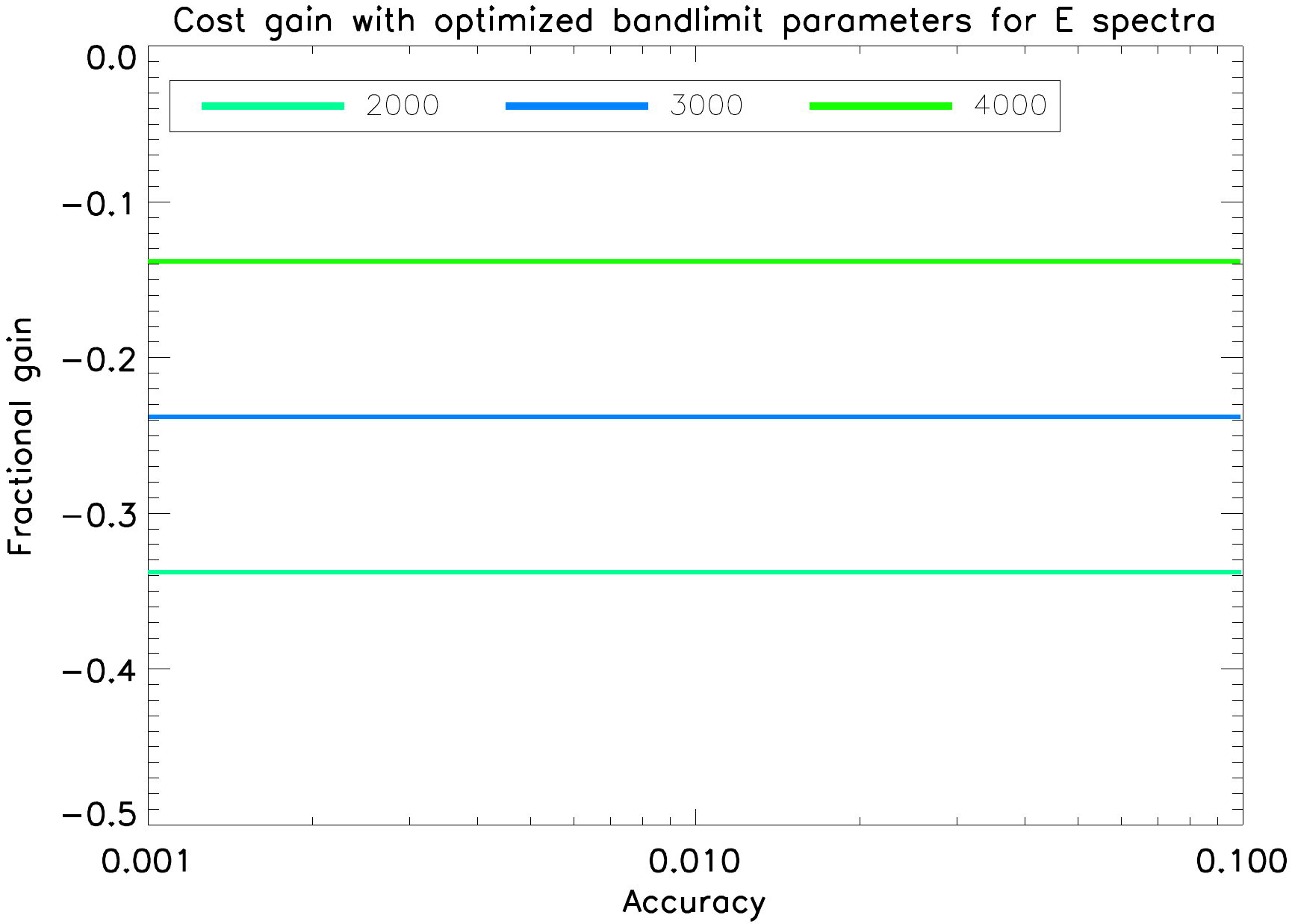}\includegraphics[width=.333\textwidth]{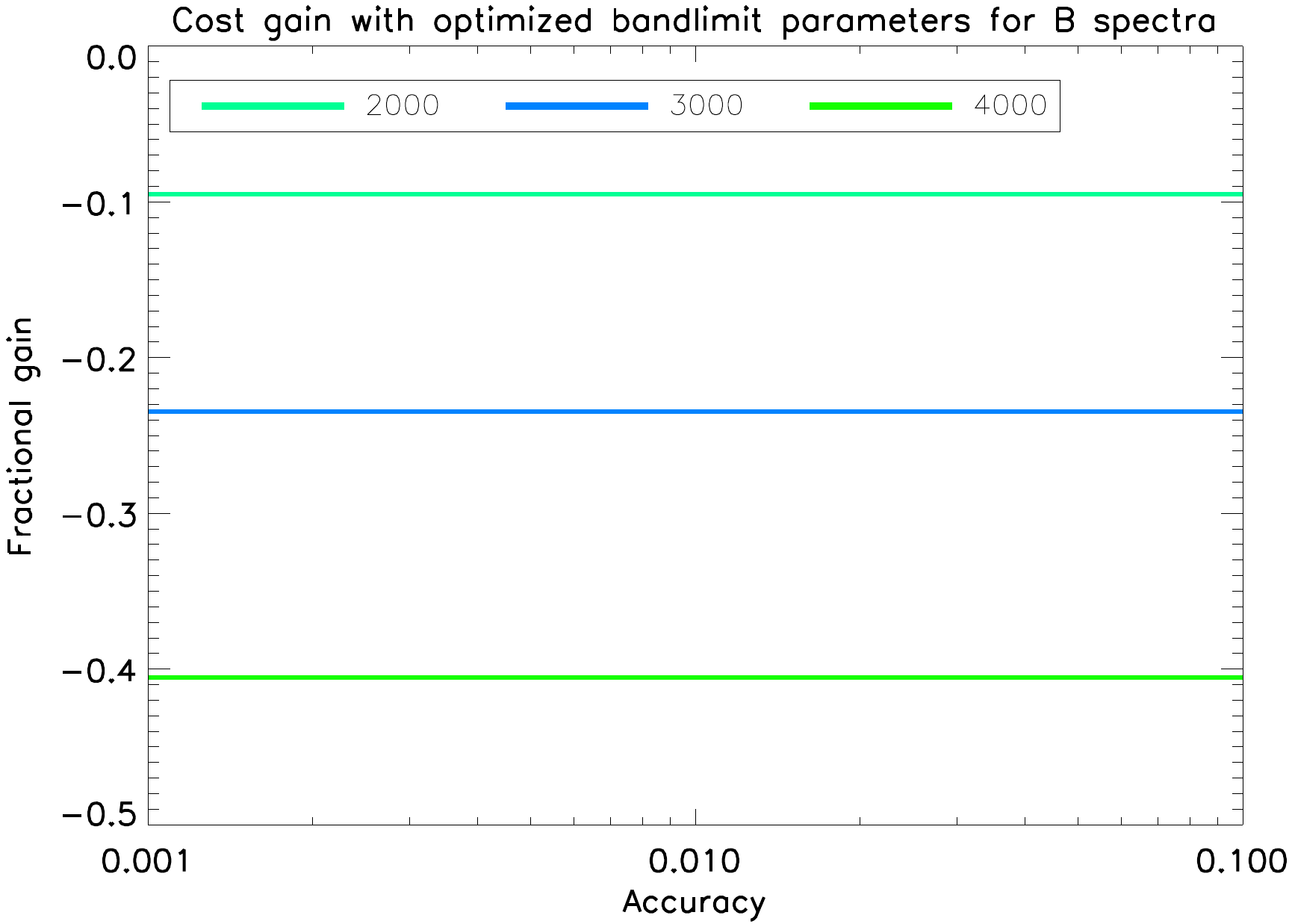}
\caption{Numerical cost gain by using the optimized set of $\ellphi,\elle$ parameters compared with assumin $\ellphi=\elle$ as a function of the accuracy of the computed spectrum for several values of the highest multipole of interest. An oversampling factor of $\kappa=8$ was assumed to compute the cost function.}
\label{fig:costgain}
\end{figure*}

\begin{figure*}[!htb]
\includegraphics[width=.5\textwidth]{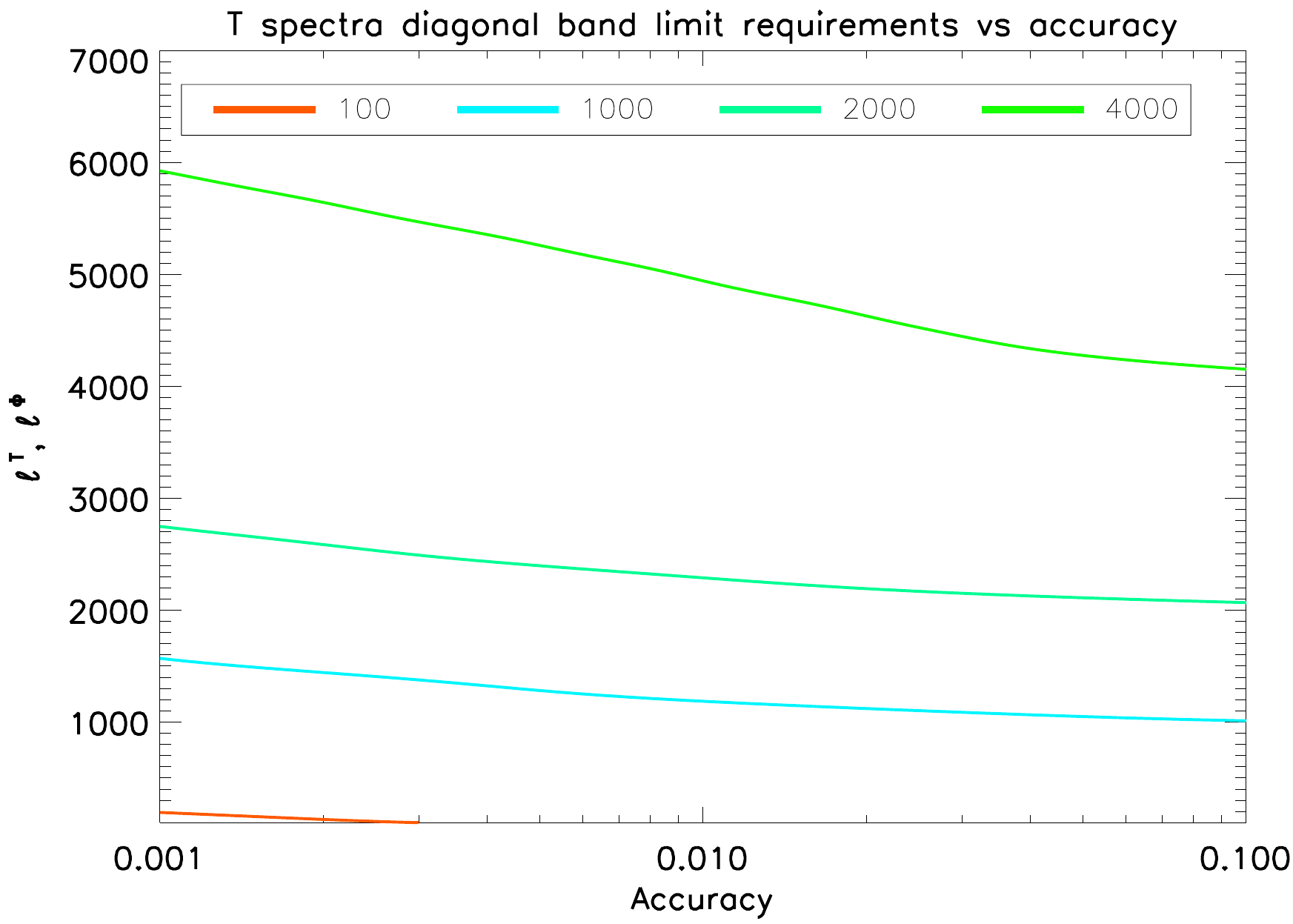}\includegraphics[width=.5\textwidth]{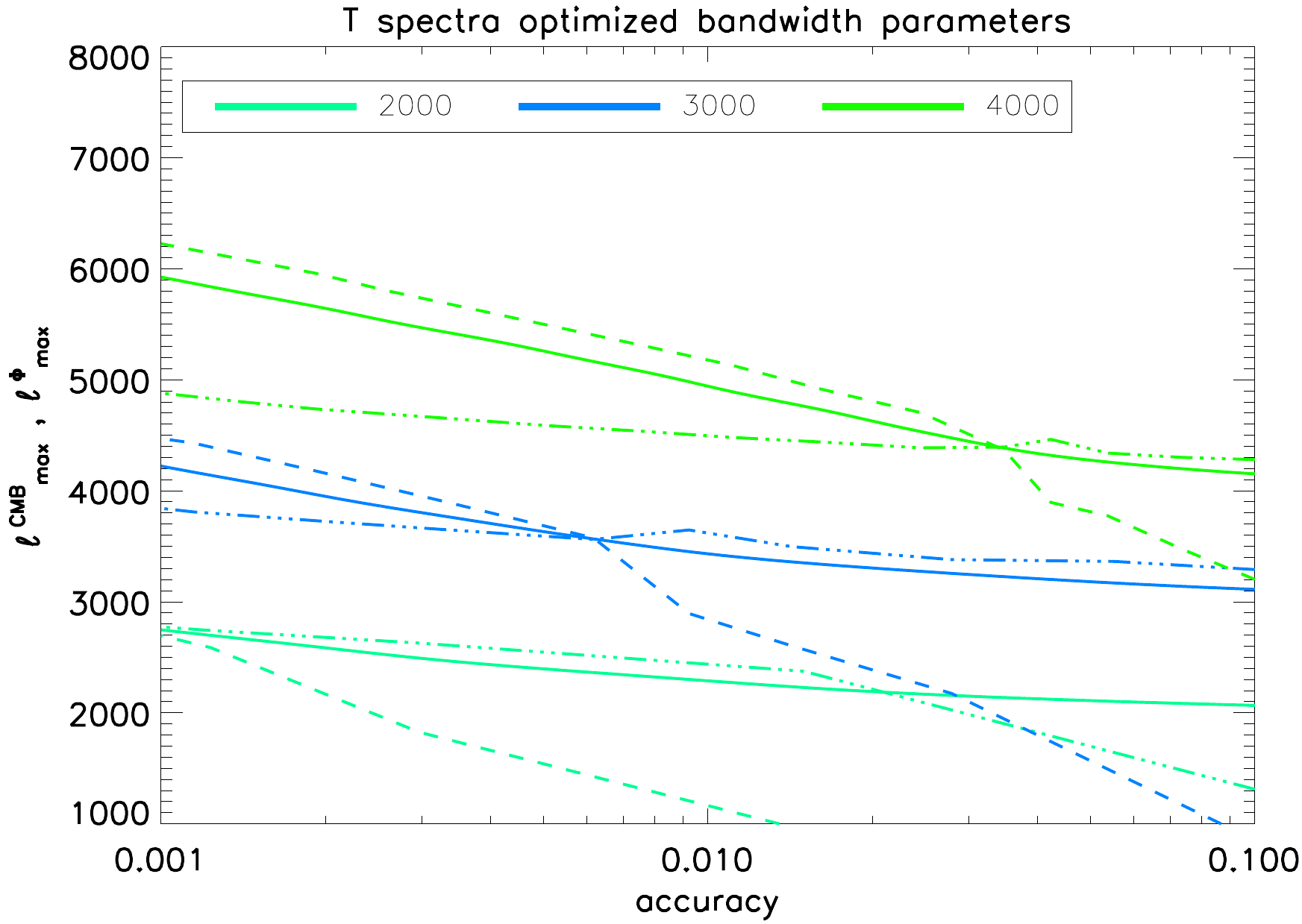}\\
\includegraphics[width=.5\textwidth]{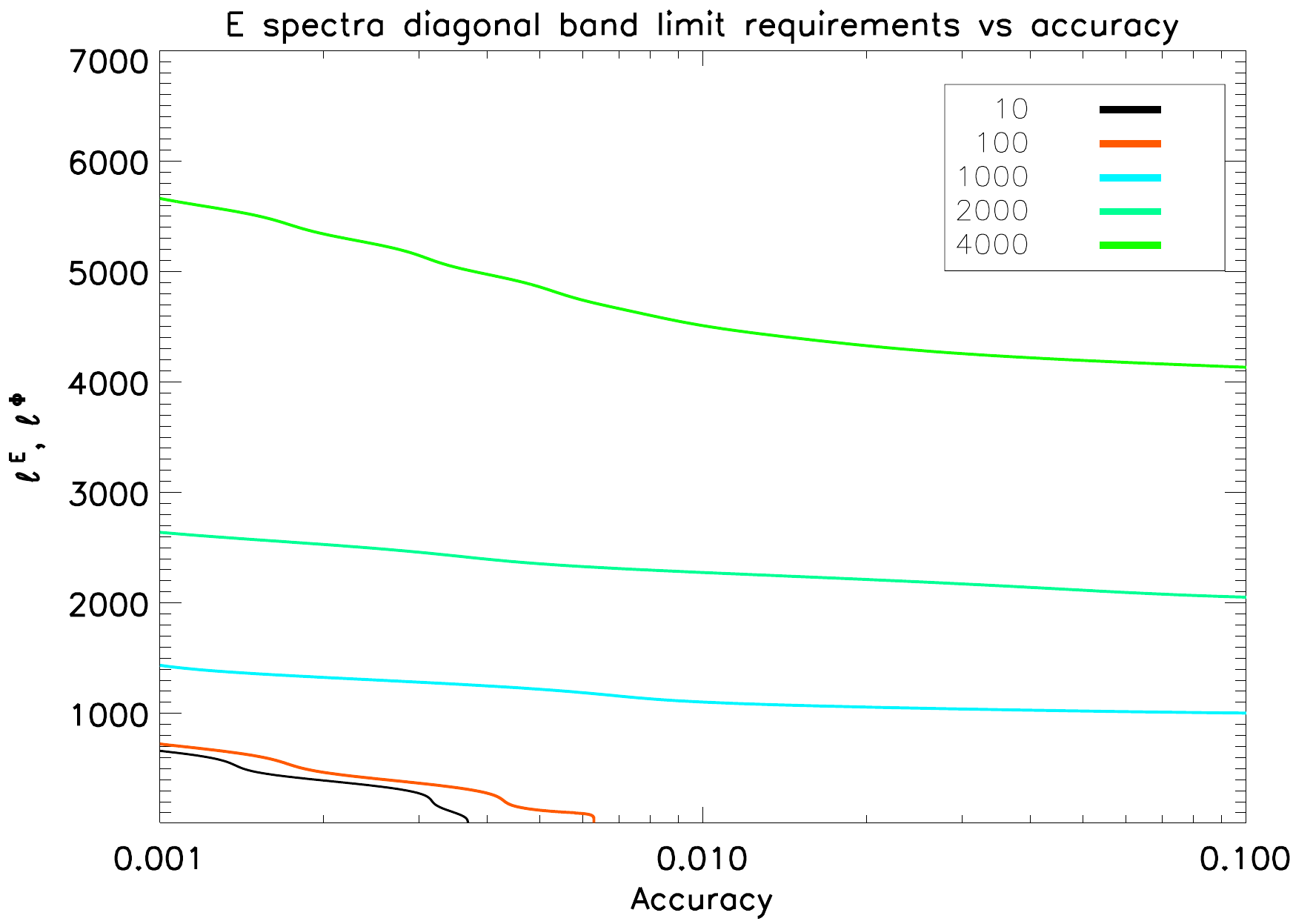}\includegraphics[width=.5\textwidth]{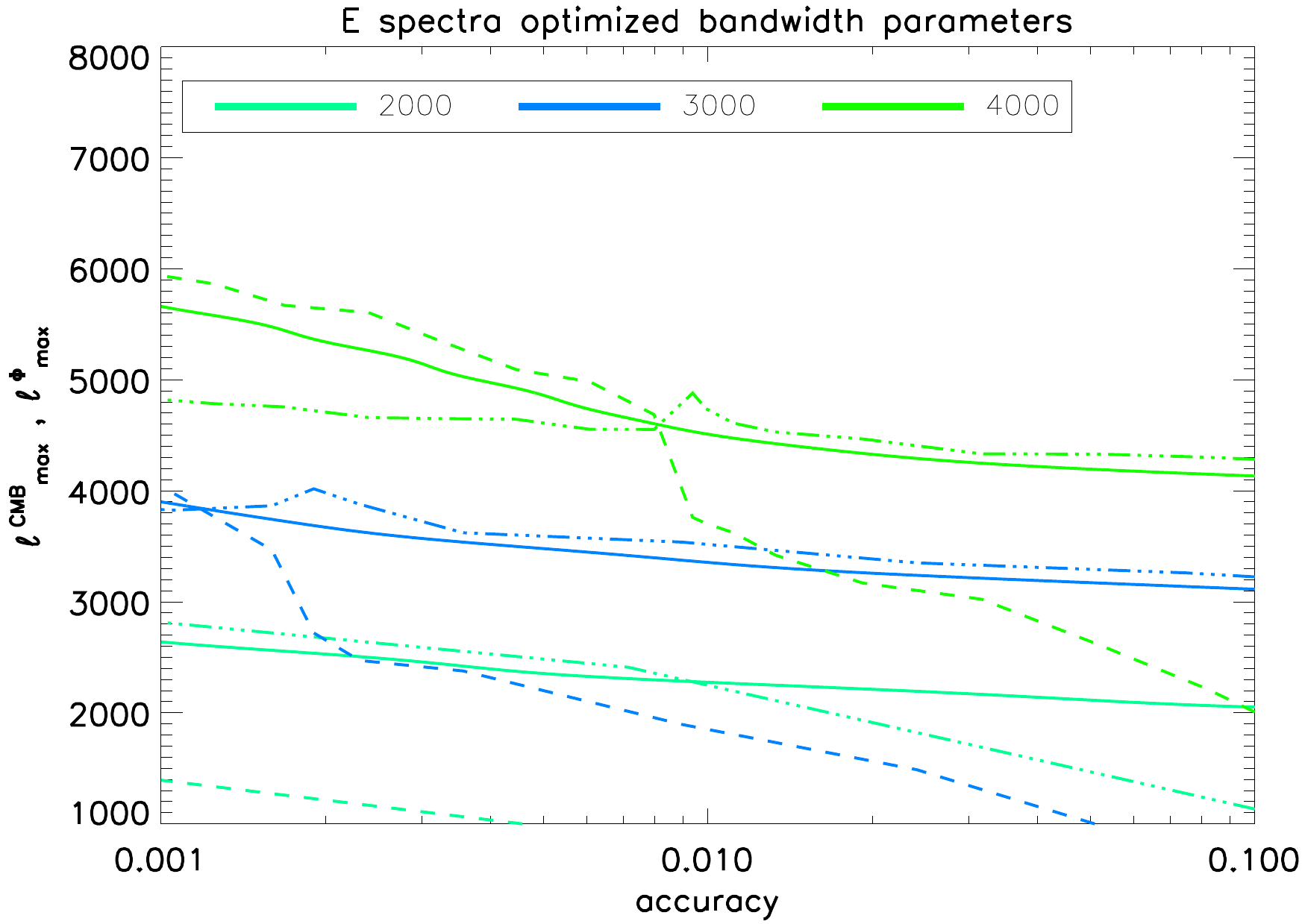}\\
\includegraphics[width=.5\textwidth]{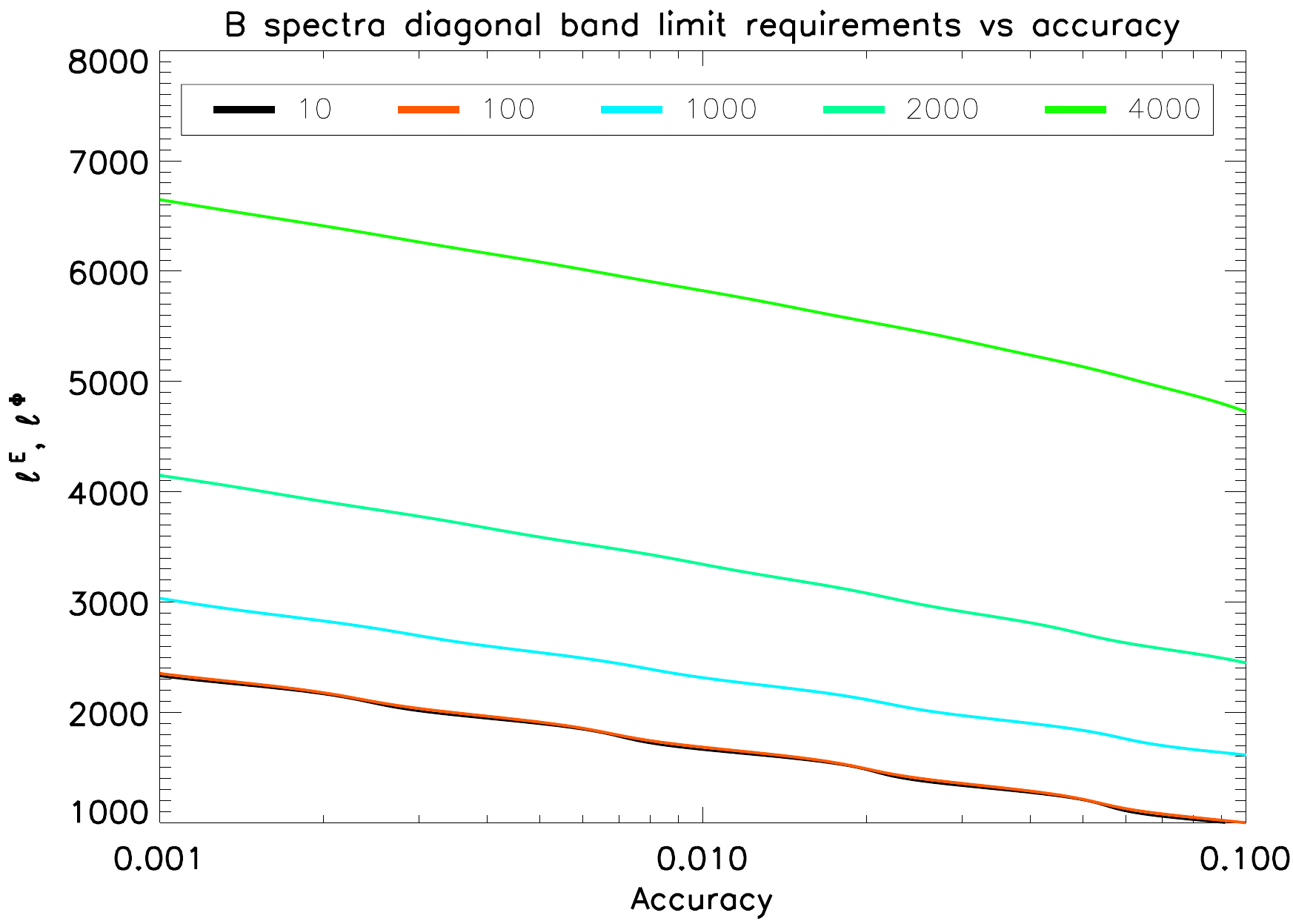}\includegraphics[width=.5\textwidth]{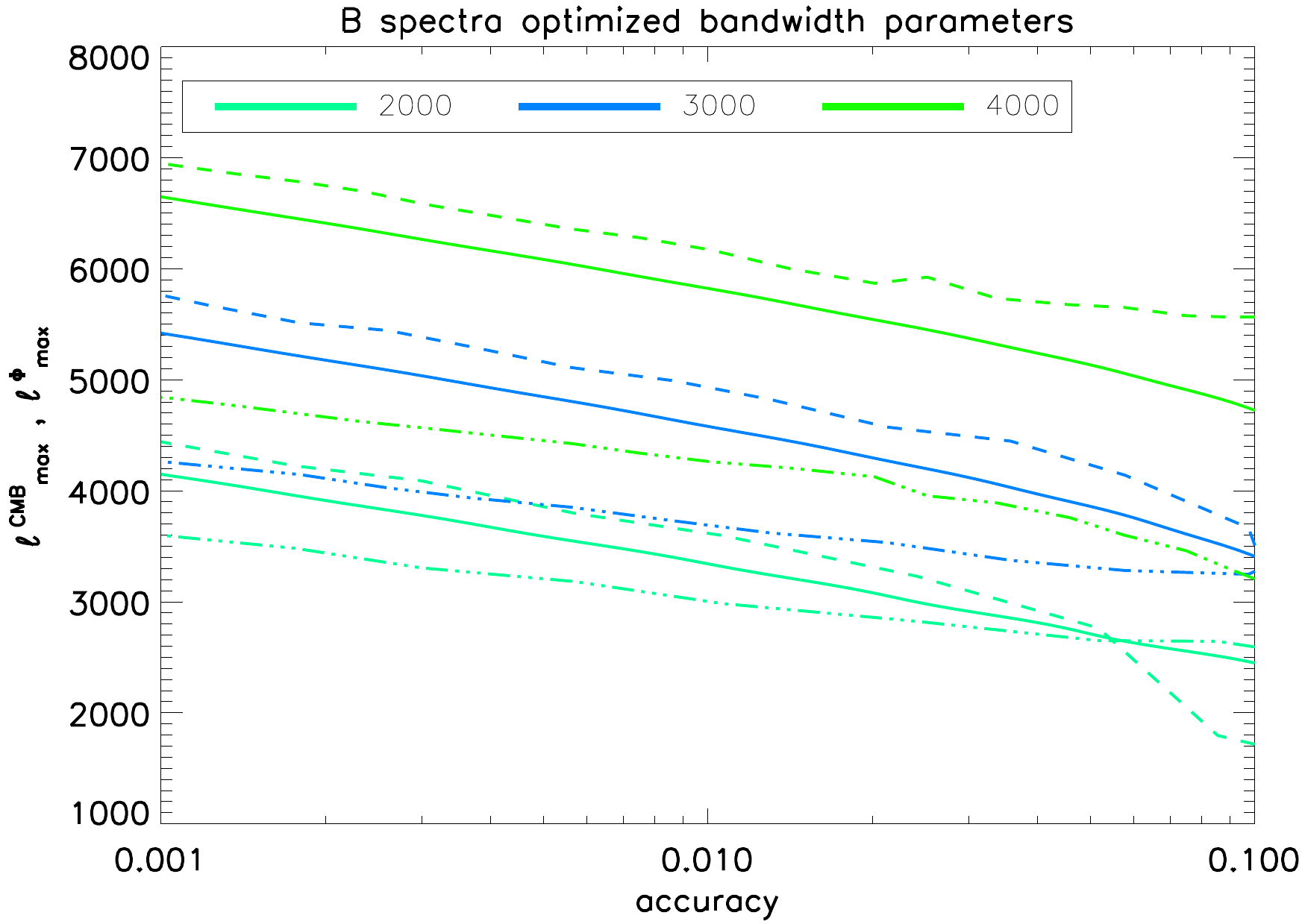}

\caption{Left column: Examples of band-limit requirements for unlensed CMB and lensing potential as a function of obtainable accuracy, assuming they are equal, on several multipoles of lensed $T$ , $E$  and $B$ spectra. Right column: summary of the optimized choice of bandwidth parameters for CMB (dot-dashed) and  $\Phi$ field (dashed) compared with the cost function of the algorithm, as defined in eq. \ref{eq:costfunc}. The diagonal bandwidth parameters are shown as a solid line for comparison.}
\label{fig:diag_bandreq}
\end{figure*}

\paragraph{Spherical harmonic transforms.}
To sidestep the problem of computing spherical harmonic transforms with a huge number of grid points and a very high band limit, \lenstwo\ resorts to parallel computers and massively 
parallel numerical applications. With these becoming quickly more ubiquitous and
affordable this solution is becoming progressively more attractive. 

Parallelization of the fast spherical harmonic transforms is difficult due to the character of
the input and output objects and the involved computations, where a calculation of each output datum requires knowledge of, and access to, all input data.
This is clearly not straightforward to achieve without extensive data redundancy, as done e.g., in \lenspix\ or parallel routines of HEALPix, or complex data exchanges between 
the CPUs involved in the computation.
 To avoid such problems in our implementation we used the publicly available 
scalable spherical harmonic transform (\stwo) library\footnote{\url{http://www.apc.univ-paris7.fr/APC\_CS/Recherche/Adamis/MIDAS09/software/s2hat/s2hat.html}}.
This library provides a set of routines designed to perform harmonic analysis of arbitrary spin fields on the sphere on distributed memory architectures (though it has an efficient performance even when working in the serial case). It has a nearly perfect memory scalability obtained via a memory distribution of all main pixel and harmonic domain objects (i.e., maps and harmonic coefficients), and ensures very good load balance from the memory
and calculation points of view. 
It is a very flexible tool that allows a simultaneous, multi-map analysis of any iso-latitude pixelization, symmetric with respect to the equator, with pixels equally distributed in the azimuthal angle, and provides support for
a number of pixelization schemes, including the above mentioned ECP; see \citet{szydlarski2011} for more details. The core of the library is written in F90 with a C interface and it uses the message passing interface (MPI) to institute distributed memory communication, which ensures its portability. The latest release of the library also includes routines suitable for general purpose graphic processing units (GPGPUs) coded in CUDA~\citep[][]{hupca2011, szydlarski2011, fabbian2012}.
\\*
We emphasize that if a sufficient resolution can be indeed attained, the approach implemented here can produce results with essentially arbitrary precision. In the following
we demonstrate that thi is indeed the case for the described code.

\subsection{Code parameters}\label{sect:codepars}

\subsubsection{Overview}

In this section we describe how we fixed the  essential parameters of the code. We first emphasize important relations between them. A detailed description of the 
procedures used to assign specific values to them, is given in the following sections.

\begin{enumerate}
\item We start by defining a target value in terms of
the highest value of the harmonic mode, $\tell{Y}_{req}$, that we aim to recover and its desired precision, $\varepsilon$. We then use the reasoning from Sect.~\ref{sect:accuracy} to translate this requirement into corresponding bandwidths, $\ellx$ and $\ellphi$, of  the relevant unlensed signals, $X$ and $\Phi$.
These ensure that the precision of all modes of the lensed signal up to $\tell{Y}_{req}$ will be not lower than $\varepsilon$, barring any unaccounted-for, numerical inaccuracies.
The values of $\ellx$ and $\ellphi$ are then used to estimate the bandwidth of the output, lensed map, $\tell{Y}_{out}$.
\item
We then simulate two unlensed maps, $\bm{m}^X$ and $\bm{m}^\Phi$, of the signal $X$ and potential field, $\Phi$, with their band limits set to $\ellx$ and $\ellphi$, 
as estimated earlier. The number of pixels of the displacement map, $\bm{m}^\Phi$, is equal to that in the output map of the lensed signal, and for the ECP grid,
equal therefore to $N^\Phi_{pix} = 4\,{{\tell{Y}_{out}}\,}^2$. The number of pixels in the $X$-signal map, $\bm{m}^X$ is then given
by $N_{pix}^X = 4\,\kappa^2\,{{\ellx}\,}^2$, where $\kappa$ is an overpixelization factor introduced in Sect.~\ref{s3ect:basics} and discussed in detail below, Sect.~\ref{sect:pix}.
For simplicity, we assume that the grid for which the unlensed field $X$ is computed is a subgrid of the grid used for $\Phi$.
\item The reassignment procedure (step 5 of the algorithm, Sect.~\ref{s3ect:basics}) is then straightforwardly performed, leading to the map containing power potentially up to $\tell{Y}_{out}$, which
maybe needed to be filtered down to the band limit of $\tell{Y}_{req}$, as initially required.
\end{enumerate}

\subsubsection{Intrinsic bandwidths}
We employ the procedure described earlier in this work in Sect.~\ref{sect:accuracy} to set the intrinsic band limits. Instead of using generic predictions, we aim at optimizing their values to ensure the lowest possible computational overhead. To do so we need to provide a  model of the cost of numerical calculations involved in \lenstwo. This is dominated by large spherical harmonic transforms, one needed to calculate the map of $\Phi$ and
the other to calculate that of signal $X$. Given the parameters introduced above and because the total cost of a spherical harmonic transform is proportional to $N_{pix}\,\ell_{max}$ we therefore obtain
\begin{eqnarray}
C\equiv C(\ellphi,\ellx) & \propto & 2\,N_{pix}^{\Phi}\ellphi+n_{stokes}\,N_{pix}^{X}\ellx \nonumber\\
& = & 8\, {\tell{Y}_{out}\,}^2 \, + \, 4\,n_{stokes}\,\kappa^2\,{\ellx\,}^2\nonumber\\
& = & 8\, \eta^2 \, (\ellphi+\ellx)^2\,\ellphi \, + \, 4\,n_{stokes}\,\kappa^2\,{\ellx\,}^3.
\label{eq:costfunc}
\end{eqnarray}
Here $n_{stokes}$ stands for the number of signal maps, that we aim to produce and is equal $1$ -- $T$-only, $2$ -- $E$ and $B$, or $3$ -- $T$, $E$, and $B$, while for the field $\Phi$ the pre-factor is fixed and equal to
$2$, reflecting the number of components of a vector field on the sphere. In deriving the last equation above we have assumed that $\tell{Y}_{out} = \eta\, (\ellphi+\ellx)$. This is justified below, as are the values that should be adopted for $\eta$ and $\kappa$. The expression above includes neither the cost of the interpolation nor reshuffling, but because in both these cases the number of involved operations is proportional to $N_{pix}$, their cost is negligible with respect to that of the transforms.
\\*

Solving for the optimized values of the bandwidths, which simultaneously ensure the desired precision,  $\varepsilon$, at a selected multipole, $\tell{Y}_{req}$, involves minimizing the cost function in Eq.~\ref{eq:costfunc}, with a constraint, $A_{\tell{Y}_{req}}^{Y}(\ellphi,\ellx) = \varepsilon$, Eqs.~\ref{Baccuracydef} and~\ref{Xaccuracydef}.
This is implemented as follows. First, we define a grid of levels of the cost function and for each level calculate the best accuracy achievable on its corresponding contour.
If this accuracy for some of the levels is close to our target, we find a corresponding pair of bandwidth values, $(\ellphi,\ellx)$, which then defines our optimized solution. If none of the accuracies is sufficiently close to 
the required precision, we take two levels for which the assigned accuracies bracket the target value and insert an intermediate level for which we calculate the corresponding best accuracy. We repeat this procedure
until the best accuracy found for the newly added contour is sufficiently close to the target one. We then use it to find the pair of the optimized bandwidths as above.
As mentioned before, in general, the two optimized bandwidth values will not be equal.  This appears to be particularly the case when simulating the CMB spectra at very high multipoles and especially in the cases involving the $B$ modes, which have broader kernels and are more demanding in terms of bandwidth requirements. 
The procedure allows one to gain a factor of nearly $40$\% in terms of runtime inthea range of accuracy of interest for lensed $B$ multipoles close to $4000$, especially if high oversampling is required. For temperature and $E$-mode polarization, where less extra power is required in $\Phi$ to obtain an accurate result, the gain can be quantified to be nearly $20$\% - $30$\%. We report  in Figure~\ref{fig:diag_bandreq} the dependence of the optimized bandwidth parameters as a function of the required accuracy imposed at different lensed multipoles of $T$, $E$, and $B$ spectra, in the right column, and contrast them with the bandwidths obtained in the case when both of them
are assumed to be equal. In Figure~\ref{fig:costgain} we show typical runtime gains as a function of the required accuracy. 

We note that here that whether we choose to optimize the bandwidths or just assume that they are equal, we find that imposing a certain accuracy level at some multipole, $\tell'$, ensures that the same accuracy requirement
will be fulfilled for all $\tell{} \le \tell'$.

\subsubsection{Lensed map band-limit}
For the resolution of the final map, we note that in an absence of
numerical effects, such as those due to the pixelization and interpolation, the lensing procedure would be described by Eq.~\ref{eq:kerb} and the bandwidth of the lensed map would 
be simply given by $\ellx + \ellphi$. In the presence of the numerical effects, the output map will have an effective bandwidth typically higher than that, which will  lead to some power-aliasing at the high-$\ell$ end if this theoretical band limit is imposed. We find this to be indeed the case in our numerical calculations. However, we also find that once the overpixelization factor is set correctly, 
 the aliasing is localized to at most
$25$\% of the bandwidth and therefore easy to deal with in post-processing, e.g., step 6 of the algorithm outline in Sect.~\ref{s3ect:basics}. Consequently, we used
$\tell{Y}_{out} = \eta\, (\ellx+\ellphi)$ in our numerical simulations, with $\eta=1.25$  as the band limit. 

It is important to emphasize that NGP is one of the sources of the aliasing, because it does not preserve the bandwidth of the interpolated function, like some of the other, ad hoc procedures proposed
 in this context. Clearly, an interpolation that preserves the function bandwidth would be a significant improvement for this type of algorithms, if it comes without prohibitive numerical cost. We leave such 
an investigation to future work.

\begin{figure}[!htbp]
\centering
\includegraphics[width=.49\textwidth]{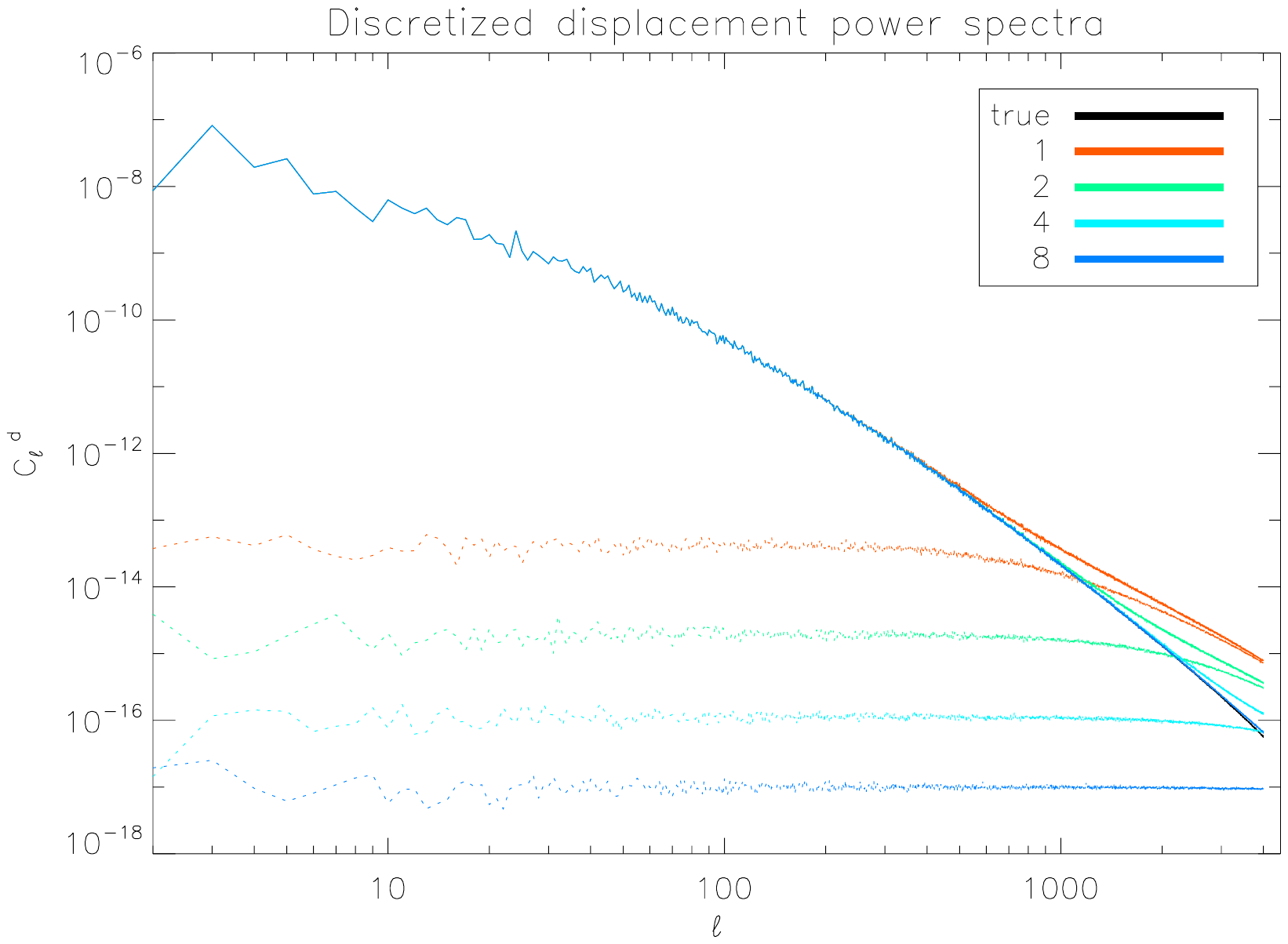}
\caption{Comparison between the $E$-modes power spectra of the input displacement field (black) and the displacement field after NGP assignment for several values of the oversampling factor $\kappa$. The input displacement is computed on an ECP grid with a number of pixel $N_{pix}=16384^{2}$ while the discretized one is the result of an NGP assignment on a grid of $\kappa^{2}N_{pix}$. With progressively higher resolution the extra power due to discretization becomes negligible and the two spectra become almost indistinguishable. The discretization-induced error power spectrum is shown as a dotted line for reference; both $E$ and $B$ modes of the discretization error have the same power spectra.}
\label{fig:discdisp}
\end{figure}

\subsubsection{Overpixelization factor}\label{sect:pix}
As we explained already our interpolation procedure consists of two steps: an overpixelization that is followed by an NGP assignment. The overpixelization involves producing maps with the sky signal sampled at
significantly higher rate than is necessary given from the signal's band limit. For the ECP grids used internally by \lenstwo, this is implemented by using $\kappa$-times more points in each of the $\phi$ and 
$\theta$ directions. The remaining problem is then to fix the appropriate value of $\kappa$. To do so, we first observe that for the overpixelized grid, the NGP assignment can be seen in two ways. Either as approximating the true
value of the sky signal, which needs to be calculated in one of the displaced directions, which are precisely computed in turn, which is the standard perspective and the only one available if a more sophisticated interpolation scheme
is applied. Or it can be seen as approximating each displaced direction by a direction pointing toward the nearest grid point, with a correct sky signal value assigned to it. This second viewpoint provides us with an independent test to check if the density of 
our overpixelized grid is sufficiently high. The involved procedure involves first calculating the approximate displacement field and  its power spectrum, which is then compared with the input power
spectrum for the gravitational potential, $\Phi$.  We note that the approximation used here can in general generate a non-zero curl and
therefore there will be two non-vanishing spectra of the approximate displacement field, corresponding to its $E$ (gradient) and $B$ (curl) components.
We then require that the recovered $B$ spectrum is significantly smaller than that of $E$, and that both the recovered $E$ spectrum and the input one
agree sufficiently well up to the angular scales, which are of interest given the $\ell$-range of the lensed spectrum we are
after and its precision.  These latter two are turned into the $\ell$-range requirement using one of the 2D kernels.  

Examples of such comparisons are shown in Fig.~\ref{fig:discdisp} for a number of values of the oversampling factor ranging from $\kappa = 1$ up to $8$. We see that for the latter value the approximate $E$ spectrum is consistent  over the entire shown range of $\ell$ values and the recovered $B$ is there significantly smaller. 
We therefore continue to use this value in the runs discussed later in this work, even if, as noted in the next section, $\kappa = 4$ could be sufficient at least for $\tell{B} \simlt 2000$. We also point out that, as it might have been expected, the departures of the recovered $E$ spectrum for the displacement from the input one are consistent with the presence of the non-zero $B$-type mode in the approximated displacement field with an amplitude similar to that of its $E$-mode spectrum, which renders our two criteria redundant.
In addition, if only $T$ and $E$ CMB spectra are of interest, then $\kappa\approx 2$ is usually sufficient to obtain accurate results on the scales of interest because the long-tails
of the displacement spectrum are less relevant in these cases.

\begin{figure}[!htbp]
\centering
\includegraphics[width=.5\textwidth]{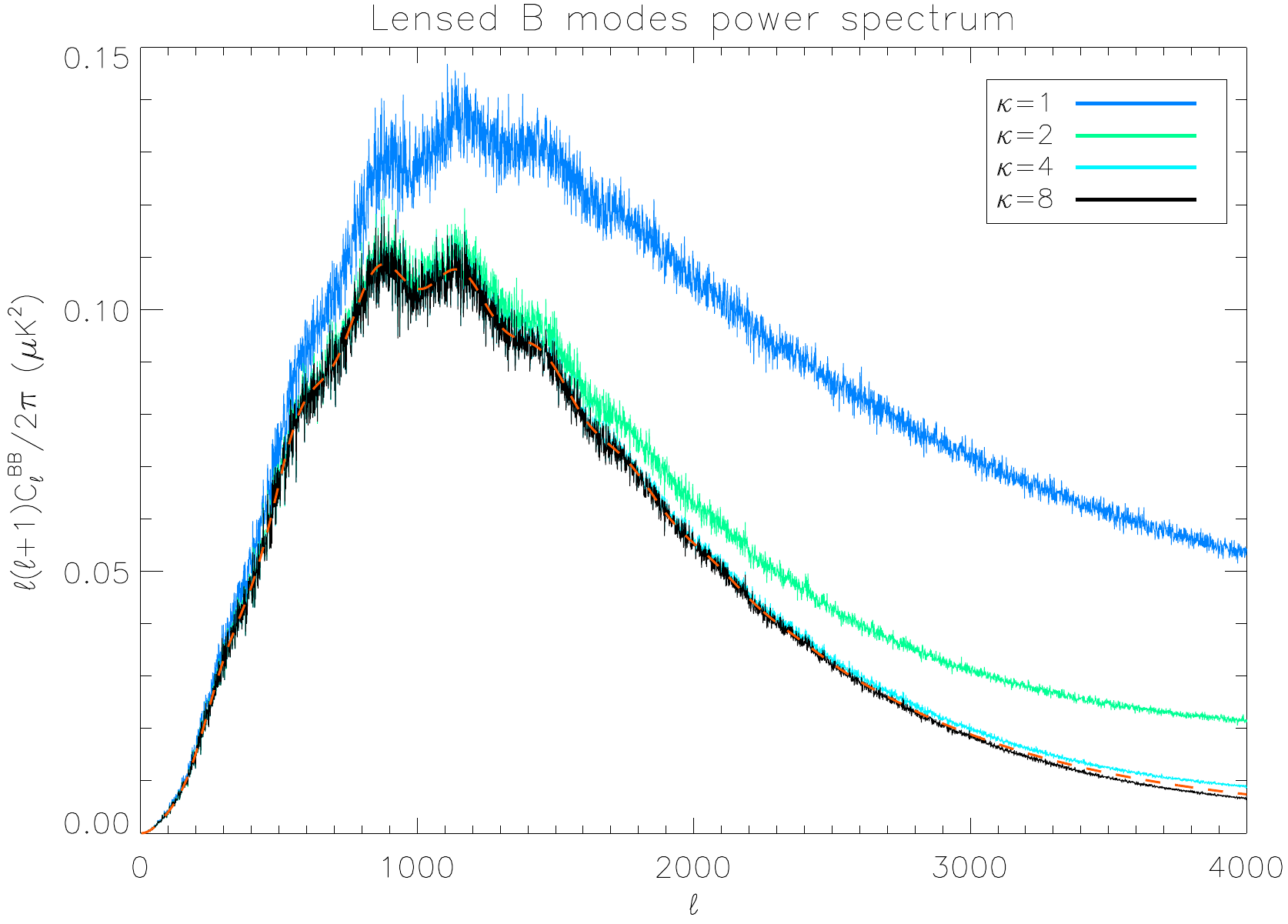}
\caption{Lensed $B$-modes spectrum computed for different values of over sampling factor compared with the lensed spectrum obtained with the analytical Boltzmann code CAMB (red dashed).}
\label{bmodes_vs_sampling}
\end{figure}

For completeness, in Fig.~\ref{bmodes_vs_sampling}  we show the relevant CMB $B$-spectra computed with the same values $\kappa$ as shown in Fig.~\ref{fig:discdisp} and aiming at a high-precision reconstruction for $\tell{B}  \le 2000$, demonstrating that both  overpixelization rates, as inferred above, ensure a satisfactory recovery of this spectrum in the targeted range of $\ell$. We provide more details about this Figure in Sect.~\ref{s3ect:simSpecs}.

\subsection{Validation and tests}
\subsubsection{Simulated kernels}\label{sect:simker}

\begin{figure*}[!htbp]
\centering
\includegraphics[width=.48\textwidth]{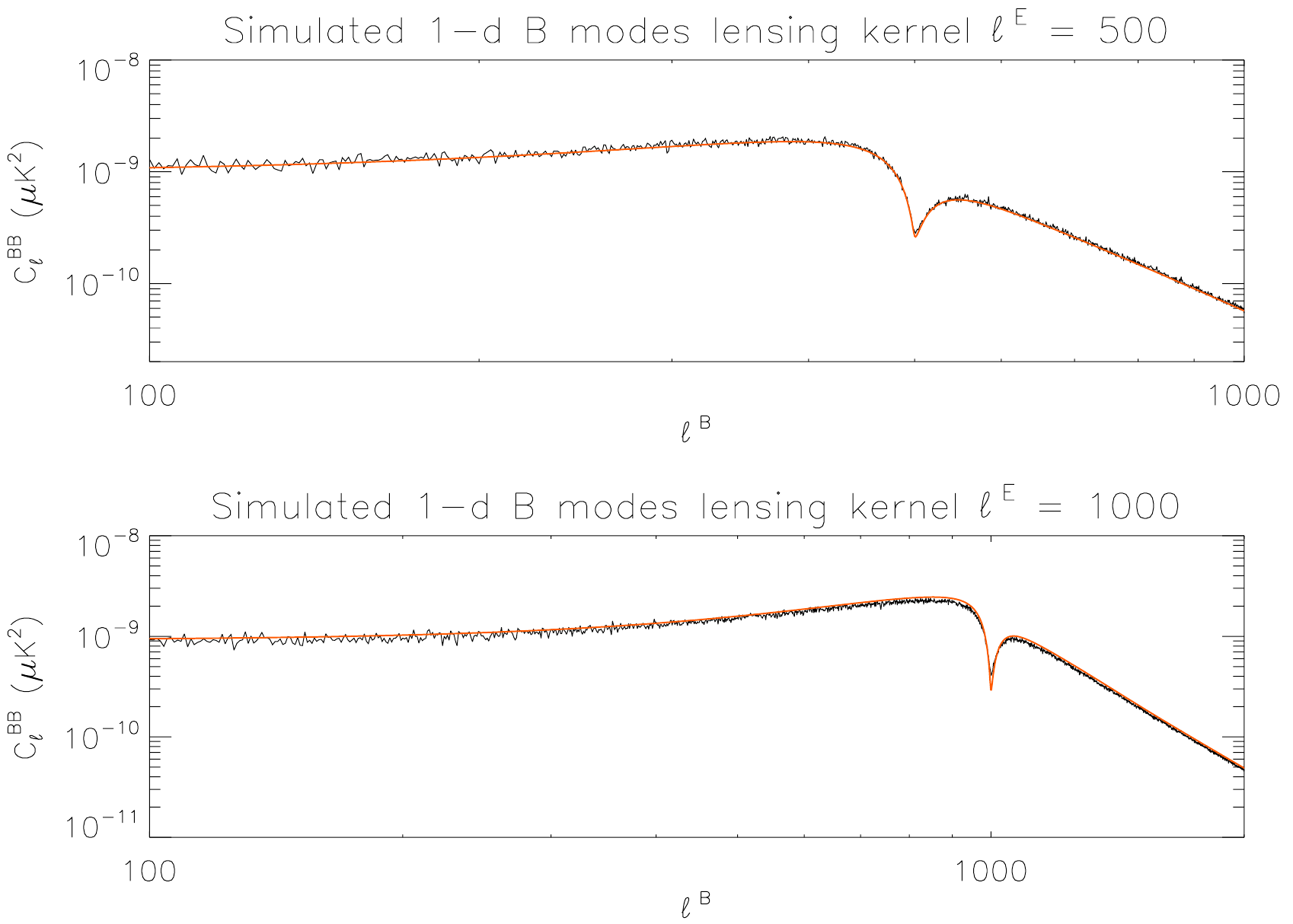}
\includegraphics[width=.48\textwidth]{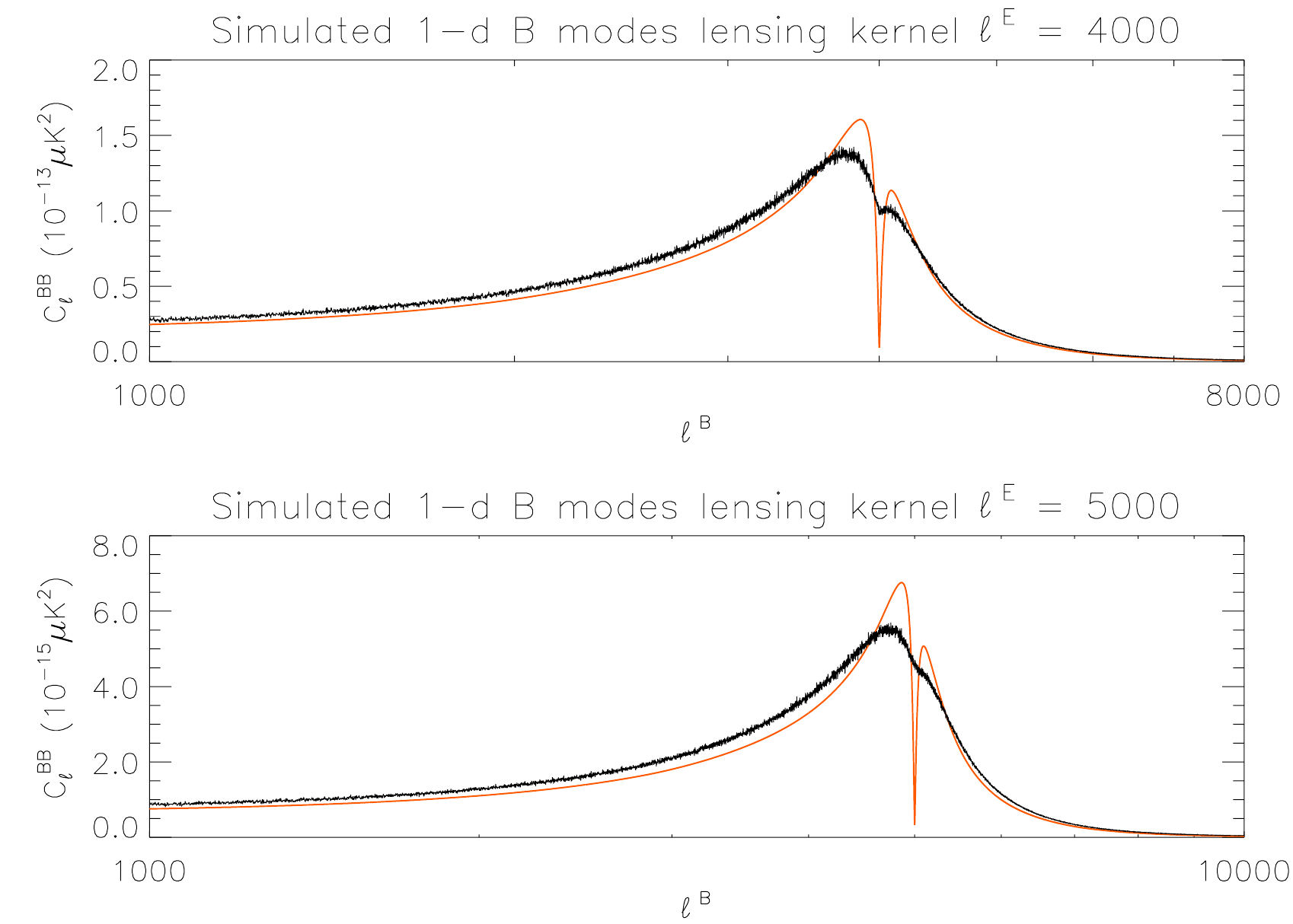}
\caption{Comparison of the simulated, solid black lines, and analytical, solid red lines, 1D $B$-modes kernels, ${\cal H}_{\elle} (\tell{B})$,  shown as a function of $\tell{B}$, and computed for the unlensed CMB $E$ power contained initially only in a single mode, $\elle = 500, 1000, 4000$ and $5000$. For low values of $\elle$, left panels, the agreement between the analytic expression, Eq.~(\ref{eq:kerb}), and numerical results is very good all the way to $\ell^B \simlt 2000$, as expected.  For higher values of $\elle$, though the agreement is poorer, it remains qualitatively very good, which justifies our semi-analytic considerations presented in Sect.~\ref{sect:kers}. } 
\label{fig:numeric-ker}
\end{figure*}

As a first step of validation of our code, we investigated whether its results agree with the prediction of the semi-analytical approach used to model convolution in the harmonic domain. We focus here on numerically feasible studies of the 1D kernels, as defined in Eq.~\ref{eq:oneDimKerDef}. For this purpose we assumed that the unlensed CMB signal, i.e., $E$-modes polarization in the case of the lensed $B$-modes spectrum, contains power only in a single harmonic mode, $\ell_0$ i.e., $C_\ell^{EE} \propto \delta^{Kronecker}_{\ell\,\ell_0}$ and computed
 the resulting lensed $B$-modes spectrum for several values of $\ell_0$ using \lenstwo. We compared them with the analytic results obtained for the same multipole and displayed in Fig~ \ref{fig:1dkers}. 
The results of this calculation are shown in Fig. \ref{fig:numeric-ker}, where we see that in a range where the analytic model is more reliable the agreement between the two curves  is excellent if only a sufficient resolution for the unlensed grid is used.
On the other hand, in the region where the analytic approximation we used is not accurate anymore because amplitudes of the CMB signal and its gradient are comparable and therefore the truncation in the series expansion introduces a non-negligible error, the discrepancy between our analytical model and the simulated 1D kernels becomes more evident. Such an approximation tends to overestimate the contribution of each single mode to its neighboring angular scales of a factor of nearly 50\% with respect to simulated kernels and to slightly underestimate the contribution of each mode to the  kernel tails, i.e., to the multipoles higher than the one in exam. 
Nevertheless, the analytically-approximated and simulated kernels are found to be qualitatively quite similar, which validates therefore our semi-analytic bandwidth requirements presented earlier.
\\
\subsubsection{Simulated spectra}\label{s3ect:simSpecs}
Another batch of performed tests consisted in comparing the spectra obtained from \lenstwo\ and those derived with Boltzmann codes such as CAMB or CLASS.
In particular, the black solid line in Fig.~\ref{bmodes_vs_sampling} shows an example of the result obtained for a simulation of lensed $B$-modes designed to reach an accuracy of up to 0.1\% at $\ell^{B} \simlt 2000$. Because no band-limit optimization is performed, and it is therefore assumed that $\ellphi=\elle=\ellmax$,  the latter value has to be at least  $\ellmax = 4000$, Fig.~\ref{fig:kerlog2}. The lensing convolution of signals with such a band limit leads to polarized maps with power up to $2\,\ell_{max}$, which means that to avoid any aliasing, we would need a grid for the lensed sky and  the displacement field with at least $N_{\theta}\approx 2\ellmax$ rings with $N_{\phi}\approx 2\ellmax$ pixels per ring, i.e., $N_{pix}\approx 16384^{2}$, where we have rounded the number of rings and pixels per rings to a power of $2$. 
Once the band limit of the signals and the respective grid for the lensed sky is set, we still need to define the overpixelization rate as required by our interpolation. 
As noted in the previous section, there seem
to be a general
reasoning based on the discretized displacement spectra, which points toward $\kappa = 8$ as a sufficient value. Because calculating the overpixelized map, albeit with a restricted band-limit,
is the most time-consuming part of the code, there may be an interest on occasion to tune $\kappa$ to be as small as possible.
In this context we find, as illustrated in Figs.~\ref{fig:discdisp} and \ref{bmodes_vs_sampling},  that if the extra power introduced by discretization of displacement field does not exceed $10$\% of  the power in the non-discretized displacement field on scales $\ell\approx 1.5 \,\ell^{B}$, an oversampling factor of $4$ is sufficient to render a power spectrum on scales $\ell\simlt \ell^{B}$  with an accuracy  as determined by the assumed bandwidth. However, the factor $4$ should be treated as a lower bound and be used with care because there will typically be a significant amount of extra power in the $B$-mode spectrum for $\ell \simgt \tell{B}$, which may need to be efficiently filtered out before the respective map can be further used.  In contrast, if the extra power found in the discretized displacement does not exceed $10$\% of  the original power for all angular scales up to $\ellphi_{max}$, then no overshooting takes place and the results remain highly accurate also beyond the scale of interest $\ell^{B}$.\\

\begin{figure*}[!htbp]
\centering
\includegraphics[width=\textwidth]{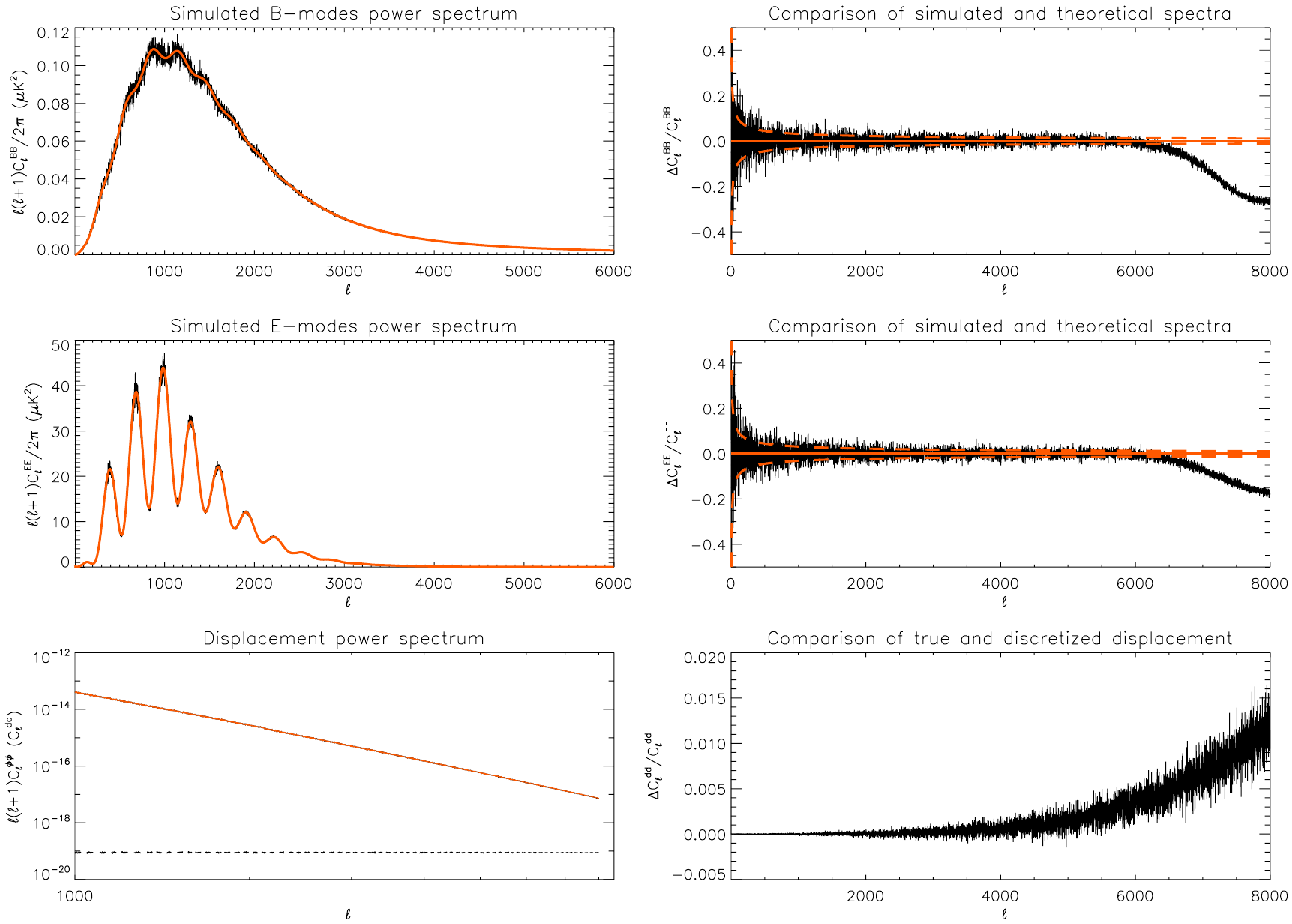}
\caption{Example of a single realization of a high-accuracy and high-precision simulation of the lensed CMB polarization spectra obtained with \lenstwo. The run required an unprecedented resolution and bandwidth and was performed as a demonstration of the code capabilities. The bandwidth parameters have been set to obtain $0.1\%$ accuracy of modes up to $\ell\approx 5500$ for both $B$ and $E$ spectra. 
For the polarization signals, top and middle panels, the left panels show the spectra recovered from the \lenstwo\ run, black solid line, and the theoretical one obtained from CAMB (red solid line). The right panels show the relative difference between these two with the dashed lines outlining the expected $1\sigma$ uncertainty due to sampling variance.
The bottom-left panel shows the $E$-modes power spectrum of the recovered displacement {\em before} the NGP assignment, red solid line, and the $E$-modes and $B$-modes power spectra of the displacement field {\em after} the NGP, black dashed lines, with the former $E$-modes spectrum overlapping the latter nearly perfectly.  The bottom-right panel shows the relative discretization error.}
\label{fig:heroicFig}
\end{figure*}

In Fig.~\ref{fig:heroicFig} we present the spectra for the two polarized components $E$ and $B$ as well as the displacement field, $\Phi$, computed in a run aiming at recovering of these signals in a band up to $\tell{X} \simlt 5500$ with precision better than to $0.1$\%. For this run we assumed the value of the required bandlimits to be  $\elle_{max}=\ellphi_{max}=8000$ . These values were extrapolated from Fig. \ref{fig:diag_bandreq}, where to obtain a $0.1$\% accuracy on $B$-modes on similar angular scales (e.g. $\tell{X}= 4000$) we needed to include power up to $\ellmax\approx \tell{X}+2500$. Following the same prescription as given for the previously detailed case of Fig. \ref{bmodes_vs_sampling}, we set the resolution of the unlensed sky and displacement field to $N_{pix}=32768^2$ while, to ensure the highest possible reliability of the result, we pushed the oversampling factor to 16. The discretization errors introduced by this setup are found to stay under the $1$\% level on all the angular scales involved in the calculation and no significant overshooting is shown (see Fig \ref{fig:heroicFig}).
Though the band limit and resolution involved may look exaggerated from the practical point of view, they simultaneously demonstrate the capability of the numerical code while illustrating our conclusions concerning the precision of these calculations.\\*
In general, we find that a simple algorithm as proposed in \lenstwo\ is  capable of simulating effects of lensing on CMB  over the range of angular scales of interest for current and foreseeable experimental efforts. Moreover, if used properly, it does so with an accuracy that on very small scales is limited rather by the precision of the input power spectrum of unlensed CMB than by the employed numerical algorithm.

\subsection{Convergence tests}
To investigate the precision and reliability of our approach it is interesting to investigate the numerical convergence of the results without relying on a direct comparison to an external Boltzmann code. Since several experiments in the future will be able to target non-Gaussianities in CMB polarization, i.e., the statistical moments beyond the power spectrum, we decided to study the convergence of the results not only on the power spectrum level, but also in the real domain, i.e., on the map level.\\*
\subsection{Power spectrum convergence}
\begin{figure*}[!htb]
\centering
\includegraphics[width=.3\textwidth]{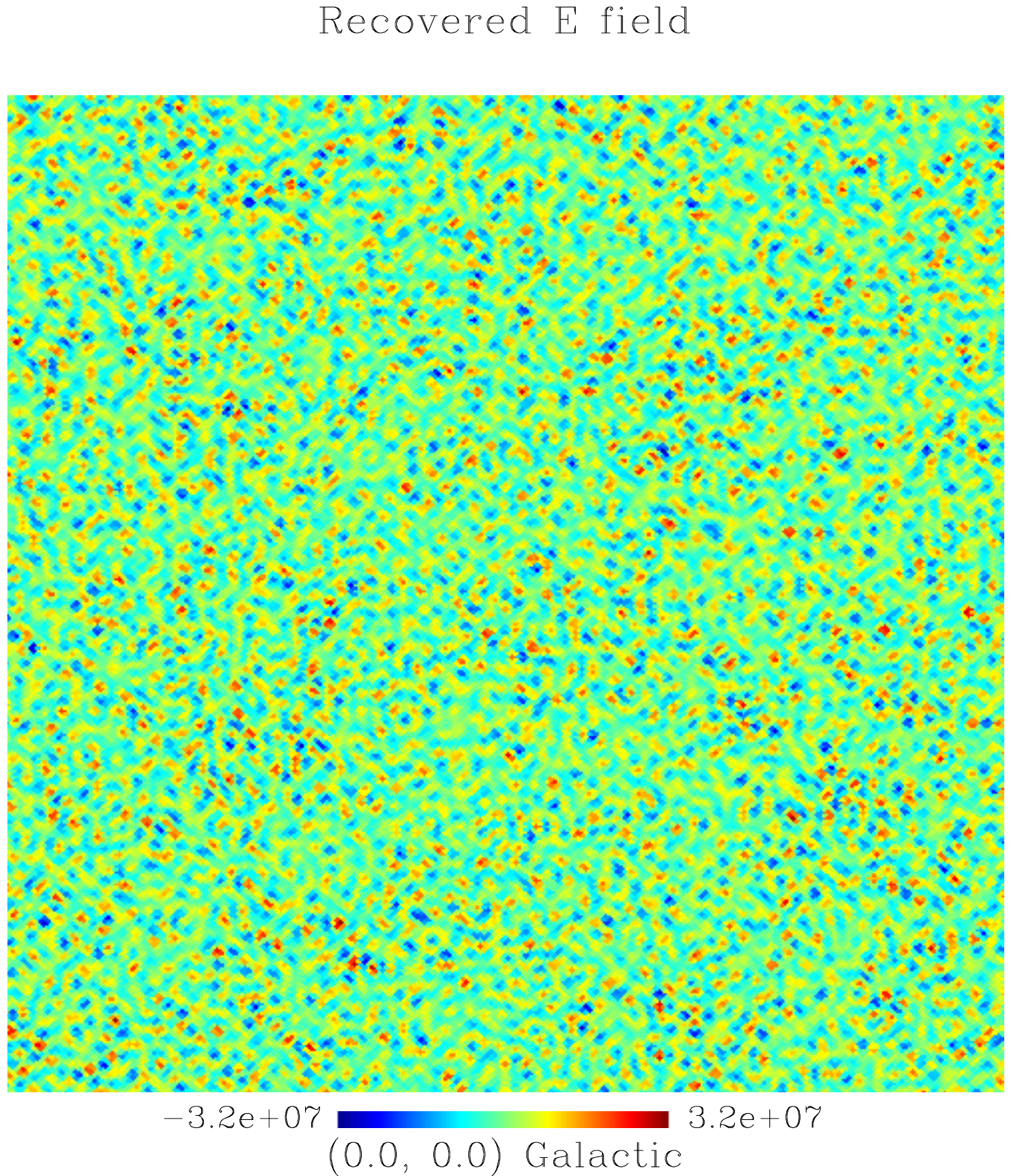}\hspace{.5cm}\includegraphics[width=.3\textwidth]{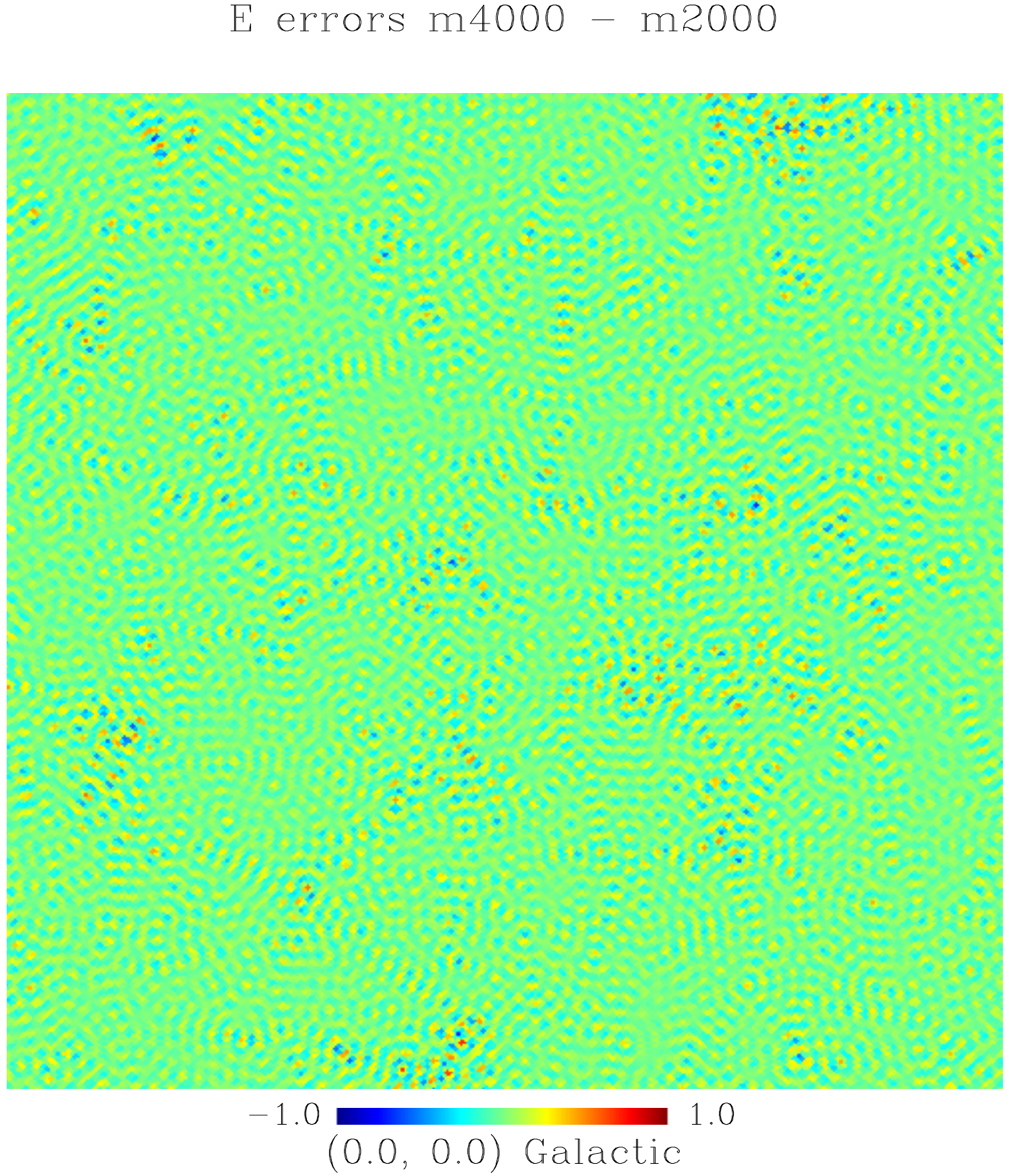}\hspace{.5cm}\includegraphics[width=.3\textwidth]{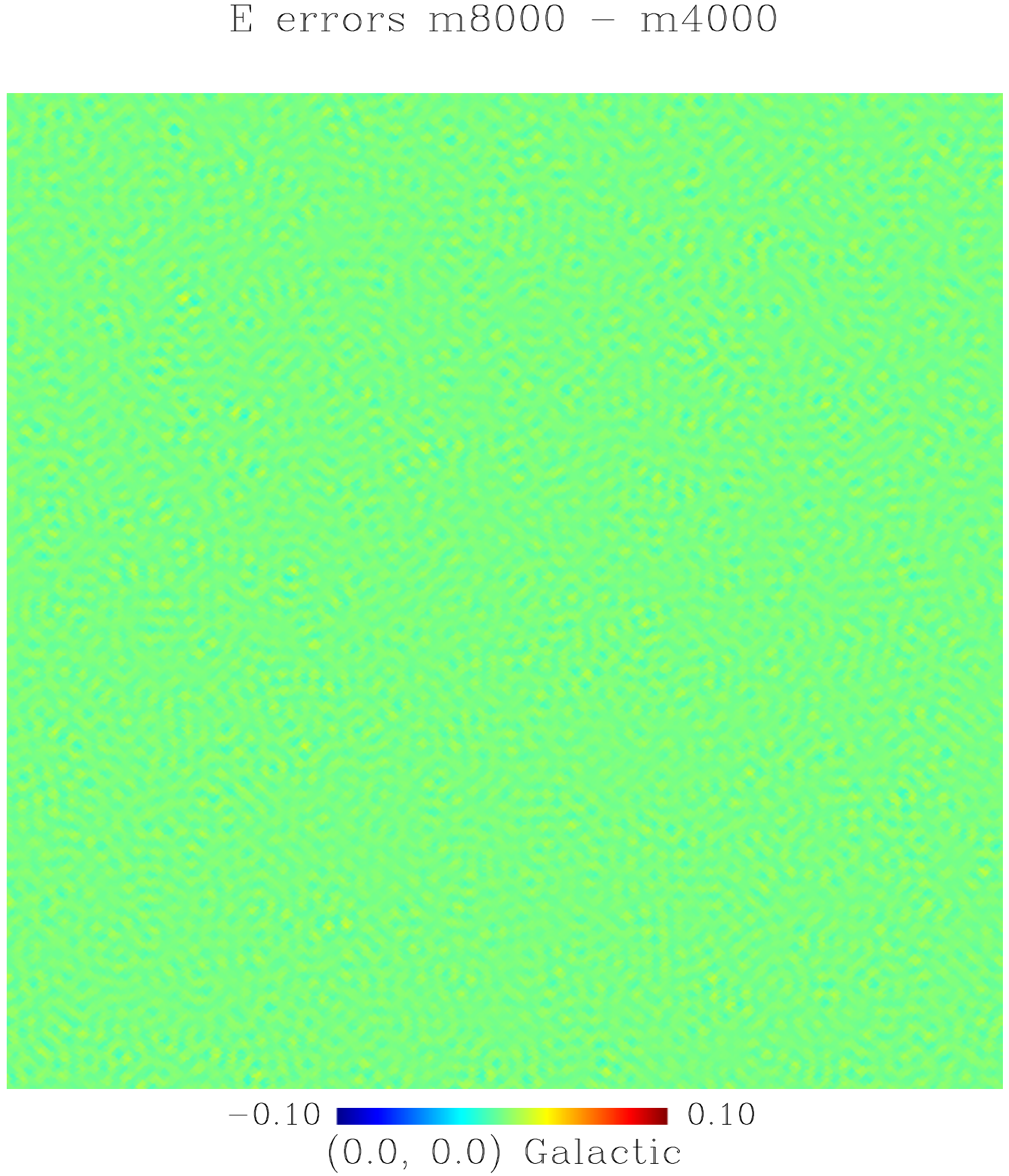}\\
\includegraphics[width=.3\textwidth]{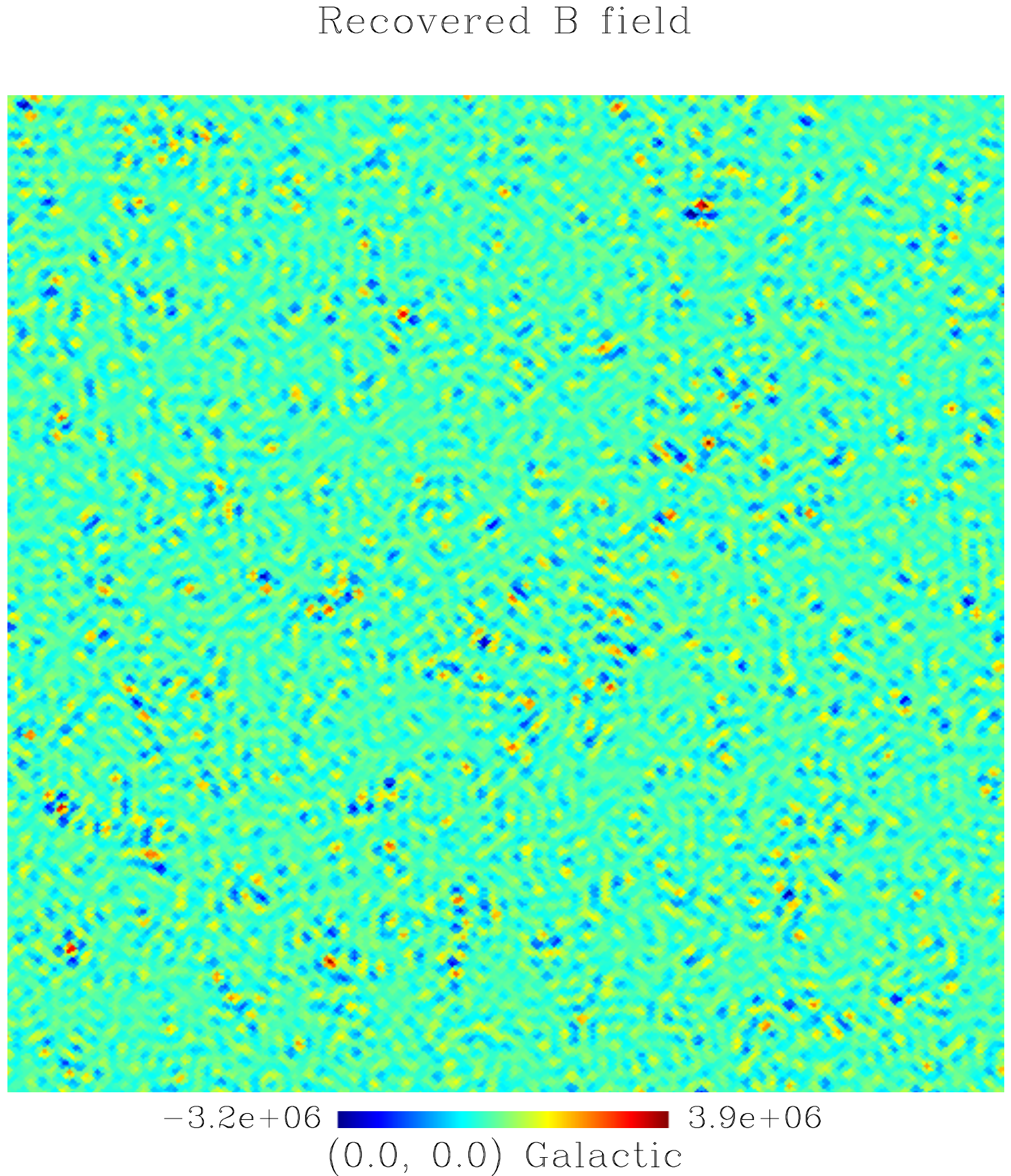}\hspace{.5cm}\includegraphics[width=.3\textwidth]{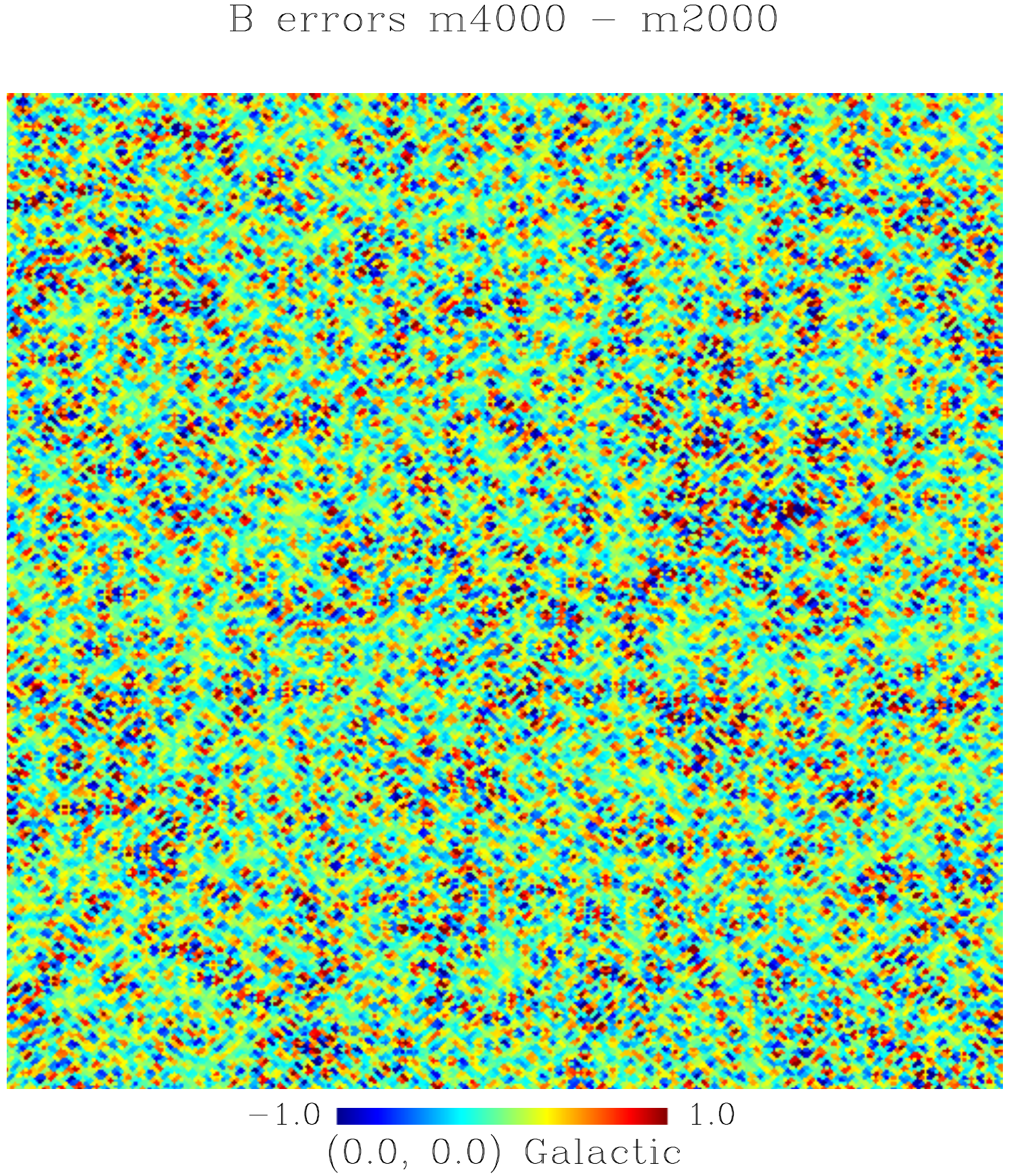}\hspace{.5cm}\includegraphics[width=.3\textwidth]{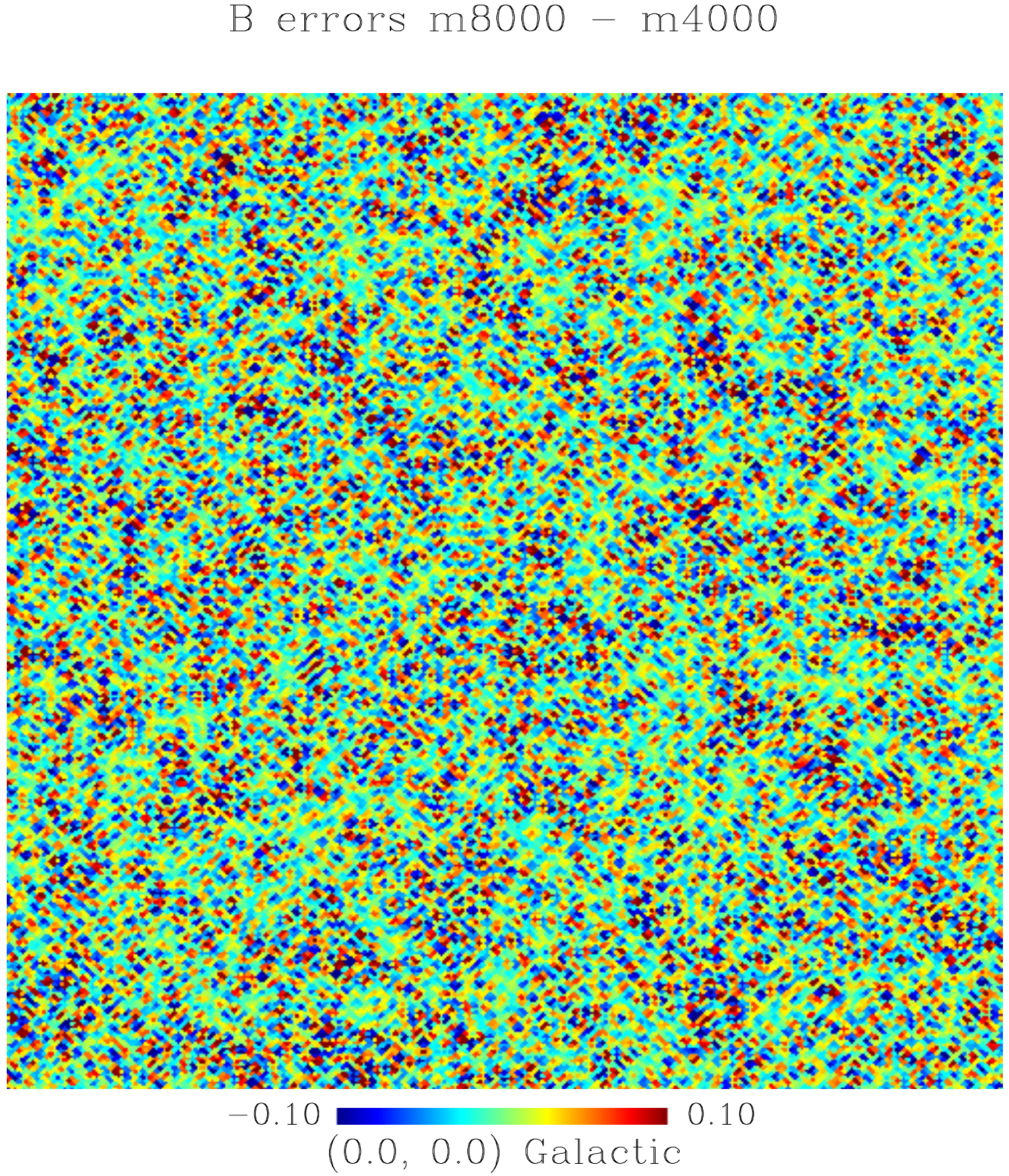}\\
\caption{Maps of the recovered $\chi^{E}$ (upper row) and $\chi^{B}$ (lower row) fields as defined in \cite{seljak-zaldarriaga} (left column) and difference maps normalized by the input map rms, $\mathscr{E}^{E/B}_{\ell_{max, 1}, \ell_{max, 2}}$, Eq.~(\ref{eq:relDiffDef}), computed for  $\l(\ell_{max, 1}, \ell_{max, 2}\r) = \l( 4000, 2000\r)$ (middle column) and  $\l(\ell_{max, 1}, \ell_{max, 2}\r) = \l( 8000, 4000\r)$ (right). There is a factor of $10$ difference between the color scales of the middle and left column.}
\label{fig:errmap2000}
\end{figure*}
We first investigate the convergence of the power spectrum up to a given scale of interest $\tell{X}$ as a function of the bandwidths. This procedure allows us to simultaneously show the precision of our code and also to indirectly prove the validity of the bandwidth requirements given in Sect. \ref{sect:lens2hat}.  For this purpose we assumed the bandwidths for CMB and $\Phi$ fields to be equal and then fixed the resolution of the grid following the prescriptions of Sect. \ref{s3ect:simSpecs} assuming $\kappa=8$. We simulated CMB maps off all three Stokes parameters $T$, $Q$, and $U$ and then computed the precision of the amplitude of the power in some multipole of interest, $\tell{X}$, recovered from the simulation. The precision is defined as the fractional difference between the amplitudes obtained from two simulations performed for two considered values of $\ellmax$. For these specific tests we verified that the random realization of the harmonic coefficients used in the simulation is the same when changing the value of the bandwidth from $\ellmax$ to a value $\ellmax^{\prime}$ for $\ell\leq\ellmax$. We report the result of the numerical convergence for $\tell{X}=2000$ in Table \ref{tab:2000}. We note that the results agree with the analytic calculation of Sect. \ref{sect:lens2hat}, where we saw that extending the band limit has no visible effect on the recovered results on the scale of interest if a proper amount of power has already been convolved. As expected, a significant fraction of $E$-modes power is converted into $B$-modes for angular scales $\ell^{E}\in\left[3000,4000\right]$ but no significant improvement is seen if power beyond $\ell^{E}=4000$ is included. We also performed a test case for $\tell{X}=4000$, i.e., in the regime where the gradient approximation is expected to be less accurate, Table~\ref{tab:4000}. The $B$-modes accuracies are consistent with those derived in Sect. \ref{s3ect:simSpecs} except for the last set of bandwidth parameters, where the fractional difference between simulated spectra seems to saturate at a level of $0.1$\%. This may be related to a small residual aliasing  due to an underestimated oversampling parameter.\\*
\begin{table}[!htbf]
\centering
\begin{tabular}{ccccc}
\toprule
$\tell{X}$&$\Delta_{2000}^{3000}$&$\Delta_{3000}^{4000}$&$\Delta_{4000}^{6000}$&$\Delta_{6000}^{8000}$\\
\midrule
TT&43\% &0.04\%&0.02\% &0.003\%\\
EE&31\%&0.01\%&0.01\%&0.005\%\\
BB&35\%&3\%&0.02\%&0.004\%\\
TE&32\%&0.04\%&0.01\% & 0.002\%\\
\bottomrule
\end{tabular}
\caption{Numerical convergence of simulated lensed CMB power spectra at multipole $\tell{X}=2000$. Each column shows a fractional change in the lensed spectra amplitude due 
to an increase of the bandwidths of both the unlensed CMB and potential field, assumed here to be equal, as denoted by super- and sub-scripts of $\Delta_{\ell_{max}}^{\ell_{max}^{\prime}}$.}
\label{tab:2000}
\end{table}

\begin{table}[!htbf]
\centering
\begin{tabular}{ccccc}
\toprule
$\tell{X}$&$\Delta_{4000}^{5000}$&$\Delta_{5000}^{6000}$&$\Delta_{6000}^{7000}$&$\Delta_{7000}^{8000}$\\
\midrule
TT&31\%& 0.2\%&  0.09\%&  0.1\%\\
EE&32\% &0.2\%&0.05\%&  0.03\%\\
BB& 7\%&   4.6\%&   0.1\% & 0.09\%\\
TE& 21\%  &0.7\%&  0.2\%&  0.2\%\\
\bottomrule
\end{tabular}
\caption{As Table~\ref{tab:2000}, but for $\tell{X}=4000$.}
\label{tab:4000}
\end{table}
\subsection{Map convergence}
\begin{figure}[!htpb]
\centering
\includegraphics[width=.48\textwidth]{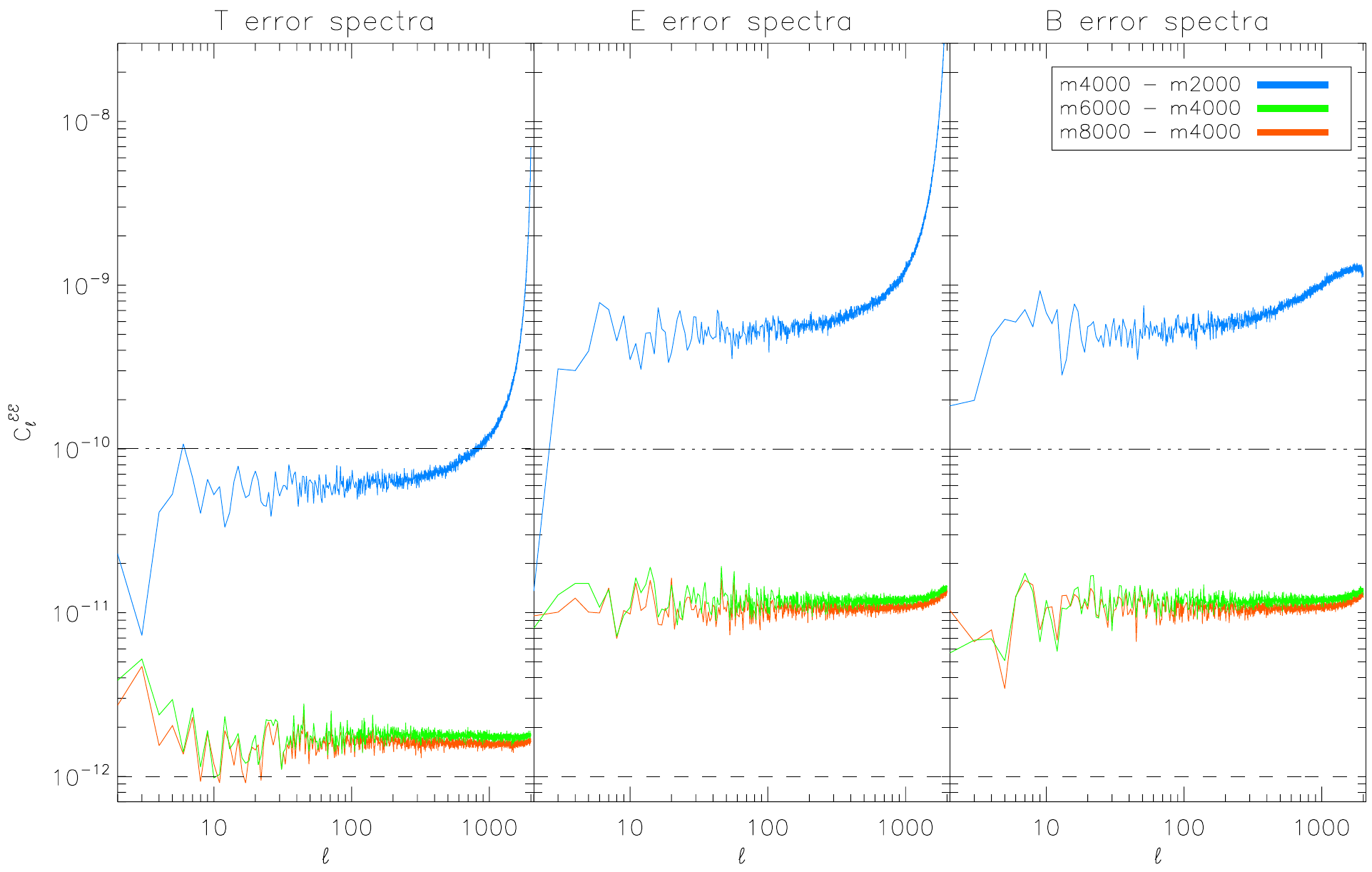}
\caption{$T$, $E$, and $B$ power spectrum of the error field obtained from maps simulated with different values of bandwidth parameters (colored lines). For reference, the dot-dashed and dashed lines show
the spectra of a white-noise process with variance equal to $0.01$ and $0.001$, respectively.
}
\label{fig:error-ps2000}
\end{figure}

After showing the convergence on the power spectrum level, which provides information on the overall variance of the simulated maps, we investigated if the convergence of our numerical result is also realized in the real domain. For this purpose we first defined an error map obtained as a difference of two maps computed assuming two different bandwidths $\ell_{max,1},\ell_{max,2}$ rescaled by the rms value of one of the two maps, i.e.,
\begin{equation}
\mathscr{E}^{X}_{\ell_{max,1},\ell_{max,2}}=\frac{\bm{m}^{X}_{\ellmax 2}-\bm{m}^{X}_{\ellmax 1}}{\sqrt{{\rm Var}\,(\bm{m}^{X}_{\ellmax 2})}} \qquad X\in\{T,Q,U\},
\label{eq:relDiffDef}
\end{equation} 
where $\bm{m}^{X}_{\ell_{max,1}}$ is a simulated map of the field $X$ obtained assuming $\ell_{max,1}$ as the bandwidth.  After deriving the harmonic coefficients with the procedure outlined in the previous section, we filtered out all modes on angular scales $\ell\geq\tell{X}$ and resampled the signal on a grid that properly samples the signal up to multipole $\tell{X}$. To take advantages of the HEALPix visualization tools, we use for this purpose an HEALPix grid having $nside=1024$ ($2048$) for $\tell{X}=2000$ ($4000$). After resampling the harmonic coefficients we computed the power spectrum of the error field $\errfield$, which demonstrates the precision obtained on the map level. In Fig. \ref{fig:errmap2000}  we report the result of this analysis for the test case $\tell{T,E,B}=2000$ . The error-field power spectrum is found to be very similar to a white noise spectra with r.m.s below $1$\%. If the power is properly resolved, an accuracy equivalent to $0.1\%$ on the map level can indeed be obtained, while the error slightly increases to $0.3\%$ for polarization (see Fig.~\ref{fig:error-ps2000}). However, this test case does not include the effect of any realistic experiment setup; in a real-life case the criterion for the convergence is set by the noise level on the pixel, if instrumental noise has to be added to the simulated maps.

The results presented above show that the systematic uncertainties inherent to the pixel-domain simulation method can be controlled with high accuracy, demonstrating that this method can provide
a sufficiently precisely framework within which to compare and study different physical assumptions entering such calculations and in particular to investigate the impact of cosmological models on the $B$-mode 
lensing predictions. We emphasize that the pixel-domain method is sufficiently general to be applicable to a range of diverse physical contexts of this kind. Even more importantly, the applicability of the considerations 
presented here goes beyond the pixel-domain method and can be straightforwardly extended to, for instance, ray-tracing approaches, which do not involve Born-approximation.

\subsection{Monte Carlo simulations}
To test whether our method produces any significant bias in the power spectrum we produced $N_{r}=100$ independent realizations of $ lensed T$, $Q$, and $U$ maps that were required to reach $0.1\%$ accuracy up to $\tell{B}=3000$ and investigated the statistical properties of the power spectra averaged over these realizations. The latter is expected to be nearly Gaussian-distributed since the non-Gaussian correlations in the lensed power- spectrum covariance induced by lensing itself are negligible for $T$, $E$ and $TE$ spectra. However, for all the power spectra including the $B$ field, we expect the latter statement to be only partially correct since the the covariance of this spectrum is non-Gaussian, especially on small angular scale. Identifying the expected scatter of the averaged spectrum with the theoretical Gaussian sample variance therefore tends to underestimate the scatter itself.\\
For each pair of the Stokes parameters, $X$ and $Y$, we define a quantity
\begin{equation}
G^{XY}_{\ell}=\sqrt{\frac{(2\ell+1)\cdot N_{r}}{\left[(\bar{C}^{XY}_{\ell})^{2}+{C}^{XX}_{\ell,th}{C}^{YY}_{\ell,th}\right]}}(\bar{C}^{XY}_{\ell}-{C}^{XY}_{\ell,th}),
\label{eq:gxy}
\end{equation}
\noindent
where the bar denotes a power spectrum averaged over $N_{r}$ realizations. The ensemble of all values of $G_{\ell}^{XY}$ is expected to be Gaussian-distributed with $0$ mean and variance $1$, which can be tested
by means of a Kolmogorov-Smirnov (KS) test. In addition, we define a reduced $\chi^{2}$ statistics, Eq.~(\ref{eq:gxy}) and following \citet{basak}, as 
\begin{equation}
\tilde{\chi}^{2}_{XY}=\sum_{\ell=2}^{\ellmax}\frac{\left | G^{XY}_{\ell}\right |^{2}}{\ellmax -1}.
\end{equation}

We report in Table~\ref{tab:statistics} the results of these two tests expressed as the significance level probability of the null hypothesis. We found that the method does not produce any significant biases on $TB$ and $EB$ cross spectra either; these were not shown in the previous analysis but are of potential interest, because they are a sensitive test of any artificially induced correlation. Moreover, the precision and accuracy of the result can be tested quite independently of analytical models by devising a custom convergence procedure as explained in the previous section.

\begin{table}
\centering
\begin{tabular}{ccc}
\toprule
$\bar{C}_{\ell}^{XY}$&Significance $P_{KS}$&Significance $ P_{\tilde{\chi}^{2}_{XY}}$\\
\midrule
TT& 0.19&0.92\\
EE& 0.97&0.65\\
BB& 0.79&0.14\\
TE& 0.58&0.84\\
TB&  0.20&0.34\\
EB&  0.71&0.67\\
\bottomrule
\end{tabular}
\caption{Results of statistical tests on the recovered lensed power spectra averaged over $100$ realizations. The significance-level probability for the null hypothesis using a Kolmogorov-Smirnov test ($P_{KS}$) and a reduced chi-square $\tilde{\chi}^{2}$ statistics ($P_{\tilde{\chi}_{XY}^{2}}$) show no bias on a statistical level.
\label{tab:statistics}
}
\end{table}

\subsection{Numerical performance and requirements}

\begin{figure*}[!ht]
\centering
\includegraphics[width=.325\textwidth]{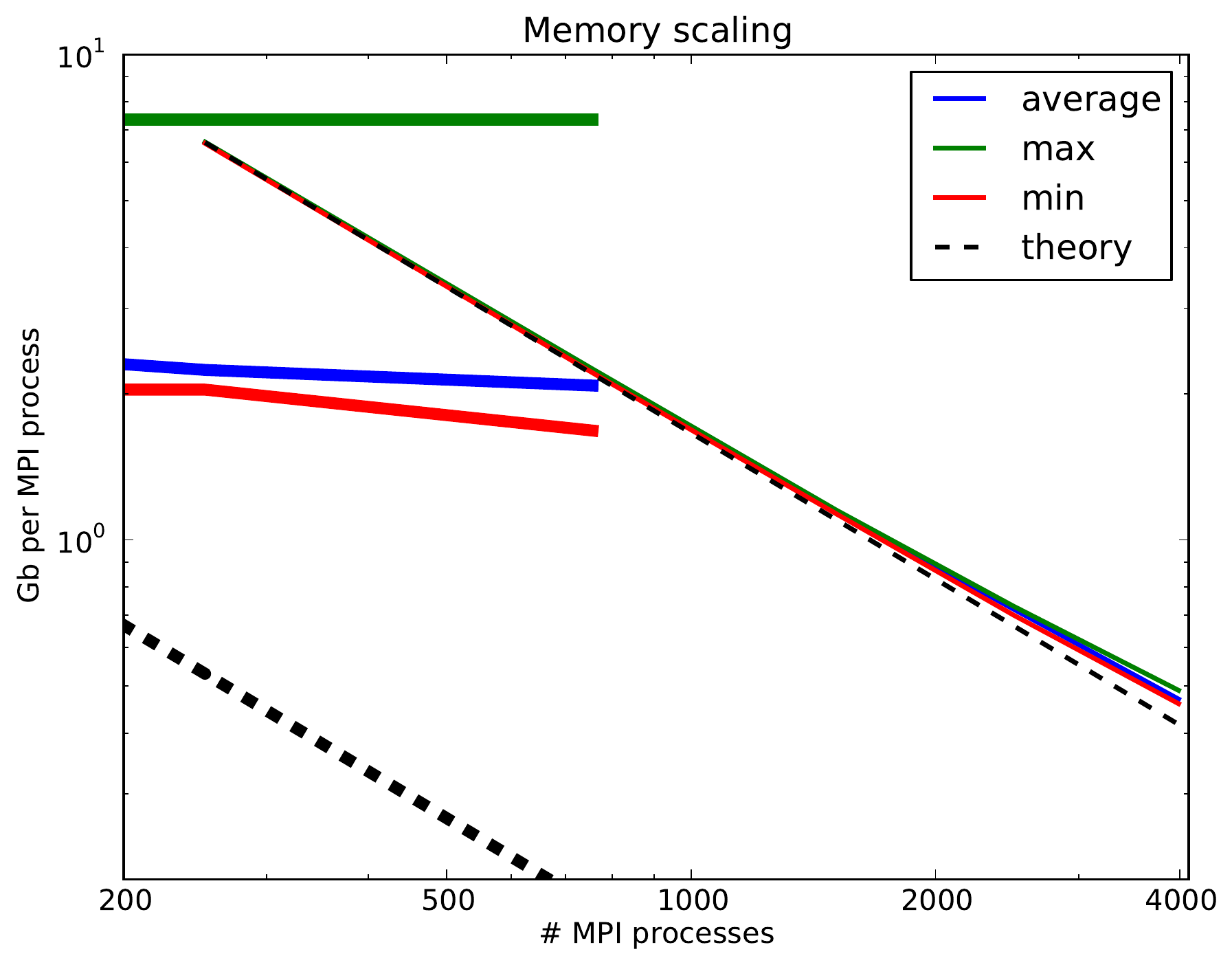}
\includegraphics[width=.325\textwidth]{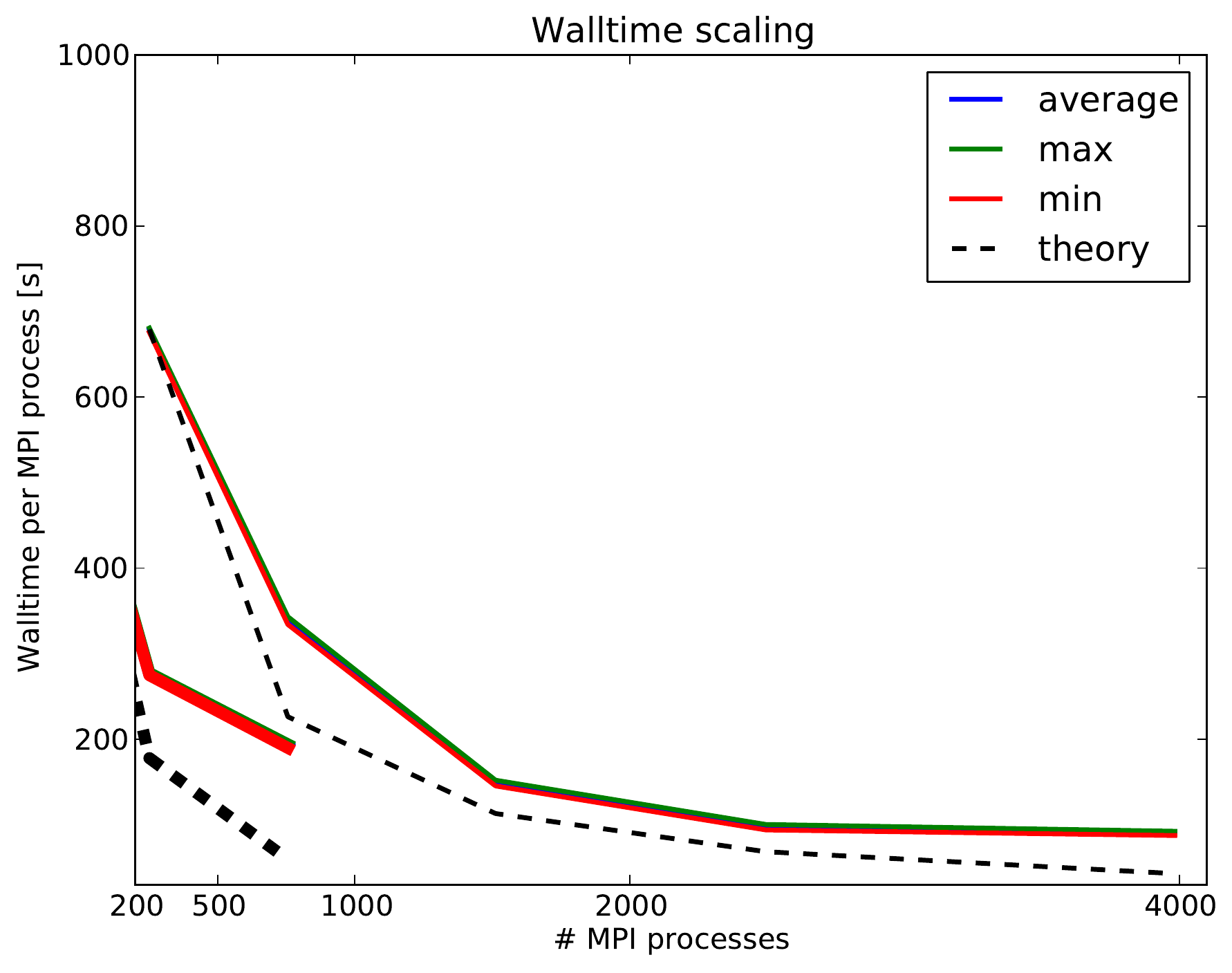}
\includegraphics[width=.325\textwidth]{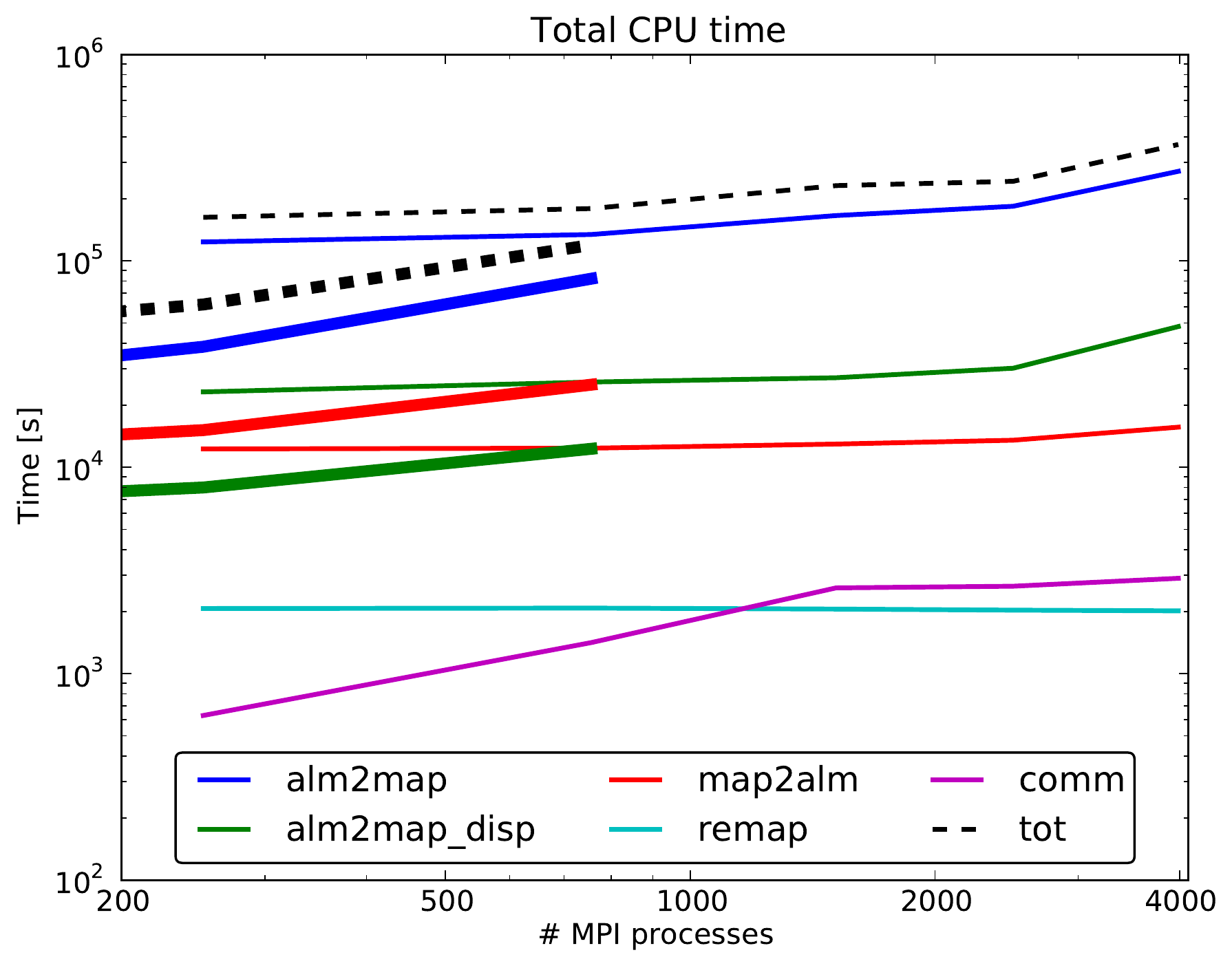}
\caption{Performance benchmarks of the \lenstwo\ code. From left to right, we show the memory, runtime, and total CPU time, summed over all MPI processes, as a function of the number of processors (or MPI tasks). In all cases the simulations have been made for three Stokes parameters and used a ECP grid of $32768^{2}$ points, $\ellmax^{CMB}=\ellmax^{\Phi}=8000$ and oversampling factor $\kappa=8$. In each panel the dashed lines show a theoretical scaling as expected in the ideal circumstances. The thick lines show the indicative results from the \lenspix\ code run on a HEALPix grid with $nside=4096$, band limit $\ellmax = 8000$ and an oversampling factor of $2$. In the right panel, the \lenspix\ {\sf alm2map} curve includes the time required to generate the oversampled ECP grid used for interpolation and the interpolation time itself. See Sect.~\ref{sec:numPerf} for a detailed discussion.}
\label{fig:scaling}
\end{figure*}

\label{sec:numPerf}

In this section we evaluate the strong scaling relations for numerical cost and memory requirements of \lenstwo, i.e., we run the code with the same parameters and test its scalability as a function of the number of MPI processes used in the calculation. For this benchmark test we used $\ellmax^{E/B}=\ellmax^{\Phi}=8000$ and a grid for lensed sky and displacement field of $32768^{2}$ pixels.\\*
The main data volume involved in the calculations is given by harmonic coefficients and maps that are evenly distributed between processors through the \stwo\ library. Their distribution is optimized for all spherical harmonics transform steps involved. The remapping method itself only depends on structures that are also distributed between processors, allowing us to preserve the scalability features inherited from the \stwo\ library. The overall memory requirements per processors for a \lenstwo\ run are on the order of $\mathcal{O}(N_{pix}/n)$, where $N_{pix}$ is the total number of pixels of both the lensed map and displacement field and $n$ is the number of MPI tasks (or processors) used for the simulation, which is assumed to be one for the physical core available on our test architecture. The prefactor varies as a function of the oversampling rate and is equal to $(3+\kappa^{2})$ for the temperature and
 to $(4+(1+\kappa^{2})n_{S\!tokes})$ for polarized cases.  We report in Figure~\ref{fig:scaling} the results of strong- and weak-scaling tests performed on the Cray XE6  Hopper cluster of the NERSC\footnote{\url{https://www.nersc.gov}} supercomputing center using the integrated performance monitoring library\footnote{\url{http://ipm-hpc.sourceforge.net/}} (IPM). The discrepancy between our model and the actual memory resource requirements per processors are due to MPI buffer allocations for collective communications and duplications of auxiliary objects describing the properties of the pixelization and observed sky patch used in the simulations as required by the \stwo\ library and remapping method. They have a size $\mathcal{O}(11(1+\kappa)\sqrt{N_{pix}})$ and accounts for nearly $25$Mb of data duplicated on each processor.  The memory-overhead of the communication part of the \lenstwo\ algorithm depends instead on the number of pixels in the local memory of each processor that is lensed on an area of the map stored on the memory of another processor. This quantity controls the size of MPI buffers, but cannot be precisely determined a priori since it depends on the specific realization of the displacement field used in the simulation.We found for this specific test case that the memory-overhead for the communication has a size of roughly 75Mb per processor.\\*
\noindent
The computational cost of our method is driven by the synthesis of the unlensed map, which is the highest-resolution object to be computed and has a number of pixels $\kappa^{2}$ higher  than the displacement field. As can also be seen from the right panel of Fig.~\ref{fig:scaling}, the runtime connected to the inverse spherical harmonic transform of the unlensed sky, despite being perfectly load-balanced, tends to flatten due to required internal communication steps and precomputations to initialize the recurrence to compute spherical harmonics. These are per se subdominant steps, but they are expected to play a role for a very fine parallelization \citep{szydlarski2011}.
The overall remapping procedure of pixel values requires a number of operation of about $\mathcal{O}(N_{pix}/n)$ and is subdominant, since it operates on a lower-resolution map, and perfectly scalable because it does not require any communication. The step involving the reconstruction of the lensed map after reshuffling the pixels (denoted as communication part in Fig.~\ref{fig:scaling}) is subdominant, but the walltime required by this step is expected to slightly grow because it potentially involves the collective communication of small amounts of data between processors and is expected to approach the latency limit for message sending.\\*

In Fig.~\ref{fig:scaling} we also mark the performance of the \lenspix\ code. However, because these two codes follow different algorithmic approaches and perform different operations to obtain a lensed map, it is not straightforward to set up a proper comparison. The presented results should therefore be viewed as merely indicative.
In this case, we have attempted to set the code parameters to obtain an accurate spectrum up to $\ell \simeq 5000$. We assumed the same bandwidth for the \lenspix\ runs as for \lenstwo, i.e.,  $\ellmax=8000$, and used the lowest resolution capable of supporting the corresponding harmonic modes on a HEALPix grid, setting $nside=4096$. This value may be somewhat on the low side given the increase of the bandwidth due to the lensing. \lenspix\ also requires as an input an oversampling parameter that defines the unlensed sky resolution in the ECP pixelization. We chose this parameter to be $2$ because this is a value commonly used and has been reported to be sufficient to produce accurate results \citep[e.g.][]{benoit-levy2012}. With this choice of the input parameters we find that  the \lenspix\ code displays a better performance in terms of the runtime for an intermediate number of employed MPI-process, but the gain is quickly offset by the superior scaling properties of the \lenstwo\ code and its ability to employ many processors. Moreover, for the sake of comparison, no bandwidth optimization
procedure was applied here, which would result in about a factor $\sim 1.4$ improvement in the \lenstwo\ runtimes. We note that the memory and communication bottlenecks prevented us from successfully running \lenspix\ on more than $\sim 800$ MPI processors of our computational platform. 
The performance of \lenstwo\ can be further improved by performing a simultaneous, multi-map analysis (see Appendix \ref{appendix:code}), made
feasible thanks to its low memory overhead and near perfect scalability of the memory requirements.
However, as they are, the two codes seem to be complementary and to address the needs of different computational platforms.

\section{Conclusions}
We have investigated and clarified details of modeling and simulations of the gravitational lensing effect on CMB. We particularly aimed at elucidating the role and  impact of bandwidths 
of considered signals on the precision of the pixel-domain approaches~\citep[e.g.,][]{lewis2005} to simulating the lensing effect on polarized anisotropies, paying special attention to recovering of
the $B$ component. These bandwidths play a crucial role in ensuring a sufficient accuracy of the produced lensed maps and need to be carefully taken into account if numerical effects such
as power aliasing are to be kept under control.
We developed a semi-analytic approach based on the formalism of \cite{hu2000} to study these effects, and with its help investigated the necessary requirements for the signal's bandwidths.
In particular, we found out that the simple convolution picture, where the convolution kernel has
a limited width of at most few hundred in $\ell$ space due to the gravitational potential, though it works very well for the total intensity, $T$, and $E$ polarization up to $\tell{T/E} \simlt 2000$, is adequate neither for much smaller 
angular scales in these two cases nor for the $B$-mode signal. Instead, the proposed semi-analytic formalism should be used to guide a selection of the simulation parameters to ensure
the final precision of the result, but also to optimize the computational time.
We point out that the accuracy considerations we presented are sufficiently generic that they should be applicable to other CMB lensing simulation techniques providing sound
guidelines for choices of suitable parameters, that these techniques involve. For the pixel-domain-based methods, our main object of study here, we find that sufficiently high precision can 
indeed be ensured and permits meaningful simulations of small effects due to different physical assumptions.

Furthermore, we validated our semi-analytic results with the help of extensive numerical computations, for which we developed a simple, massively-parallel lensing
simulation code, \lenstwo. The code uses an extremely simple but robust approach to the interpolation, involving sky overpixelization and a simple NGP assignment scheme, which, as we showed, leads to easily understandable and controllable numerical effects. These effects are minimized because the code, thanks to its efficient parallelization, permits analyses of extremely large sky maps with very dense sky grids/pixelization. In this way the simulated sky power can be resolved all the way to its actual bandwidths, which are carefully kept track of in the code.

The developed code, \lenstwo, is suitable for massively parallel computational platforms, with either shared or distributed memory. It displays nearly perfect scalability in terms of runtime and 
allocated memory per processor up to the maximal number of CPUs it can employ. This last is determined by the lowest value of the band-limit parameters for either the CMB or the displacement field that is to be used in the runs, $n_{proc}^{max} = \min(\elle_{max},\ellphi_{max})/2$. It therefore permits extensive simulations involving hundreds of simulated maps in a reasonable time. The major bottleneck of the code performance is due to the need
of calculating a single inverse spherical harmonic transform which is required to obtain the overpixelized map of the unlensed signal. This can certainly be alleviated further by using better algorithms
and/or numerical implementations, e.g., capitalizing on hardware accelerators such as GPGPU~\citep[][]{hupca2011, szydlarski2011, fabbian2012, reinecke2013}. We leave these code optimizations for future work.
The code and its forthcoming version will be publicly available.

\begin{acknowledgements}
This work has been supported in part by French National Research Agency (ANR) through COSINUS program (project MIDAS no. ANR-09-COSI-009).
GF acknowledges the support of Universit\`a degli Studi di Milano Bicocca and CARIPLO Foundation through the EXTRA program for the preliminary phase of this work.
This research used resources of the National Energy Research Scientific Computing Center, which is supported by the Office of Science of the U.S. Department of 
Energy under Contract No. DE-AC02-05CH11231. We acknowledge use of the publicly available software tools and libraries HEALPix, SPRNG, IPM, S$^2$HAT, CAMB, and LensPix.
\end{acknowledgements}

\bibliographystyle{aa}
\bibliography{citations}

\appendix
\section{Precision of the accuracy formula}\label{appendix:accuracy_precision}
The harmonic approximation of \cite{hu2000}, which we used throughout the paper, is known to reproduce the lensed CMB spectra with an accuracy of only a few percent~\citep{challinor2005}. 
In this appendix we discuss the validity of the definition of the approximate accuracy function, Eq.~\ref{Xaccuracydef}, which is based on this approximation.\\* 
Assuming that we have access to the exact 2D convolution kernels instead of the one derived with the gradient approximation, such that
\begin{equation}
C_{\ell}^{X,exact}= \sum_{\ell_{*}^{\Phi}=0}^{\infty}\sum_{\ell_{*}^{X}=0}^{\infty}{\cal K}^{exact}_{\tell{X}}(\ell_*^{\,X}, \ell_*^{\,\Phi}) ,
\end{equation}
we can express the exact accuracy formula as
\begin{equation}
A_{\ell}^{X,true}(\ellphi,\ellx)=1-\frac{\alpha^{\ell}_{\ellx, \ell^{\Phi} }+\zeta_{\ellx, \ell^{\Phi}}}{\alpha^{\ell}_{\ell_{max}^{X}, \ell^{\Phi}_{max} } +\epsilon},
\end{equation}
where  $\alpha^{\ell}_{\ellx, \ell^{\Phi}}$ and $ \alpha^{\ell}_{\ellx_{max}, \ell^{\Phi}_{max}}$ correspond to the numerator and dominator in Eq.~\ref{Xaccuracydef},
and the two extra terms that quantify the corresponding, effective errors, which themselves may depend on the cutoff assumed in the computation of both the numerator and denominator, are given as
\begin{eqnarray}
\epsilon&=&\sum_{\ell_{*}^{\Phi}=0}^{\infty}\sum_{\ell_{*}^{X}=0}^{\infty}{\cal K}^{exact}_{\tell{X}}(\ell_*^{\,X}, \ell_*^{\,\Phi}) -\alpha^{\ell}_{\ellx_{max}, \ell^{\Phi}_{max}}\\
\zeta_{\ellx, \ell^{\Phi}} &=& \sum_{\ell_{*}^{\Phi}=0}^{\ell^{\Phi}}\sum_{\ell_{*}^{X}=0}^{\ellx}{\cal K}^{exact}_{\tell{X}}(\ell_*^{\,X}, \ell_*^{\,\Phi})-\alpha^{\ell}_{\ellx, \ell^{\Phi}}.
\end{eqnarray} 
Hereafter, we assume that the absolute cutoff for the CMB and lensing potential, $\ellmax$, is chosen sufficiently high so that all the relevant
power is included when computing the considered lensed multipole.
Because the accuracy of the harmonic expansion $\beta\equiv \epsilon/\alpha^{\ell}_{\ell_{max}^{X}}$ is on the order of percent, we can Taylor-expand the previous expression to the first order in $\beta$, i.e.,
\begin{eqnarray}
A_{\ell}^{X,true}(\ellphi,\ellx)&\approx&1- \frac{(\alpha^{\ell}_{\ellx, \ell^{\Phi} }+\zeta_{\ellx, \ell^{\Phi}})}{\alpha^{\ell}_{\ell_{max}^{X}, \ell^{\Phi}_{max} }} (1-\beta) +\mathcal{O}(\beta^{2})\\ \nonumber
&=&A_{\ell}^{X}(\ellphi,\ellx) +\beta A_{\ell}^{X}(\ellphi,\ellx) +\beta -\frac{\zeta_{\ellx, \ell^{\Phi}}}{\alpha^{\ell}_{\ell_{max}^{X}, \ell^{\Phi}_{max} }}.
\end{eqnarray}
From now on, we denote the last term on the rhs as $\delta_{\ellx,\ell^{\Phi}}$. We can then rewrite the precision of  the accuracy function as
\begin{eqnarray}
\left |\frac{\Delta A_{\ell}^{X}(\ellphi,\ellx)}{A_{\ell}^{X}(\ellphi,\ellx)}\right | &\approx&\left |  \frac{1}{A_{\ell}^{X}(\ellphi,\ellx)}\left(\beta\left(1- A_{\ell}^{X}(\ellphi,\ellx) \right)- \delta_{\ellx,\ell^{\Phi}}\right) \right |  \nonumber\\
&\approx&\left |\frac{\left(+\beta - \delta_{\ellx,\ell^{\Phi}}\right) + \mathcal{O}(\beta^{2})}{A_{\ell}^{X}(\ellphi,\ellx)}\right |,
\label{eq:accuracy}
\end{eqnarray}
where we have assumed that the accuracy function is at most $\mathcal{O}(\beta)$ in the regime of interest here.\\*
The overall precision of the accuracy function, as expressed by Eq.~\ref{eq:accuracy}, is then driven by the difference between the two terms, which by construction tend to cancel each other because $\zeta_{\ellx, \ell^{\Phi}}$ 
goes to $\epsilon$ as $\ellx, \ell^{\Phi}$ approach $\ellmax$. The formula therefore becomes more and more accurate as we approach the cutoff limit.

\section{\lenstwo\ code}\label{appendix:code}
The code outline follows the general simulation guidelines discussed in Sect. \ref{s3ect:basics}, but we detail here several features of potential interest of the code structure.
\begin{enumerate}
\item When generating a Gaussian random realization of a harmonic coefficients for the unlensed CMB and the displacement field, both the correlation between temperature or $E$-modes and the displacement field generated by the Sachs-Wolfe effect can be taken into account if requested. However, since both are negligible for most multipoles, we neglected them in the runs performed for this paper. We do not expect this correlation to affect the results of our analysis, especially in the high $\ell$ tail of the spectrum because the correlation is confined to large angular scales. \\*
\item The effects of nonlinear LSS evolution, which consequently affect the lensing potential, are naturally taken into account in the code if they are included in an effective lensing potential power spectrum. Even though nonlinear evolution of matter perturbations induces non-Gaussianities in the matter power spectrum, the contributions of higher-order statistical moments to the lensing potential have been proven to be on the subpercent level \citep{merkel2010}.The assumption of a purely Gaussian lensing potential is thus well applicable and usually sufficient for this kind of simulations. As an alternative, the code can accept pre-computed maps of the potential on the input, which can be therefore arbitrarily non-Gaussian, and which will be used to produce the displacement field.\\*
\item Since harmonic coefficients are distributed between processors and generated directly in a distributed way, we used the scalable parallel random number generator library\footnote{\url{http://sprng.cs.fsu.edu}} (SPRNG) to avoid correlation between random number streams on each processors. \\*

\item The computation of the displaced coordinates and the remapping of the pixel locations are managed by two separate routines, one optimized for grids with equidistant rings (e.g. ECP) and the other developed for any pixelization conforming with the requirements imposed by \stwo. Since maps are distributed between processors, it can happen that the remapping procedure on a given processor identifies the required displaced pixel to be located in a region of the sky map that is not stored in the local memory. For this reason the code has to manage pixel indexing using two different indexing scheme (one for the full-sky map and one for the chunk of the full map stored locally) and has to be able to switch from one to the other. To performed this operation efficiently we have to allocate on each processor an auxiliary bi-dimensional array, that encodes the indices of the pixels required by the processors and that are not present in its memory and the processors on which these are effectively located. The total volume of this structure is therefore equal to that of the part of the lensed sky stored locally and constitutes the only memory overhead required by the remapping procedure.\\*

\item A collective {\tt MPI\_All2allv} communication step is performed to redistribute the information on pixels, that are needed by a processor to build the final lensed map, but is not stored in its local memory. This pattern ensures an even distribution of memory between all cores and a very good scalability up to several thousand MPI processes. On the numerical level the communication time is subdominant, although it can in principle be further optimized with nonblocking MPI local communication calls or by exploiting an hybrid MPI/OpenMP approach.\\*

\item  The code can perform simulations in an arbitrary pixelization scheme that meets the \stwo\ requirements. Though we have found that ECP is preferred for the internal computation, the output results can be delivered
 on a grid selected by the user, e.g. the HEALPix grid.\\*

\item The code supports simultaneous multi-map analysis on the spherical harmonics step of the algorithm, when memory available for a given processor is sufficient. In this case, the gain in the runtime of the code is roughly equal to the number of maps processed at the same time. This option makes the code particularly appealing for the data analysis steps involving massive use of Montecarlo realizations of lensed CMB maps.
\end{enumerate}
\end{document}